\documentclass[11pt, a4paper]{article}

\usepackage[affil-it]{authblk}

\usepackage[utf8x]{inputenc}
\usepackage{amsmath}
\usepackage{amsthm}
\usepackage{amssymb}
\usepackage{graphicx}
\usepackage{subfigure}
\usepackage[colorlinks=false]{hyperref}
\usepackage[nodisplayskipstretch]{setspace}
\usepackage{color}
\usepackage{verbatim} 
\usepackage{setspace}
\usepackage{extsizes}
\usepackage{anysize} 
\usepackage{chngcntr}
\usepackage{lineno}
\usepackage{fancyhdr}
\marginsize{1.5cm}{1.5cm}{0.8cm}{0.8cm}
\counterwithin{equation}{section}

\newtheorem{proposition}{Proposition}[section]
\newtheorem{remark}{Remark}[section]
\newtheorem{theorem}{Theorem} 

\providecommand{\keywords}[1]{\textbf{\textit{Index terms---}} #1}

\theoremstyle{plain} 
\newcommand{\thistheoremname}{}
\newtheorem{genericthm}[theorem]{\thistheoremname}



\begin{document}
\title{Final-state, Open-loop Control of Parabolic PDEs with Dirichlet Boundary Conditions\footnote{This is a revised version of \url{
https://doi.org/10.48550/arXiv.2209.00683} which encompasses a broader class of PDEs.}}

\author{Gilberto O. Corr\^ea%
  \thanks{email: \texttt{gilberto\emph{{@}}lncc.br} - Corresponding author.}}
\affil{Laborat\'{o}rio Nacional de Computa\c{c}\~{a}o Cient\'{i}fica -- LNCC/MCTIC,
Petr\'{o}polis, RJ, Brazil}
{\author{Marlon M. L\'{o}pez-Flores%
  \thanks{email: \texttt{mmlf\emph{{@}}econ.puc-rio.br.}}}
\affil{Dep. de Economia, Pontificia Universidade Cat\'{o}lica do Rio de Janeiro -- PUC-RJ, Brazil.}}
\author{Alexandre L. Madureira%
  \thanks{email: \texttt{alm\emph{{@}}lncc.br}.}}
\affil{Laborat\'{o}rio Nacional de Computa\c{c}\~{a}o Cient\'{i}fica -- LNCC/MCTIC,
Petr\'{o}polis, RJ, Brazil}

\date{}

\maketitle

\begin{abstract}
In this paper, a quadratic optimal control problem is considered
for second-order parabolic PDEs with homogeneous Dirichlet
boundary conditions, in which the ``point'' control function (depending only on
time) constitutes a source term. These problems involve choosing a
control function (with or without ``peak-value'' constraints) to
approximately steer the solution of the PDE in question to a desired
function at the end of a prescribed (finite) time-interval. To compute
approximations to the desired optimal control functions, semi-discrete,
Galerkin approximations to the equation involved are introduced
and the corresponding (approximating) control problems are tackled. It is
shown that the sequences of solutions to both the constrained and
unconstrained approximating (finite-dimensional) control problems
converge, respectively, to the optimal solutions of the control problems
involving the original initial/boundary value problem. The solution to
the unconstrained approximating problem can be quite explicitly
characterized, with the main numerical step for its computation
requiring only the solution of a Lyapunov equation. Whereas approximate
solutions to the constrained control problems can be obtained on the
basis of Lagrangian duality and piecewise linear multipliers. These
points are worked out in detail and illustrated by numerical examples involving the heat equation (HEq).
\end{abstract}



\section{Introduction}
Control problems for systems described by linear evolution equations (\emph{essentially}, equations involving partial derivatives relative to time and spatial coordinates) have received considerable attention (see for example ([4], [11], [12]) and their references).

In particular, the basic objective of approximately reaching a desired final state from a given initial state has given rise to various control problems (in open loop) for parabolic equations. Such problems can encompass different types of boundary conditions (Dirichlet, Neumann or Robin) and control signals acting on the boundary of the spatial domain involved or as a source term inside it.

Usually, these problems aim at obtaining a control function defined both in a prespecified time interval and in the spatial domain in which the equation is defined ,\emph{i.e.}, at each instant the control ``signal'' assumes as ``value"  a function defined in the entire spatial domain in question.

On the other hand, with a view to potential applications, it is interesting to consider the
case of ``point control'', \emph{i.e.}, of time-dependent control functions which 
assume as ``value" \ a point in $\mathbb{R}^{^{m}}$ (for some fixed $m$) at each 
instant, whose spatial action is defined by the ``actuators" \ used. 

In this work, a quadratic optimal control problem is considered for a class of second-order parabolic PDEs with a homogeneous Dirichlet boundary condition, in which
the control function (depending only on time) appears in a source term.

More specifically, the problem of approximate positioning of the final state on a 
finite time interval is examined. Minimization of a quadratic cost involving the final-state approximation error is 
considered with and without a constraint on the maximum magnitude of the control
functions. To compute approximate solutions to such control problems, approximate 
versions of them are tackled which are obtained from finite-dimensional 
approximations for the control-to-final-state operator. 

This report is organized as follows. In Section \ref{sec:02}, the basic control problem is 
introduced. In Section \ref{sec:03}, its optimal solution is characterized. In Section \ref{sec:04}, 
approximate solutions to the basic, unconstrained control problem are derived, on the basis of Galerkin approximations to the equation involved, and their convergence to the solution of the original problem is established. In
Section \ref{sec:05}, ``peak value'' constraints are added to the basic problem and both the 
original and approximate versions of it are discussed, including the use of 
Lagrangian duality to obtain approximate solutions. Finally, in Section \ref{sec:06}, two 
simple numerical examples are presented to illustrate the main points previously 
discussed. Proofs are to be found in the Appendix.

\section*{Notation}
\begin{itemize}
\item $q$ - dimension of the spatial domain.
\item $\Omega \subset \mathbb{R}^{^{q}}$ - spatial domain.
\item $\underline{k}=\left(k_{_{1}},\cdots,k_{q}\right)$ - multiindex.
\item $\nu_{\underline{k},i}=\left(\dfrac{k_{_{i}}\pi}{L_{x_{_{i}}}}\right)^{^{2}}$.
\item $\underline{\theta}_{r}: \Omega\rightarrow \mathbb{R}$ - desired final state.
\item $t_{_{F}}$ - final instant.
\item $\boldsymbol{u}: \left[0, t_{_{F}}\right]\rightarrow \mathbb{R}^{^{m}}$ - control signal, $m \in \mathbb{Z}_{+}$.
\item $L_{2}\left(\Omega\right)$ - set of real square-integrable functions defined on $\Omega$.
\item $L_2\left(0,t_{_F}\right)^{^m}$ - set of square-integrable functions defined on $\left(0, t_{_F}\right)$ taking values in $\mathbb{R}^{^m}$.
\item $H_{0}^{1} \left(\Omega\right)$ - the set of all locally summable functions $f:\Omega \to \mathbb{R}$ such that $f\in L_{2}(\Omega)$, for $i=1, \mathellipsis,m$, $f_{x_{i}}$ exists in the weak sense, $f_{x_{i}} \in L_{2}(\Omega)$ and $f(\boldsymbol{x})=0$ on the boundary of $\Omega$ ([3], Section 5.2.2).
\item $\underline{\theta}\left(t\right)$ or $\theta\left(\cdot,t\right)$ - real functions defined in $\Omega$ (for every $t \in (0, t_{_{F}}]$).
\item $\mathcal{T}_{_{\theta}}\left[\boldsymbol{u}\right]: \Omega \rightarrow \mathbb{R}$ - final state reached by the action of $\boldsymbol{u }$ from state zero.
\item $\mathbb{R}$ and $\mathbb{Z}_{+}$ - set of real numbers and positive integers, respectively.
\end{itemize}
\section{Background and Problem Formulation}\label{sec:02}

Linear-quadratic optimal control problems for partial differential equations have been extensively studied -- see, for example ([10], [12]) and references therein. Very often, general parabolic equations and more general cost-functionals involving state values along the whole of $[0, t_{_{F}}]$ are considered. To cope with such general set-ups, results tend to concentrate on showing existence of optimal controls and establishing ``abstract''\ optimality conditions (rather that computational schemes to compute control signals). This is often achieved invoking advanced general methods such as the so called \emph{Hilbert Uniqueness Method} (HUM for short) devised by Lions in ([6], [7]).

In contrast, the main objective here is to exploit a simpler set-up (``point'' control functions and final-state control) to obtain, by elementary
means, explicit characterizations of approximate optimal control signals which would only involve relatively simple computational tasks -- with the end result that the desired ``approximately-optimal''\ control signals could be effectively generated.

To this aim, consider a initial/boundary condition problem for the parabolic equation given (``in its  classical form") by
\begin{eqnarray}
\frac{\partial \theta(\boldsymbol{x},t)}{\partial t} &+&L[\theta(\cdot, \cdot)](\boldsymbol{x}, t)=f(\boldsymbol{x},t) \ \ \ \ \ \ \ \ \ \ \
 \ \ \ \ \ \ \ \ \ \ \ \ \  \forall \ \boldsymbol{x}\in \Omega,  \ \forall \ t\in (0,t_{F} )\ \label{ch4:eq-01}\\
 \theta (\boldsymbol{x},t)&=&0 \ \ \ \ \ \ \ \  \  \ \ \ \  \ \   \ \ \ \  \ \ (\text{Boundary Conditions}) \ \ \forall  \ t\in [0,t_{F} ],
 \ \forall \boldsymbol{x}\in \partial \Omega \label{ch4:eq-02}\\
\theta (\boldsymbol{x},0)&=&g(\boldsymbol{x}) \ \ \ \ \ \ \ \ \ \ \ \ \ \  \ \ \ \ \ \  \ (\text{Initial Condition})   \ \ \  \ \ \ \ \  \ \ \ \ 
\ \ \  \ \ \ \ \ \ \ \ \forall \boldsymbol{x}\in \Omega \label{ch4:eq-03}
\end{eqnarray}
where $\Omega\in  \mathbb{R}^{m_{_{x}}}$ is a bounded, open and connected set, $L[\theta]=-\sum_{i,j=1}^{m_{\boldsymbol{x}}}(a^{ij}\theta_{x_{i}})_{x_{j}}+\sum_{i=1}^{m_{\boldsymbol{x}}}b^{i}\theta_{x_{i}}+c\theta$, $a^{ij}$,$b^{i}$, $c$, $f$ and $g$ are given functions and $\alpha \in \mathbb{R}_{+}$.
The ``weak'' (or variational) version of this problem is then formulated as follows:

Given $\alpha \in \mathbb{R}_{+}$, $f \in L_{2}(\Omega \times (0, t_{F}])$, $g\in L_{2}(\Omega)$, $a^{ij}$, $b^{i}$ and $c\in L_{\infty}(\Omega)$, $a^{ij}=a^{ji}$, find $\underline{\theta}:[0, t_{_{F}}]\rightarrow H_{0}^{1}(\Omega)$ such that $\forall \phi \in H_{0}^{1}(\Omega)$ and $\forall t \in (0, t_{_{F}})$
\begin{eqnarray}
 \left\langle\frac{d\underline{\theta}}{dt}(t), \phi\right\rangle&+&
 \mathbf{B}[\phi, \psi]= \langle \underline{f}(t), \phi\rangle,\label{ch4:eq-04}\\
 \langle\underline{\theta}(0), \phi \rangle&=&\langle g, \phi\rangle,\label{ch4:eq-05}
\end{eqnarray}
where $\mathbf{B}[u,v]=\displaystyle\int_{\Omega}\{\sum_{i,j=1}^{m_{\boldsymbol{x}}}a^{ij}u_{x_{i}}v_{x_{j}}+\sum_{i=1}^{m_{\boldsymbol{x}}}b^{i}u_{x_{i}}v+cuv\}d\boldsymbol{x}$, for $u,v \in H_{0}^{1}(\Omega)$, $\underline{f}(t)=f(\cdot, t)$. The existence and uniqueness of solutions to this problem follows from the result of ([3], Theorem 7.1.3, p. 356).

Given that the main interest here is the final-state control problem, the semigroup representation of the solution to
(\ref{ch4:eq-01})--(\ref{ch4:eq-03}), see ([1], pp. 13--52), will be exploited. To bring in such a representation, let the closed operator $A:\operatorname{dom}(A)\rightarrow L_{2}(\Omega)$ be defined by
\begin{equation}\label{ch4:eq-06}
\forall \phi \in \operatorname{dom}(A), \forall \psi \in H_{0}^{1}(\Omega),\ \ \ \langle -A[\phi ],\psi \rangle = \mathbf{B}[\phi , \psi],
\end{equation}
where  $\operatorname{dom}(A)=\left\{\phi \in H_{0}^{1}(\Omega): \mathbf{B}[\phi, \cdot]\in L_{2}(\Omega)\right\}$, $\langle\cdot, \cdot\rangle$ denotes the inner product of $L_{2}(\Omega)$ and  $\operatorname{dom}(A)$ stands for the domain of $A$. It is assumed that $\mathbf{B}[\cdot, \cdot]$ statisfies Garding's inequality (see, [10], Section 5.2). The operator $A$ so defined is the infinitesimal generator of a $C_{o}-$semigroup
$S_{_{A}}(t):L_{2}(\Omega)\rightarrow L_{2}(\Omega)$,
$t\geq0$ on the basis of which $\underline{\theta}(\cdot)$ is given by
\begin{equation}\label{ch4:eq-09}
\underline{\theta}(t;f,g)=S_{_{A}}(t)[g]+\int^{t}_{0}S_{_{A}}(t-\tau )[\underline{f}(\tau)]d\tau, \ \ \forall t\in [0,t_{_{F}}],
\end{equation}
(see [1], pp. 101--107).

It is now assumed that $f(\boldsymbol{x},t)=f_{_{S}}(\boldsymbol{x},t)+
\boldsymbol{\beta}_{_{\boldsymbol{S}}}(\boldsymbol{x})^{^{\mathrm{T}}}\boldsymbol{u}(t)$,
where $f_{_{S}}:\Omega\times[0, t_{_{F}}]\rightarrow \mathbb{R}$ and $\boldsymbol{\beta}_{_{\boldsymbol{S}}}:\Omega\rightarrow\mathbb{R}^{^{m}}$ are given 
functions, $\boldsymbol{\beta}_{_{\boldsymbol{S}}}(\boldsymbol{x})=[\boldsymbol{\beta}_{_{\boldsymbol{S}1}}(\boldsymbol{x})\ \cdots\ \boldsymbol{\beta}_{_{\boldsymbol{S}m}}(\boldsymbol{x})]^{^{\mathrm{T}}}$,   where $f_{_{S}}$ would model ``disturbances''\ (\emph{i.e.}, control-independent heat sources) and 
$\boldsymbol{u}:[0,t_{_{F}}]\rightarrow \mathbb{R}^{^{m}}$  is a control signal to be chosen in such a way as to make\ \
$\underline{\theta}(t_{_{F}};f,g)$ ``close''\ to a prescribed 
$\theta_r\in L_{2}(\Omega)$, with $\boldsymbol{\beta}_{_{\boldsymbol{S}}}^{^{\mathrm{T}}}(\boldsymbol{x})$ representing 
the spatial effects (and position) of the ``point control'' function $\boldsymbol{u}(\cdot)$. The control 
function\ $\boldsymbol{u}$ \ is such that $\boldsymbol{u}(t) \in \mathbb{R}^{^{m}}$, \emph{i.e.}, 
$\boldsymbol{u}(t)=\left(u_{_{1}}(t), \ldots, u_{_{m}}(t)\right)$, 
where each\ $u_{_{i}}$\ is a control signal for the individual source given by
$\boldsymbol{\beta}_{_{\boldsymbol{S}i}}(\boldsymbol{x})u_{_{i}}(t), i=1, \ldots,m$. 
 
Now, let $\boldsymbol{u}\in L_{2}(0,t_{_{F}})^{^{m}}$, $\rho_{_{\boldsymbol{u}}}\in \mathbb{R}_{+}$ and define the cost functional
\begin{equation}\label{4eq:08}
\mathcal{J}(\boldsymbol{u})\triangleq \|\underline{\theta}(t_{_{F}};f,g)-\theta_{r}\|^{^{2}}_{_{L_{2}(\Omega)}} 
+ \rho_{_{\boldsymbol{u}}}\|\boldsymbol{u}\|^{^{2}}_{_{L_{2}(0,t_{_{F}})^{^{m}}}} 
\end{equation}
(from now on, the ``space"\  subindices of norms and inner products will be omitted whenever context information makes them redundant). 

The term $\left\|\underline{\theta}(t_{_{F}};f,g)-\theta_{r}\right\|_{_{L_{2}\left(\Omega\right)}}^{^{2}}$
measures the proximity of the system's optimal final state under the effect of the control and the desired 
state (objective) which is to be approximated. 

The ``energy'' that the control $\boldsymbol{u}$ requires to take the system to the desired final state in a finite 
interval of time $(0, t_{_{F}})$ is measured by $\left\|\boldsymbol{u}\right\|^{^{2}}_{_{L_{2}(0, t_{_{F}})^{^{m}}}}$.
By varying the parameter $\rho_{_{\boldsymbol{u}}}$ that penalizes this term, different ``trade-offs'' between ``cost of control'' (or regularization ``level'') and approximation quality can be pursued.

A control signal is to be chosen on the basis of the optimization problem
\begin{equation}\label{4eq:09}
\text{\underline{\emph{Prob.\ $I:$}}} \displaystyle\min_{\boldsymbol{u}\in L_{2}(0,t_{_{F}})^{^{m}}}\mathcal{J}(\boldsymbol{u}).
\end{equation}
Moreover, the cost functional $\mathcal{J}(\boldsymbol{u})$ is a
\emph{convex, continuous and coercive functional}, This, together with the fact that $L_{2}(0, t_{_{F}})^{^{m}}$ is closed and convex guarantees the existence
of a function $\boldsymbol{u}$  that minimizes it ([2], pp. 35--36). 
\section{Final State Positioning with Source Control}\label{sec:03}

In this section, optimality conditions are presented for \emph{Prob. I} on the basis of which its solution can be explicitly 
characterized. To this effect, note first that due to the 
linearity of $\underline{\theta}(\cdot ;f,g)$ on $(f,g)$,
\begin{equation}\label{4eq:10}
\underline{\theta}(\cdot ;f,g)=\underline{\theta}(\cdot ;f_{_{S}},g)+
\underline{\theta}(\cdot ;f_{_{\boldsymbol{u}}},0),\ \mbox{where} \ f_{_{\boldsymbol{u}}}(t)=
\boldsymbol{\beta}_{_{\boldsymbol{S}}}^{^{\mathrm{T}}}(\cdot)\boldsymbol{u}(t),
\end{equation}
\emph{i.e.},
\begin{equation}\label{4eq:11}
\underline{\theta}(\cdot ;f,g)=\underline{\theta}(\cdot ;f_{_{S}},g) +\check{{\mathcal{T}}}_{_{\theta}}[\boldsymbol{u}](\cdot),
\end{equation}
where $\check{{\mathcal{T}}}_{_{\theta}}:L_{2}(0,t_{_{F}})^{^{m}}\rightarrow \left\{\underline{h}:[0, t_{_{F}}] \rightarrow H_{0}^{1}(\Omega)\right\}$
\begin{equation}\label{4eq:12}
\check{\mathcal{T}}_{_{\theta}}[\boldsymbol{u}](t)\triangleq
\int^{t}_{0} S_{_{A}}(t-\tau )[f_{_{\boldsymbol{u}}}(\tau )]d\tau.
\end{equation}
Note also that $\mathcal{J}(\boldsymbol{u})$ can be rewritten as
\begin{equation}\label{4eq:13}
\mathcal{J}(\boldsymbol{u})=\| \mathcal{T}_{\theta }[\boldsymbol{u}]-\theta_{r\mathrm{o}}\|^{^{2}}_{_{L_{2}(\Omega)}}+\rho_{_{\boldsymbol{u}}}\| \boldsymbol{u}\|^{^{2}}_{_{L_{2}(0,t_{_{F}})^{^{m}}}},
\end{equation}
where $\theta_{r\mathrm{o}}\triangleq \theta_{r} - \underline{\theta}(t_{_{F}};f_{_{s}},g)$ and
$\mathcal{T}_{_{\theta}}:L_{2}(0, t_{_{F}})^{^{m}}\rightarrow L_{2}(\Omega)$ is defined by 
$\mathcal{T}_{_{\theta}}[\boldsymbol{u}]=\check{\mathcal{T}}_{_{\theta}}[\boldsymbol{u}](t_{_{F}})$.

Exploiting the specific nature of the cost functional, the existence of an optimal solution to \emph{Prob. I} can be ascertained
by means of a basic result on minimum-distance problems 
pertaining to closed convex  sets ([9], p. 69), as stated in the next proposition in which the optimal solution is also 
characterized.

\begin{proposition}\label{prop:01}
There exists $\boldsymbol{u}_{_{\mathrm{o}}} \in L_{2}(0,t_{_{F}})^{^{m}}$ such that $\forall \boldsymbol{u}\in L_{2}(0, t_{_{F}})^{^{m}}$, 
$\boldsymbol{u}\neq \boldsymbol{u}_{_{\mathrm{o}}}$, $\mathcal{J}(\boldsymbol{u}_{_{\mathrm{o}}}) < \mathcal{J}(\boldsymbol{u})$.

Moreover, $\boldsymbol{u}_o$ is the unique solution of the linear equation
\begin{equation}\label{4eq:14}
\rho_{_{\boldsymbol{u}}}\boldsymbol{u}_{_{\mathrm{o}}}+\mathcal{T}_{_{\theta}}^{*}\cdot \mathcal{T}_{_{\theta}}[\boldsymbol{u}_{_{\mathrm{o}}}]-
\mathcal{T}^{*}_{_{\theta}}[\theta_{r\mathrm{o}}]=0,
\end{equation} 
\emph{i.e.},
\begin{equation}\label{4eq:15}
\boldsymbol{u}_{_{\mathrm{o}}}=\left[\rho_{_{\boldsymbol{u}}}I+\mathcal{T}^{*}_{_{\theta}}\cdot 
\mathcal{T}_{_{\theta}}\right]^{-1}\left[\mathcal{T}^{*}_{_{\theta}}[\theta_{r\mathrm{o}}]\right],
\end{equation}
where
$\mathcal{T}^{*}_{_{\theta}}\ :\ L_{2}(\Omega)\rightarrow L_{2} (0,t_{_{F}})^{^{m}}$ is the adjoint of $\mathcal{T}_{_{\theta}}$. \hfill $\nabla$
\end{proposition}

\begin{remark}\label{remark:4.2}
The final-state error achieved with a given control signal, namely,
\[
\| \underline{\theta}(t_{_{F}};f_{_{S}}+\boldsymbol{\beta}_{ \boldsymbol{S}}^{^{\mathrm{T}}} \boldsymbol{u},g)-\theta_r\|_{_{2}}^{^{2}}=
\| \mathcal{T}_{_{\theta}}[\boldsymbol{u}]-\theta_{r\mathrm{o}}\|_{_{2}}^{^{2}}
\]
can be written as $$\|\mathcal{T}_{_{\theta}}[\boldsymbol{u}]-\hat{\theta}_{r\mathrm{o}}\|_{_{2}}^{^{2}}+
\| \theta_{r\mathrm{o}}-\hat{\theta}_{r\mathrm{o}}\|_{_{2}}^{^{2}},$$
where $\hat{\theta}_{r\mathrm{o}}$ denotes the $L_{2}(\Omega)$--orthogonal projection of $\theta_{r\mathrm{o}}$ on the closure of $\mathcal{T}_{_{\theta}}[L_{2}(0,t_{_{F}})^{^{m}}]$ in $L_{2}(\Omega)$. 
Thus, by appropriately choosing control signals, the final-state error can be made arbitrarily close to
\[
\inf \left\{\| \mathcal{T}_{_{\theta}}[\boldsymbol{u}]-\hat{\theta}_{r\mathrm{o}}\|_{_{2}}^{^{2}}\ :\ \boldsymbol{u}\in  L_{2}(0,t_{_{F}})^{^{m}}\right\}+
\| \theta_{r\mathrm{o}}-\hat{\theta}_{r\mathrm{o}}\|^{^{2}}_{_{2}}=\| \theta_{r\mathrm{o}}-\hat{\theta}_{r\mathrm{o}}\|^{^{2}}_{_{2}}.
\]
In fact, this can be done with the optimal $\boldsymbol{u}_{_{\mathrm{o}}}(\rho_{_{\boldsymbol{u}}})$ of \emph{Prob. I}, for decreasing values of $\rho_{_{\boldsymbol{u}}}$. 
Indeed, taking 
$\varepsilon > 0$ and $\boldsymbol{u}_{_{\varepsilon}}\in L_{2}(0,t_{_{F}})^{^{m}}$ such that
\[
\|\mathcal{T}_{_{\theta}}[\boldsymbol{u}_{_{\varepsilon}}]-\hat{\theta}_{r\mathrm{o}}\|_{_{2}}^{^{2}}\leq \varepsilon \ , \ \mbox{the fact that}\ 
\mathcal{J}(\boldsymbol{u}_{_{\mathrm{o}}}(\rho_{_{\boldsymbol{u}}});\rho_{_{\boldsymbol{u}}})\leq \mathcal{J}(\boldsymbol{u}_{_{\varepsilon}};\rho_{_{\boldsymbol{u}}})
\]
implies that
\[
\rho_{_{\boldsymbol{u}}}\| \boldsymbol{u}_{_{\mathrm{o}}}(\rho_{_{\boldsymbol{u}}})\|_{_{L_{2}(0,t_{_{F}})^{^{m}} }}^{^{2}}+\|\mathcal{T}_{_{\theta}}[\boldsymbol{u}_{_{\mathrm{o}}}(\rho_{_{\boldsymbol{u}}})]-\hat{\theta}_{r\mathrm{o}}\|_{_{2}}^{^{2}}\leq
\rho_{_{\boldsymbol{u}}}\|\boldsymbol{u}_{_{\varepsilon}}\|_{_{L_{2}(0,t_{_{F}})^{^{m}}}}^{^{2}}+\varepsilon.
\]
Thus,
\[
\forall \varepsilon >0\ , \forall  \rho_{_{\boldsymbol{u}}}>0\ ,
\ \|\mathcal{T}_{_{\theta}}[\boldsymbol{u}_{_{\mathrm{o}}}(\rho_{_{\boldsymbol{u}}})]-\hat{\theta}_{r\mathrm{o}}\|
\leq \rho_{_{\boldsymbol{u}}}\|\boldsymbol{u}_{_{\varepsilon}}\|_{_{L_{2}(0, t_{_{F}})^{^{m}}}}^{^{2}}
+\varepsilon
\]
and, hence, $\displaystyle\lim_{\rho_{_{\boldsymbol{u}}} \rightarrow 0}\| \mathcal{T}_{_{\theta}}[\boldsymbol{u}_{_{\mathrm{o}}}(\rho_{_{\boldsymbol{u}}})]-
\hat{\theta}_{r\mathrm{o}}\|_{_{2}}^{^{2}}=0$.\hfill $\nabla$
\end{remark}


Proposition \ref{prop:01} above characterizes the optimal solution $\boldsymbol{u}_{_{\mathrm{o}}}$ in terms of the linear operators 
$\mathcal{T}_{_{\theta}}$ and  $\mathcal{T}^{*}_{_{\theta}}$. However, computing $\boldsymbol{u}_{_{\mathrm{o}}}$  involves finding ways of computing the
operator $(\rho_{_{\boldsymbol{u}}}\mathbf{I}+\mathcal{T}_{_{\theta}}^{*}\circ\mathcal{T}_{_{\theta}})^{^{-1}}$ as well as applying the result to
$\mathcal{T}_{_{\theta}}^{*}[\theta_{r\mathrm{o}}]$.  To do so, it is natural to search for explicit approximations to $\boldsymbol{u}_{_{\mathrm{o}}}$, which are to 
be obtained by considering finite-dimensional approximations to the operator $\mathcal{T}_{_{\theta}}$ and $\mathcal{T}_{_{\theta}}^{*}$ and the 
corresponding version of equation (\ref{4eq:14}). This is the theme of the next section.

\section{Approximate Solutions}\label{sec:04}

In this section, a sequence $\{\boldsymbol{u}_{_{K}}\}$ is introduced which is defined on the basis of finite-dimensional
approximations to the operator $\mathcal{T}_{_{\theta}}$. It is then shown that under appropriate conditions this sequence converges to 
$\boldsymbol{u}_{_{\mathrm{o}}}$ in the $L_{2}(0,t_{_{F}})^{^{m}}$--norm.

To this effect, let $\{X_{_{K}}\}$  be a sequence of finite-dimensional subspaces of $H^1_0(\Omega)$ with the approximability property, 
\emph{i.e.}, such that 
$\forall \psi \in H^1_0(\Omega)$  there exists a sequence  $\{\psi_{_{K}}\}\subset H^1_0(\Omega)$  such that $\psi_{_{K}} \in X_{_{K}}$ and
\begin{equation}\label{4eq:16}
\lim_{K\rightarrow \infty}\| \psi - \psi_{_{K}}\|_{H^{1}_{0}(\Omega)}=0.
\end{equation}

Let $\mathcal{A}_{_{K}}: X_{_{K}} \rightarrow X_{_{K}}$ be such that
\[
\forall \phi \in X_{_{K}},\forall \psi \in X_{_{K}}, \ \ \ \ \langle \mathcal{A}_{_{K}}[\phi], \psi \rangle = - \mathbf{B}[\phi, \psi] 
\]
or, equivalently, for an $L_{2}-$orthonormal basis $\{\phi_{_{1}}, \mathellipsis, \phi_{_{n_{_{K}}}}\}$  of $X_{_{K}}$,
\[
\forall \phi \in X_{_{K}},\ \ \ \ \mathcal{A}_{_{K}}[\phi] = -\sum_{k=1}^{n_{_{K}}}\mathbf{B}[\phi, \phi_{_{k}}]\phi_{_{k}}\ \ \ \Leftrightarrow\ \ \
\forall \ell=1, \mathellipsis, n,\ \ \ \mathcal{A}_{_{K}}[\phi_{_{\ell}}]=-\sum_{k=1}^{n_{_{K}}}\mathbf{B}[\phi_{_{\ell}}, \phi_{_{k}}]\phi_{_{k}}.
\]
Let then $\mathbf{A}_{_{K}} \in \mathbb{R}^{n_{_{K}} \times n_{_{K}}}$ be defined by $\{\mathbf{A}_{_{K}}\}_{\ell k}= -\mathbf{B}[\phi_{_{\ell}}, \phi_{_{k}}]$, 
\emph{i.e.}, $\mathbf{A}_{_{K}}$ is the matrix representation of $\mathcal{A}_{_{K}}$ in the basis $\{\phi_{_{1}}, \mathellipsis, \phi_{n_{_{K}}}\}$ so that for
$\phi=\sum_{k=1}^{n_{_{K}}}\gamma_{_{k}}\phi_{_{k}}$,\ $\mathcal{A}_{_{K}}^{^{\ell}}[\phi]=\sum_{k=1}^{n_{_{K}}}\gamma_{_{k}}^{^{\ell}}\phi_{_{k}}$, \ \ where\ \
$\mathcal{A}_{_{K}}^{^{\ell}}[\phi]$\ \  is the\ \ $\ell$th-power of\ \ $\mathcal{A}_{_{K}}[\phi]$\ \ and\ \
$\bar{\gamma}^{^{\ell}}=\mathbf{A}_{_{K}}^{^{\ell}}\bar{\boldsymbol{\gamma}}$, \ $\bar{\boldsymbol{\gamma}}=[\gamma_{_{1}}\ \cdots\ \gamma_{n_{_{K}}}]^{^{\mathrm{T}}}$\ \  and\ \ 
$\bar{\boldsymbol{\gamma}}^{^{\ell}}=[\gamma_{_{1}}^{^{\ell}}\ \cdots\ \gamma_{_{K}}^{^{\ell}}]^{^{\mathrm{T}}}$.

\begin{remark}
 By way of example, consider the one-dimensional heat equation -- in this case\ \ $\Omega=(0, L_{_{x}})$ \ \ and \ \  
 $\mathbf{B}[\phi, \psi]=\alpha\left\langle\frac{\partial \phi}{\partial x}, \frac{\partial \psi}{\partial x}\right\rangle$ \ \ and let \ \
 $\phi_{_{k}}=\sqrt{\frac{2}{L_{_{x}}}}\sin\left[\frac{k\pi x}{L_{_{x}}}\right]$ and  $X_{K}=\operatorname{span}\{\phi_{k}:k=1, \mathellipsis, K\}$.\ \  Thus, \ \
 $\{A_{_{K}}\}_{\ell k}=-\alpha\left\langle\frac{\partial \phi_{_{\ell}}}{\partial x}, \frac{\partial \phi_{_{k}}}{\partial x}\right\rangle$,\ \ 
 \emph{i.e.},\ \
 $\{A_{_{K}}\}_{\ell k}=\alpha\left[\frac{\ell \pi}{L_{_{x}}}\right]\left[\frac{k \pi}{L_{_{x}}}\right]
 \left\langle \sqrt{\frac{2}{L_{_{x}}}}\cos\left[\frac{\ell \pi x}{L_{_{x}}}\right], \sqrt{\frac{2}{L_{_{x}}}}\cos\left[\frac{k \pi x}{L_{_{x}}}\right]\right\rangle$, \ \ 
 so that\\
 $A_{_{K}}=diag\left( -\alpha\left[\frac{\pi}{L_{_{x}}}\right]^{^{2}}\ \cdots \   -\alpha\left[\frac{K\pi}{L_{_{x}}}\right]^{^{2}}\right)$. \hfill$\nabla$
 \end{remark}

\noindent 
The corresponding approximation $\mathcal{T}_{_{\theta}}^{^{K}}$ of $\mathcal{T}_{_{\theta}}$ is introduced in the next proposition. 

\begin{proposition}\label{prop:4.1}
Let $P_{_{K}}$ be the orthogonal projection from $L_{2}(\Omega)$ onto $X_{_{K}}$ and define
\[
\mathcal{T}^{^{K}}_{_{\theta}}\ :\ L_{2}(0,t_{_{F}})^{^{m}}\rightarrow X_{_{K}}\ \ \mbox{by}\ \ \mathcal{T}_{_{\theta}}^{^{K}}[\boldsymbol{u}]
\triangleq \left[\int^{t_{_{F}}}_{0}S_{_{K}}(t_{_{F}}-\tau)\left[P_{_{K}}\left[\boldsymbol{\beta}_{_{\boldsymbol{S}}}^{^{\mathrm{T}}}\boldsymbol{u}(\tau )\right]\right]d\tau \right],
\]
where $S_{_{K}}$ is the semigroup generated by $\mathcal{A}_{_{K}}$, \emph{i.e.}, 
$S_{_{K}}(t)= \sum_{\ell=0}^{\infty}\mathcal{A}_{_{K}}^{^{\ell}}t^{^{\ell}}/\ell!$. $\mathcal{T}_{_{\theta}}^{^{K}}$ is given by
$\mathcal{T}_{_{\theta}}^{^{K}}[\boldsymbol{u}]=\sum_{q=1}^{n_{_{K}}}c_{q}(t_{_{F}}, \boldsymbol{u})\phi_{q}$, where 
$\underline{\boldsymbol{c}}_{_{K}}(t; \boldsymbol{u})=[c_{_{1}}(t; \boldsymbol{u}),\mathellipsis, c_{n_{_{K}}}(t; \boldsymbol{u})]^{^{\mathrm{T}}}$ is given by 
$\underline{\boldsymbol{c}}_{_{K}}(t; \boldsymbol{u})=
\displaystyle\int_{0}^{t}\exp[\mathbf{A}_{_{K}}(t-\tau)]\mathbf{M}_{_{\boldsymbol{\beta}}}^{^{K}}\boldsymbol{u}(\tau)d\tau$,\ \ \ 
$\boldsymbol{\beta}_{_{\boldsymbol{S}}}^{^{\mathrm{T}}}=[\boldsymbol{\beta}_{_{\boldsymbol{S}1}}\ \cdots\ \boldsymbol{\beta}_{_{\boldsymbol{S}m}}]$\ \ \ and 
\[
\mathbf{M}_{_{\boldsymbol{\beta}}}^{^{K}}\triangleq \begin{bmatrix}
                                               \langle\boldsymbol{\beta}_{_{\boldsymbol{S}1}}, \phi_{_{1}}\rangle& \cdots & \langle\boldsymbol{\beta}_{_{\boldsymbol{S}m}}, \phi_{_{1}}\rangle\\
                                               \vdots &  & \vdots\\
                                                \langle\boldsymbol{\beta}_{_{\boldsymbol{S}1}}, \phi_{n_{_{K}}}\rangle& \cdots &  \langle\boldsymbol{\beta}_{_{\boldsymbol{S}m}}, \phi_{n_{_{K}}}\rangle\\
                                              \end{bmatrix}.
\]
\hfill$\nabla$ 
\end{proposition}

The corresponding version of \emph{Prob. I} is then defined by
\begin{equation}\label{4eq:17}
\text{\underline{\emph{Prob.\ $I_{_{K}}:$}}} \displaystyle\min_{\boldsymbol{u}\in L_{2}(0,t_{_{F}})^{^{m}}}\mathcal{J}_{_{K}}(\boldsymbol{u}),
\end{equation}
where
\[
\mathcal{J}_{_{K}}[\boldsymbol{u}]\triangleq \|\mathcal{T}_{_{\theta}}^{^{K}}[\boldsymbol{u}]-\theta_{r\mathrm{o}}\|_{_{L_{2}(\Omega)}}^{^{2}}+
\rho_{_{\boldsymbol{u}}}\| \boldsymbol{u}\|_{_{L_{2}(0,t_{_{F}})^{^{m}}}}^{^{2}}.
\]

Similarly to what happens in the case of \emph{Prob. I}, \emph{Prob. $I_{_{K}}$} has a unique solution $\boldsymbol{u}_{_{K}}$ which 
is obtained from the optimality condition
\begin{equation}\label{4eq:18}
\rho_{_{\boldsymbol{u}}} \boldsymbol{u}_{_{K}}+(\mathcal{T}_{_{\theta}}^{^{K}})^{*}[\mathcal{T}_{_{\theta}}^{^{K}}[\boldsymbol{u}_{_{K}}]-\theta_{r\mathrm{o}}]=0,
\end{equation}
where the adjoint operator $(\mathcal{T}_{_{\theta}}^{^{K}})^{*}: L_{2}(\Omega)\rightarrow L_{2}(0, t_{_{F}})^{^{m}}$ is such that 
\[
\forall \boldsymbol{u} \in L_{2}(0, t_{_{F}})^{^{m}}, \ \forall \phi \in L_{2}(\Omega), \ \langle \phi, 
\mathcal{T}_{_{\theta}}^{^{K}}[\boldsymbol{u}]\rangle=\langle(\mathcal{T}_{_{\theta}}^{^{K}})^{*}[\phi], \boldsymbol{u}\rangle
\]
\[
\Leftrightarrow \ \ \langle(\mathcal{T}_{_{\theta}}^{^{K}})^{*}[\phi],\boldsymbol{u}\rangle=\sum_{k=1}^{n}\langle\phi, \phi_{_{k}}\rangle c_{_{k}}(t_{_{F}}; \boldsymbol{u})=
\bar{ \boldsymbol{\phi}}_{_{K}}^{^{\mathrm{T}}}\underline{\boldsymbol{c}}_{_{K}}(t_{_{F}}; \boldsymbol{u})=
\int_{0}^{t_{_{F}}}(\mathbf{F}_{_{K}}(\tau)\bar{\boldsymbol{\phi}}_{_{K}})^{^{\mathrm{T}}}\boldsymbol{u}(\tau)d\tau
\]
so that $(\mathcal{T}_{_{\theta}}^{^{K}})^{*}[\phi]=\mathbf{F}_{_{K}}(\tau)\bar{\boldsymbol{\phi}}_{_{K}}$, where
$\bar{\boldsymbol{\phi}}_{_{K}}^{^{\mathrm{T}}}\triangleq[ \langle\phi, \phi_{_{1}}\rangle \cdots \langle\phi, \phi_{_{K}}\rangle]$ and
\begin{equation}\label{4eq:19}
 \mathbf{F}_{_{K}}(\tau)\triangleq (\mathbf{M}_{_{\boldsymbol{\beta}}}^{^{K}})^{^{\mathrm{T}}}\exp[\mathbf{A}_{_{K}}^{^{\mathrm{T}}}(t_{_{F}}-\tau)].
\end{equation}
The unique solution of $Prob\ I_{_{K}}$ is now explicitly characterized.

\begin{proposition}\label{prop:4.2new}
The unique solution of \eqref{4eq:18} is given by 
\begin{equation}\label{4eq:24}
\boldsymbol{u}_{_{K}}(\tau)=\mathbf{F}_{_{K}}(\tau )(\rho_{_{\boldsymbol{u}}}\mathbf{I}+\mathbf{G}_{_{K}})^{^{-1}}\bar{ \boldsymbol{\theta}}_{r\mathrm{o}}^{^{K}}, \ \ \tau \in [0,t_{_{F}}], 
\end{equation}
where $\mathbf{G}_{_{K}}\triangleq \displaystyle\int_{0}^{t_{_{F}}}\mathbf{F}_{_{K}}(\tau )^{^{\mathrm{T}}}\mathbf{F}_{_{K}}(\tau )d\tau$ and $\bar{\boldsymbol{\theta}}_{r\mathrm{o}}^{^{K}}\triangleq[\langle\phi_{_{1}}, \theta_{r\mathrm{o}}\rangle\ \cdots\ \langle\phi_{_{n_{K}}}, \theta_{r\mathrm{o}}\rangle ]^{^{\mathrm{T}}}$.
\end{proposition}

\begin{remark}
It is interesting to notice that $\mathbf{G}_{_{K}}$ can be computed from a linear equation in $\mathbb{R}^{n_{_{K}}\times n_{_{K}}}$. Indeed, from
(\ref{4eq:22}), it can be seen that $\mathbf{G}_{_{K}}$\ can be expressed as\\
 $\mathbf{G}_{_{K}}=\displaystyle\int_{0}^{t_{_{F}}}\exp\left[\mathbf{A}_{_{K}}(t_{_{F}}-\tau)\right]\mathbf{M}_{_{\boldsymbol{\beta}}}^{^{K}}
 \left(\exp\left[\mathbf{A}_{_{K}}(t_{_{F}}-\tau)\right]\mathbf{M}_{_{\boldsymbol{\beta}}}^{^{K}}\right)^{^{\mathrm{T}}}d\tau.$
 Thus, letting $\omega=t_{_{F}}-\tau$, it follows that $\mathbf{G}_{_{K}}=\displaystyle\int_{0}^{t_{_{F}}}\check{\mathbf{H}}_{_{K}}(\omega)d\omega$, where
$\check{\mathbf{H}}_{_{K}}(\omega)\triangleq\exp\left[\mathbf{A}_{_{K}}(\omega)\right]\mathbf{M}_{_{\boldsymbol{\beta}}}^{^{K}}
 \left(\exp\left[\mathbf{A}_{_{K}}(\omega)\right]\mathbf{M}_{_{\boldsymbol{\beta}}}^{^{K}}\right)^{^{\mathrm{T}}}$.

Moreover,
 \begin{eqnarray*}
 \frac{d}{d\omega}\check{\mathbf{H}}_{_{K}}(\omega)&=&\mathbf{A}_{_{K}}\{\exp\left[\mathbf{A}_{_{K}}(\omega)\right]\mathbf{M}_{_{\boldsymbol{\beta}}}^{^{K}}
 \left(\exp[\mathbf{A}_{_{K}}(\omega)]\mathbf{M}_{_{\boldsymbol{\beta}}}^{^{K}}\right)^{^{\mathrm{T}}}\}\\
 &&+\{\exp\left[\mathbf{A}_{_{K}}(\omega)\right]\mathbf{M}_{_{\boldsymbol{\beta}}}^{^{K}}
 \left(\exp[\mathbf{A}_{_{K}}(\omega)]\mathbf{M}_{_{\boldsymbol{\beta}}}^{^{K}}\right)^{^{\mathrm{T}}}\}\mathbf{A}_{_{K}}^{^{\mathrm{T}}}
 =\mathbf{A}_{_{K}}\check{\mathbf{H}}_{_{K}}(\omega)+\check{\mathbf{H}}_{_{K}}(\omega)\mathbf{A}_{_{K}}^{^{\mathrm{T}}}.
 \end{eqnarray*}
So that integrating both sides from $0$ to $\omega$ it follows that for all $\omega \in [0, t_{_{F}}]$, 
$$ \check{\mathbf{H}}(\omega)-\check{\mathbf{H}}(0)=\int_{0}^{\omega}\{\mathbf{A}_{_{K}}\check{\mathbf{H}}_{_{K}}(\sigma)+\check{\mathbf{H}}_{_{K}}(\sigma)\mathbf{A}_{_{K}}^{^{\mathrm{T}}}\}d\sigma\\
=\mathbf{A}_{_{K}}\int_{0}^{\omega}\check{\mathbf{H}}_{_{K}}(\sigma)d\sigma+\int_{0}^{\omega}\check{\mathbf{H}}_{_{K}}(\sigma)d\sigma\mathbf{A}_{_{K}}^{^{\mathrm{T}}}.$$
Then, taking $\omega=t_{_{F}}$ and letting $\check{\mathbf{M}}_{_{K}}=\check{\mathbf{H}}(t_{_{F}})-\check{\mathbf{H}}(0)=\exp[\mathbf{A}_{_{K}}t_{_{F}}]\mathbf{M}_{_{\boldsymbol{\beta}}}^{^{K}}\left(\exp[\mathbf{A}_{_{K}}t_{_{F}}]\mathbf{M}_{_{\boldsymbol{\beta}}}^{^{K}}\right)^{^{\mathrm{T}}}
-\mathbf{M}_{_{\boldsymbol{\beta}}}^{^{K}}(\mathbf{M}_{_{\boldsymbol{\beta}}}^{^{K}})^{^{\mathrm{T}}}$, $\mathbf{G}_{_{K}}$ can be obtained as the unique solution of the \emph{Lyapunov equation} $\mathbf{A}_{_{K}}\mathbf{G}_{_{K}}+\mathbf{G}_{_{K}}\mathbf{A}_{_{K}}^{^{\mathrm{T}}}=\check{\mathbf{M}}_{_{K}}$, see ([5], pp. 144 -- 148), 
([11], pp. 71 -- 72).\hfill$\nabla$
\end{remark}

\begin{remark}\label{remark:u_K}
 Note that $\boldsymbol{u}_{_{K}}:[0, t_{_{F}}]\rightarrow\mathbb{R}^{^{m}}$ is explicitly given by (\ref{4eq:24}) in terms of\
 $\exp[\mathbf{A}_{_{K}}^{^{\mathrm{T}}}(t_{_{F}}-\tau)]$. Note also that $\boldsymbol{u}_{_{K}}$ can be obtained from the solution of the linear ordinary differential equation\break $\dot{\boldsymbol{x}}_{_{\boldsymbol{u}}}(\tau)=-\mathbf{A}_{_{K}}^{^{\mathrm{T}}}\boldsymbol{x}_{_{\boldsymbol{u}}}(\tau),\ \tau\geq 0$ with
 the initial condition $\boldsymbol{x}_{_{\boldsymbol{u}}}(0)=\exp[\mathbf{A}_{_{K}}^{^{\mathrm{T}}}t_{_{F}}]
\left(\rho_{_{\boldsymbol{u}}}\mathbf{I}+\mathbf{G}_{_{K}}\right)^{^{-1}}\bar{\boldsymbol{\theta}}_{r\mathrm{o}}^{^{K}}$, \emph{i.e.},\break
$\boldsymbol{u}(\tau)=(\mathbf{M}_{_{\boldsymbol{\beta}}}^{^{K}})^{^{\mathrm{T}}}\boldsymbol{x}_{_{\boldsymbol{u}}}(\tau)$.\hfill$\nabla$
 \end{remark} 

The next step is to analyze the question of whether the  sequence $\{\boldsymbol{u}_{_{K}}\}$ of approximate solutions to the 
optimal control problem converges to the solution $\boldsymbol{u}_{_{\mathrm{o}}}$ of the original problem. To this effect, the uniform convergence of $\{\mathcal{T}_{_{\theta}}^{^{K}}[\cdot]\}$ on $L_{2}(0, t_{F})^{m}$ - balls is established in the next proposition.

\begin{proposition}\label{prop:02}
There exists a real sequence $\{\eta^{^{K}}_{_{\mathcal{T}}}:K\in \mathbb{Z}_+\}$such that\\
\emph{\textbf{(a)}} $\forall \boldsymbol{u} \in L_{2}(0,t_{_{F}})^{^{m}}$, \ \ \ $\|\mathcal{T}_{_{\theta}}[\boldsymbol{u}]-
\mathcal{T}_{_{\theta}}^{^{K}}[\boldsymbol{u}]\|_{_{L_{2}(\Omega)}}\leq \eta_{_{\mathcal{T}}}^{^{K}}\|\boldsymbol{u}\|_{L_{2}(0,t_{_{F}})^{^{m}}}$.\\
\emph{\textbf{(b)}} $\{\eta_{_{\mathcal{T}}}^{^{K}}\}$ converges to zero. \hfill $\nabla$
\end{proposition}
The following corollary of Proposition \ref{prop:02} can now be stated.\\

\noindent
\underline{\textbf{Corollary 2.1:}} $\mathcal{J}(\boldsymbol{u}_{_{K}})\rightarrow\mathcal{J}(\boldsymbol{u}_{_{\mathrm{o}}})$. \hfill$\nabla$\\

Moreover, as $\{\boldsymbol{u}_{_{K}}\}$ is bounded and $\mathcal{J}(\boldsymbol{u}_{_{K}})\rightarrow \mathcal{J}(\boldsymbol{u}_{_{\mathrm{o}}})$, the desired convergence
of the approximate solutions $\{\boldsymbol{u}_{_{K}}\}$  can be established, as stated in the following proposition.

\begin{proposition}\label{prop:03}
The sequence  $\{\boldsymbol{u}_{_{K}}:K\in \mathbb{Z}_{+}\}$ of solutions to the approximate problems \emph{Prob. $I_K$} converges to 
the solution $\boldsymbol{u}_{_{\mathrm{o}}}$ of \emph{Prob. I} in the sense of the  $L_{2}(0,t_{_{F}})^{^{m}}$--norm. \hfill $\nabla$
\end{proposition}

A summary is now presented of the steps required to compute the approximate solution $\boldsymbol{u}_{_{K}}$
for the problem $\displaystyle\min_{\boldsymbol{u}\in L_{2}(0, t_{_{F}})^{^{m}}}\mathcal{J}(\boldsymbol{u})$ where $\mathcal{J}(\boldsymbol{u})$ is 
given by (\ref{4eq:08}).

Given the problem data $(f, g, \theta_{r}, \rho_{_{\boldsymbol{u}}}, \boldsymbol{\beta}_{_{\boldsymbol{S}}})$ and the family $\{X_{_{K}}\}$ of subspaces each with the orthonormal
basis $\{\phi_{_{1}} \ \mathellipsis\ \phi_{_{n_{_{K}}}}\}$:

\begin{itemize}
 \item[\textbf{(1) }]Compute $\bar{\boldsymbol{\theta}}_{r\mathrm{o}}^{^{K}}=[\langle\theta_{r\mathrm{o}}, \phi_{_{1}}\rangle\ \ \mathellipsis\ 
\langle \theta_{r\mathrm{o}}, \phi_{_{n_{_{K}}}} \rangle]^{^{\mathrm{T}}}$, where $\theta_{r\mathrm{o}}=\theta_{r}-\underline{\theta}(t_{_{F}}; f, g)$ and
$\underline{\theta}(t_{_{F}}; f, g)$ is given by (\ref{ch4:eq-09}).

\item[\textbf{(2) }]Compute $\mathbf{M}_{_{\boldsymbol{\beta}}}^{^{K}}\in \mathbb{R}^{n_{_{K}}\times m}$, where $\{\mathbf{M}_{_{\boldsymbol{\beta}}}^{^{K}}\}_{k i}=\langle\boldsymbol{\beta}_{_{\boldsymbol{S}i}}, \phi_{_{k}}\rangle$.

\item[\textbf{(3) }]For $\mathbf{A}_{_{K}}\in \mathbb{R}^{n_{_{K}}}$ such that 
$\{\mathbf{A}_{_{K}}\}_{\ell k}=-\mathbf{B}[\phi_{\ell}, \phi_{k}]$ compute $\mathbf{G}_{_{K}}$ solving the Lyapunov equation\break $\mathbf{A}_{_{K}}\mathbf{G}_{_{K}}+\mathbf{G}_{_{K}}\mathbf{A}_{_{K}}^{^{\mathrm{T}}}=\check{\mathbf{M}}_{_{K}}$, where\ $\check{\mathbf{M}}_{_{K}}=\exp[\mathbf{A}_{_{K}}t_{_{F}}]\mathbf{M}_{_{\boldsymbol{\beta}}}^{^{K}}\left(\exp[\mathbf{A}_{_{K}}t_{_{F}}]\mathbf{M}_{_{\boldsymbol{\beta}}}^{^{K}}\right)^{^{\mathrm{T}}}
-\mathbf{M}_{_{\boldsymbol{\beta}}}^{^{K}}(\mathbf{M}_{_{\boldsymbol{\beta}}}^{^{K}})^{^{\mathrm{T}}}$.
\item[\textbf{(4) }]$\boldsymbol{u}_{_{K}}$ can then be obtained from (\ref{4eq:24}) (see also Remark \ref{remark:u_K}).
\end{itemize}


Thus, the computations required to obtain $\boldsymbol{u}_{_{K}}$ involve solving the Lyapunov equation in \textbf{(3)} above, carrying out the numerical evaluation of the integrals \
$\langle\theta_{r}, \phi_{k}\rangle$, \
$\langle g, \phi_{k}\rangle$ and \
$\langle\boldsymbol{\beta}_{_{\boldsymbol{S}i}}, \phi_{k}\rangle$ over the spatial domain $\Omega$, and of the exponential function over the time-interval $(0, t_{_{F}})$.


Finally, it should be noted that the approximation error on the final state for a given control signal $\boldsymbol{u}$ is given by 
$\mathcal{T}_{_{\theta}}[\boldsymbol{u}]-\theta_{r\mathrm{o}}=\boldsymbol{e}_{_{K}}[\boldsymbol{u}]+\check{\boldsymbol{e}}_{_{K}}[\boldsymbol{u}]$, where
$\boldsymbol{e}_{_{K}}[\boldsymbol{u}]\triangleq\mathcal{T}_{_{\theta}}^{^{K}}[\boldsymbol{u}]-\theta_{r\mathrm{o}}^{^{K}}$ (error projection on\
$span\{\phi_{_{1}}, \mathellipsis, \phi_{_{K}}\}$) and
$\check{\boldsymbol{e}}_{_{K}}[\boldsymbol{u}]=\{\mathcal{T}_{_{\theta}}[\boldsymbol{u}]-\mathcal{T}_{_{\theta}}^{^{K}}[\boldsymbol{u}]\}-\{\theta_{r\mathrm{o}}-\theta_{r\mathrm{o}}^{^{K}}\}$.

To get an upper bound on $\|\mathcal{T}_{_{\theta}}[\boldsymbol{u}]-\theta_{r\mathrm{o}}\|_{_{2}}$ note that
\begin{eqnarray}
\|\mathcal{T}_{_{\theta}}[\boldsymbol{u}]-\theta_{r\mathrm{o}}\|_{_{2}}^{^{2}}&=&\|\boldsymbol{e}_{_{K}}[\boldsymbol{u}]\|_{_{2}}^{^{2}}+\|\check{\boldsymbol{e}}_{_{K}}[\boldsymbol{u}]\|_{_{2}}^{^{2}},\label{eq:33}\\
\|\boldsymbol{e}_{_{K}}[\boldsymbol{u}]\|_{_{2}}^{^{2}}&=&\|\bar{\boldsymbol{c}}_{_{K}}(t_{_{F}}; \boldsymbol{u})-\bar{\boldsymbol{\theta}}_{r\mathrm{o}}^{^{K}}\|_{_{E}}^{^{2}},\label{eq:34}\\
\|\check{\boldsymbol{e}}_{_{K}}[\boldsymbol{u}]\|_{_{2}}&\leq&\|\mathcal{T}_{_{\theta}}[\boldsymbol{u}]-\mathcal{T}_{_{\theta}}^{^{K}}[\boldsymbol{u}]\|_{_{2}}
+\|\theta_{r\mathrm{o}}-\theta_{r\mathrm{o}}^{^{K}}\|_{_{2}}.\label{eq:35}
 \end{eqnarray}

Now, with $\rho_{_{F}}=\rho_{_{\boldsymbol{u}}}^{-1}$ as $\bar{\boldsymbol{c}}_{_{K}}(t_{_{F}}; \boldsymbol{u}_{_{K}})=\rho_{_{F}}
\mathbf{G}_{_{K}}(\mathbf{I}+\rho_{_{F}}\mathbf{G}_{_{K}})^{^{-1}}\bar{\boldsymbol{\theta}}_{r\mathrm{o}}^{^{K}}=
\left\{\mathbf{I}-(\mathbf{I}+\rho_{_{F}}\mathbf{G}_{_{K}})^{^{-1}}\right\}\bar{\boldsymbol{\theta}}_{r\mathrm{o}}^{^{K}}$ it follows that
\begin{equation}\label{eq:36a}
 \|\boldsymbol{e}_{_{K}}[\boldsymbol{u}_{_{K}}]\|_{_{2}}^{^{2}}=\|(\mathbf{I}+\rho_{_{F}}\mathbf{G}_{_{K}})^{^{-1}}\bar{\boldsymbol{\theta}_{r\mathrm{o}}}^{^{K}}\|_{_{2}}^{^{2}},
\end{equation}
or, equivalently
$$\|\boldsymbol{e}_{_{K}}[\boldsymbol{u}_{_{K}}]\|_{_{2}}^{^{2}}=\rho_{_{\boldsymbol{u}}}^{^{2}}\|(\rho_{_{\boldsymbol{u}}}\mathbf{I}+\mathbf{G}_{_{K}})^{^{-1}}\bar{\boldsymbol{\theta}_{r\mathrm{o}}}^{^{K}}\|_{_{2}}^{^{2}}.$$

Thus, for a fixed $K$ and as long as $\mathbf{G}_{_{K}}$ is non-singular, the first error term $\|\boldsymbol{e}_{_{K}}[\boldsymbol{u}_{_{K}}]\|_{_{2}}^{{2}}$ goes to zero with $\rho_{\boldsymbol{u}}$ whereas, for a fixed $\rho_{_{\boldsymbol{u}}}$, the second error term $\|\check{\boldsymbol{e}}_{_{K}}[\boldsymbol{u}_{_{K}}]\|_{_{2}}^{^{2}}$ diminishes as
$K$ increases.

\begin{remark}
It is often necessary to make $\rho_{_{F}}=\rho_{_{\boldsymbol{u}}}^{^{-1}}$ large in order to achieve an acceptably small value for the final-state error
norm (see Remark \ref{remark:4.2}). This might make the conditioning number (with respect to inversion) of the matrix 
$(\mathbf{I}+\rho_{_{F}}\mathbf{G}_{_{K}})$ very large which, in-turn, could give rise to numerical difficulties in the process of computing its inverse. 
To cope with this potential  problem using the available tools from MATLAB\textsuperscript{\textregistered}, the symmetric structure of 
$(\mathbf{I}+\rho_{_{F}}\mathbf{G}_{_{K}})$ can be exploited in the following way:
\begin{itemize}
\item[\emph{\textbf{(a)}}] Choose $\delta_{_{S}}>0$ and put $\check{\mathbf{G}}_{_{K\rho}}=(1+\delta_{_{S}})\mathbf{I}+
\rho_{_{F}}\mathbf{G}_{_{K}}$ (note that
the eigenvalues of $\check{\mathbf{G}}_{_{K\rho}}$ are not smaller than $1+\delta_{_{S}}$).
\item[\emph{\textbf{(b)}}] Take a SVD decomposition
$\check{\mathbf{G}}_{_{K\rho}}=\mathbf{V}_{_{K\rho}}\check{\mathbf{\Sigma}}_{_{K\rho}}\mathbf{U}_{_{K\rho}}^{^{\mathrm{T}}}$ of\
$\check{\mathbf{G}}_{_{K\rho}}$ (note that as $\check{\mathbf{G}}_{_{K\rho}}$ is symmetric and positive $\mathbf{V}_{_{K\rho}}=\mathbf{U}_{_{K\rho}}$,\ 
$\mathbf{V}_{_{K\rho}}$ and $\check{\mathbf{\Sigma}}_{_{K\rho}}$ are the eigenvector and eigenvalue matrices of $\check{\mathbf{G}}_{_{K\rho}}$,
respectively).
\item[\emph{\textbf{(c)}}] As $\check{\mathbf{G}}_{_{K\rho}}=\mathbf{V}_{_{K\rho}}\check{\mathbf{\Sigma}}_{_{K\rho}}\mathbf{V}_{_{K\rho}}^{^{\mathrm{T}}}$,
$\mathbf{G}_{_{K\rho}}=\check{\mathbf{G}}_{_{K\rho}}-\delta_{_{S}}\mathbf{I}
=\mathbf{V}_{_{K\rho}}\left[\check{\mathbf{\Sigma}}_{_{K\rho}}-\delta_{_{S}}\mathbf{I}\right]\mathbf{V}_{_{K\rho}}^{^{\mathrm{T}}}$.
\item[\emph{\textbf{(d)}}] Put 
$\mathbf{G}_{_{K\rho}}^{^{-1}}=\mathbf{V}_{_{K\rho}}\left(\check{\mathbf{\Sigma}}_{_{K\rho}}-\delta_{_{S}}\mathbf{I}\right)^{^{-1}}\mathbf{V}_{_{K\rho}}^{^{\mathrm{T}}}$ (note that as
$\left(\check{\mathbf{\Sigma}}_{_{K\rho}}-\delta_{_{S}}\mathbf{I}\right)$ is diagonal computing its inverse is reduced to computing the
inverse of real numbers). \hfill$\nabla$
\end{itemize}
\end{remark}

\section{Peak-value Constraints on Control Signals}\label{sec:05}

In this section, the main concern is that upper bounds on the magnitudes of the control signals $\boldsymbol{u}_{_{i}}$ have to be imposed in
connection with potential applications to engineering problems. Thus, although setting the coefficient $\rho_{_{\boldsymbol{u}}}$ at 
different values may indirectly contribute to such an objective, it is natural to directly impose upper bound constraints on the optimal 
control problem at stake. Accordingly, a constrained optimization problem is formulated in (\ref{5eq:28}) for which optimality conditions are 
then presented. Then a truncated version is introduced in (\ref{5eq:29}) to generate approximate solutions to the original constrained 
problem. The latter can then be tackled on the basis of the duality results in ([9]). To obtain approximate solutions to the dual problem, a 
class of piecewise-linear continuous Lagrange multipliers is introduced. The dual functional can be explicitly written as a quadratic 
functional of the ``free''\ 
parameters of this class of multipliers which are their values at a grid on $[0, t_{_{F}}]$. Obtaining approximate solutions to the dual 
problem is then reduced to maximizing this quadratic functional under non-negativeness constraints.

A summary is then provided of the computational steps required to obtain the desired control signals which satisfy the prescribed peak-value constraints.

Initially, a version of \emph{Prob. I} with pointwise (with respect to $t$) constraints is formulated as follows
\begin{eqnarray}\label{5eq:28}
&&\text{\underline{\emph{Prob. $II:$}}} \displaystyle\min_{\boldsymbol{u}\in L_{2}(0,t_{_{F}})^{^{m}}}\mathcal{J}(\boldsymbol{u})\nonumber\\
&&\ \ \ \ \  \ \ \ \ \ \ \text{subject to:}\ \forall i=1, \mathellipsis, m,\ \forall t\ \text{a.e.}\ \text{in}\ 
[0, t_{_{F}}], -\mu_{_{i}}\leq \boldsymbol{u}_{_{i}}(t)\leq \mu_{_{i}},
\end{eqnarray}
where $\mu_{_{i}}\in \mathbb{R}_{+}$.

The existence of an optimal solution to \emph{Prob. II} can be ascertained by means of an argument entirely similar to the one
used in connection with \emph{Prob. I}. This leads to the next proposition.

\begin{proposition}\label{prop:05}
Let $\mathcal{I}_{_{Fi}}(t)\triangleq [-\mu_{_{i}},\mu_{_{i}}]$ and
\[
S_{_{\boldsymbol{u}F}}\triangleq \left\{\boldsymbol{u}\in L_{2}(0,t_{_{F}})^{^{m}}:\forall i=1, \mathellipsis, m,\ \forall t \ \mbox{a.e. in}\
[0, t_{_{F}}],\ \boldsymbol{u}_{_{i}}(t)\in \mathcal{I}_{_{Fi}}(t)\right\}
\]
There exists $\boldsymbol{u}_{c}\in S_{_{\boldsymbol{u}F}}$ such that $\forall \boldsymbol{u}\in S_{_{\boldsymbol{u}F}}$,
$\boldsymbol{u}\neq \boldsymbol{u}_{c}$, $\mathcal{J}(\boldsymbol{u}_{c})<\mathcal{J}(\boldsymbol{u})$.\hfill $\nabla$
\end{proposition}

The problem of computing (approximations to) $\boldsymbol{u}_{c}$  is now tackled following the approach pursued in connection with the unconstrained problem
$\mathcal{T}_{_{\theta}}$.

To this effect, let $\mathcal{J}_{_{K}}( \boldsymbol{u})\triangleq \rho_{_{\boldsymbol{u}}}\|\boldsymbol{u}\|_{_{L_{2}(0, t_{_{F}})^{^{m}}}}^{^{2}}+\|\mathcal{T}_{_{\theta}}^{^{K}}[\boldsymbol{u}]-\theta_{r\mathrm{o}}\|_{_{2}}^{^{2}}$ and consider

\begin{equation}\label{5eq:29}
 \text{\underline{\emph{Prob.\ $II_{_{K}}:$}}}\  \displaystyle\min_{\boldsymbol{u}\in S_{_{\boldsymbol{u}F}}}\mathcal{J}_{_{K}}(\boldsymbol{u}).
\end{equation}

Approximate solutions to \emph{Prob. II} can be obtained on the basis of \emph{Prob. $II_{_{K}}$}, as stated in the following proposition (a proof of which is presented
 in the Appendix).

\begin{proposition}\label{prop:07}
\textbf{\emph{(a)}} $\forall K \in \mathbb{Z}_{+}$ there exists $\boldsymbol{u}_{c}^{^{K}}\in S_{_{\boldsymbol{u}F}}$ such that
$\forall \boldsymbol{u}\in S_{_{\boldsymbol{u}F}}$, $\boldsymbol{u}\neq \boldsymbol{u}_{c}^{^{K}}$, 
$\mathcal{J}_{_{K}}(\boldsymbol{u}_{c}^{^{K}})<\mathcal{J}_{_{K}}(\boldsymbol{u})$.\\
\textbf{\emph{(b)}} $\boldsymbol{u}_{c}^{^{K}}\rightarrow \boldsymbol{u}_{c}$ in $L_{2}(0,t_{_{F}})^{^{m}}$, as $K\rightarrow\infty$. \hfill $\nabla$
\end{proposition}

One possible approach to computing approximate solutions to \emph{Prob. $II_{_{K}}$} (\emph{i.e.}, $\boldsymbol{u}_{c}^{^{K}}$)
 is to rely on Lagrangian duality ([9]). This amounts to introducing the Lagrangian functional
\begin{equation}\label{5eq:31}
Lag_{_{K}}(\boldsymbol{u},\boldsymbol{\lambda} )\triangleq \mathcal{J}_{_{K}}(\boldsymbol{u})+2\langle\boldsymbol{\lambda}_{a},\boldsymbol{u}_{a}-\boldsymbol{u}\rangle_{L_{2}(0, t_{_{F}})^{^{m}}}
+2\langle \boldsymbol{\lambda}_{b},\boldsymbol{u}-\boldsymbol{u}_b\rangle_{L_{2}(0, t_{_{F}})^{^{m}} } \ ,
\end{equation}
where\  $\boldsymbol{u}_{a}=-\operatorname{diag}(\{\mu_{_{i}}\})[1\ \cdots\ 1]^{^{\mathrm{T}}}$,\ 
$\boldsymbol{u}_{b}=-\boldsymbol{u}_{a}$, \  $\boldsymbol{\lambda}_{a}\in L_{2}(0,t_{_{F}})^{^{m}}$, \
$\boldsymbol{\lambda}_{b}\in L_{2}(0,t_{_{F}})^{^{m}}$\  and \ $\boldsymbol{\lambda} =(\boldsymbol{\lambda}_{a},\boldsymbol{\lambda}_{b})$, and the dual functional
$$\varphi_{_{\mathbf{D}}}^{^{K}}(\boldsymbol{\lambda_{_{a}}}, \boldsymbol{\lambda_{_{b}}})=\inf\{Lag_{_{K}}(\boldsymbol{u}, \boldsymbol{\lambda}): \boldsymbol{u}\in L_{_{2}}(0, t_{_{F}})^{^{m}}\}.$$

Once approximate solutions $\hat{\boldsymbol{\lambda}}=(\hat{\boldsymbol{\lambda}}_{a}, \hat{\boldsymbol{\lambda}}_{b})$ are obtained for the dual problem
$$\underline{Prob.\ D^{^{K}}:}\ \max_{\boldsymbol{\lambda}}\varphi_{D}^{^{K}}(\boldsymbol{\lambda}_{a}, \boldsymbol{\lambda}_{b})\ \ 
\text{subject to}\ \ \boldsymbol{\lambda}_{a}(t)\geq 0, \boldsymbol{\lambda}_{b}(t)\geq0\ \ \text{a.e. in}\ [0,t_{_{F}}],$$
the corresponding approximate solution to $\boldsymbol{u}_{c}^{^{K}}$, namely,
$\check{\boldsymbol{u}}_{c}^{^{K}}(\hat{\boldsymbol{\lambda}})=\displaystyle \operatorname{arg\ min}_{\boldsymbol{u}}Lag_{_{K}}(\boldsymbol{u},\hat{\boldsymbol{\lambda}})$ is given by the following proposition.

\begin{proposition}\label{prop:5.3}
\noindent
\textbf{\emph{(a)}} The unique solution of $\min_{_{\boldsymbol{u}}}\ Lag_{_{K}}(\boldsymbol{u}, \boldsymbol{\lambda})$ is given by
\[
\boldsymbol{u}_{c}^{^{K}}[\boldsymbol{\lambda} ]=\boldsymbol{u}_{_{K}}
-\mathbf{F}_{_{K}}(\rho_{_{\boldsymbol{u}}}^{^{-1}}\mathbf{I}-(\rho_{_{\boldsymbol{u}}}\mathbf{I}+\mathbf{G}_{_{K}})^{^{-1}})
\bar{\boldsymbol{\alpha}}^{^{K}}_{_{\boldsymbol{\lambda}}}+\rho_{_{\boldsymbol{u}}}^{^{-1}}(\boldsymbol{\lambda}_{a}-\boldsymbol{\lambda}_{b}),
\]
where\
$\boldsymbol{\lambda}_{ab}=\boldsymbol{\lambda}_{a}-\boldsymbol{\lambda}_{b},\ \  \boldsymbol{\xi}^{^{K}}_{_{\boldsymbol{\lambda}}}\triangleq \displaystyle\int_{0}^{t_{_{F}}}\mathbf{F}^{^{\mathrm{T}}}_{_{K}}(\tau )(\boldsymbol{\lambda}_{a}(\tau)-\boldsymbol{\lambda}_{b} (\tau ))d\tau
\ \ \mbox{and}\ \ \mathbf{G}_{_{K}}\bar{\boldsymbol{\alpha}}^{^{K}}_{_{\boldsymbol{\lambda}}}=\boldsymbol{\xi}^{^{K}}_{_{\boldsymbol{\lambda}}}.$\hfill$\nabla$
\end{proposition}

It  should be noted that the dual functional $\varphi_{_{D}}^{^{K}}(\boldsymbol{\lambda}_{a}, \boldsymbol{\lambda}_{b})$ can be written as
$$\varphi_{_{D}}^{^{K}}(\boldsymbol{\lambda}_{a}, \boldsymbol{\lambda}_{b})
= \mathcal{J}_{_{K}}(\boldsymbol{u}_{_{K}})
+\hat{\varphi}_{_{D}}^{^{K}}(\boldsymbol{\lambda}_{a}, \boldsymbol{\lambda}_{b}),$$
where 
$$
\hat{\varphi}_{_{D}}^{^{K}}(\boldsymbol{\lambda}_{a}, \boldsymbol{\lambda}_{b})=
-\rho_{_{\boldsymbol{u}}}^{^{-1}}\langle\boldsymbol{\lambda}_{ab}, \boldsymbol{\lambda}_{ab}\rangle
+ \rho_{_{\boldsymbol{u}}}^{^{-1}}\left\langle(\mathbf{I}+\rho_{_{F}}\mathbf{G}_{_{K}})^{^{-1}}\boldsymbol{\xi}_{_{\boldsymbol{\lambda}}}^{^{K}},
\boldsymbol{\xi}_{_{\boldsymbol{\lambda}}}^{^{K}}\right\rangle_{_{E}}
-2\langle\boldsymbol{\xi}_{\lambda}^{^{K}}, \bar{\boldsymbol{\alpha}}_{_{K}}\rangle_{_{E}}
+2\langle\boldsymbol{\lambda}_{a}, \boldsymbol{u}_{a}\rangle
-2\langle\boldsymbol{\lambda}_{b}, \boldsymbol{u}_{b}\rangle,$$
so that $Prob.\ D^{^{K}}$ can be replaced by
$$\displaystyle\max_{\boldsymbol{\lambda}}\hat{\varphi}_{_{D}}^{^{K}}(\boldsymbol{\lambda}_{a}, \boldsymbol{\lambda}_{b})\ \ \text{subject to}\ \ \boldsymbol{\lambda}_{a}(t)\geq0,  \boldsymbol{\lambda}_{b}(t)\geq0\ \ \text{a.e. in}\ \ [0, t_{_{F}}].$$

\begin{remark}\label{remark:01a}
 Recall that $\boldsymbol{u}_{_{K}}=\mathbf{F}_{_{K}}\left(\rho_{_{\boldsymbol{u}}}\mathbf{I}+\mathbf{G}_{_{K}}\right)^{^{-1}}\bar{\boldsymbol{\theta}}_{r\mathrm{o}}^{^{K}}$ 
 and note that (since $\mathbf{M}(\mathbf{I}+\mathbf{M})^{^{-1}}=\mathbf{I}-(\mathbf{I}+\mathbf{M})^{^{-1}}=(\mathbf{I}+\mathbf{M})^{^{-1}}\mathbf{M}$) 
 $$(\rho_{_{\boldsymbol{u}}}\mathbf{I}+\mathbf{G}_{_{K}})^{^{-1}}=\rho_{_{\boldsymbol{u}}}^{^{-1}}(\mathbf{I}+\rho_{_{\boldsymbol{u}}}^{^{-1}} \mathbf{G}_{_{K}})^{^{-1}}
 =\rho_{_{\boldsymbol{u}}}^{^{-1}}\left\{\mathbf{I}-(\mathbf{I}+\rho_{_{\boldsymbol{u}}}^{^{-1}}\mathbf{G}_{_{K}})^{^{-1}}\rho_{_{\boldsymbol{u}}}^{^{-1}}\mathbf{G}_{_{K}}\right\}$$
 so that
 \begin{eqnarray*}\boldsymbol{u}_{c}^{^{K}}[\boldsymbol{\lambda}]&=&\boldsymbol{u}_{_{K}}
 -\mathbf{F}_{_{K}}\rho_{_{\boldsymbol{u}}}^{^{-1}}(\mathbf{I}
 +\rho_{_{\boldsymbol{u}}}^{^{-1}}\mathbf{G}_{_{K}})^{^{-1}}\rho_{_{\boldsymbol{u}}}^{^{-1}}\mathbf{G}_{_{K}}\bar{\boldsymbol{\alpha}}_{_{\boldsymbol{\lambda}}}^{^{K}}+
 \rho_{_{\boldsymbol{u}}}^{^{-1}}(\boldsymbol{\lambda}_{a}-\boldsymbol{\lambda}_{b})\\
 \Leftrightarrow\ \ \ \boldsymbol{u}_{c}^{^{K}}[\boldsymbol{\lambda}]&=&\boldsymbol{u}_{_{K}}
 -\mathbf{F}_{_{K}}\rho_{_{\boldsymbol{u}}}^{^{-1}}(\rho_{_{\boldsymbol{u}}}\mathbf{I}
 +\mathbf{G}_{_{K}})^{^{-1}}\mathbf{G}_{_{K}}\bar{\boldsymbol{\alpha}}_{_{\boldsymbol{\lambda}}}^{^{K}}+
 \rho_{_{\boldsymbol{u}}}^{^{-1}}(\boldsymbol{\lambda}_{a}-\boldsymbol{\lambda}_{b})\\
 \Leftrightarrow\ \ \ \boldsymbol{u}_{c}^{^{K}}[\boldsymbol{\lambda}]&=&\mathbf{F}_{_{K}}(\rho_{_{\boldsymbol{u}}}\mathbf{I}+\mathbf{G}_{_{K}})^{^{-1}}
 \{\bar{\boldsymbol{\theta}}_{r\mathrm{o}}^{^{K}}-\rho_{_{\boldsymbol{u}}}^{^{-1}}\boldsymbol{\xi}_{_{\boldsymbol{\lambda}}}^{^{K}}\}
 +\rho_{_{\boldsymbol{u}}}^{^{-1}}(\boldsymbol{\lambda}_{a}-\boldsymbol{\lambda}_{b})\\
\Leftrightarrow\ \ \ \boldsymbol{u}_{c}^{^{K}}[\boldsymbol{\lambda}]&=&\mathbf{F}_{_{K}}(\rho_{_{\boldsymbol{u}}}\mathbf{I}+\mathbf{G}_{_{K}})^{^{-1}}
 \bar{\boldsymbol{\theta}}_{r}^{^{K}}(\boldsymbol{\lambda})+\rho_{_{\boldsymbol{u}}}^{^{-1}}(\boldsymbol{\lambda}_{a}-\boldsymbol{\lambda}_{b}),\\
 \boldsymbol{u}_{c}^{^{K}}[\boldsymbol{\lambda}]&=& \check{\boldsymbol{u}}_{c}^{^{K}}[\boldsymbol{\lambda}]+\rho_{_{\boldsymbol{u}}}^{-1}(\boldsymbol{\lambda}_{_{\boldsymbol{a}}}-\boldsymbol{\lambda}_{_{\boldsymbol{b}}}),
\end{eqnarray*}
where $\check{\boldsymbol{u}}_{c}^{^{K}}[\boldsymbol{\lambda}] =\mathbf{F}_{_{K}}(\rho_{_{\boldsymbol{u}}}\mathbf{I}+\mathbf{G}_{_{K}})^{-1}\bar{\boldsymbol{\theta}}_{r}^{^{K}}(\boldsymbol{\lambda})$ and $\bar{\boldsymbol{\theta}}_{r}^{^{K}}(\boldsymbol{\lambda})=\bar{\boldsymbol{\theta}}_{r\mathrm{o}}^{^{K}}
 -\rho_{_{\boldsymbol{u}}}^{^{-1}}\boldsymbol{\xi}_{_{\boldsymbol{\lambda}}}^{^{K}}$. It can thus be seen the optimal solution 
 $\boldsymbol{u}_{c}^{^{K}}[\boldsymbol{\lambda}_{_{K}}]$ of the constrained problem is obtained by adding a ``correction term''\ 
 $\rho_{_{\boldsymbol{u}}}^{^{-1}}(\boldsymbol{\lambda}_{a}^{^{K}}-\boldsymbol{\lambda}_{b}^{^{K}})$ to the output of a linear autonomous system $\check{\boldsymbol{u}}_{c}^{^{K}}[\boldsymbol{\lambda}](\tau)
=(\mathbf{M}_{_{\boldsymbol{\beta}}}^{^{K}})^{^{\mathrm{T}}}\boldsymbol{x}_{_{\boldsymbol{u}}}^{c}(\tau)$ where $\boldsymbol{x}_{_{\boldsymbol{u}}}^{c}$ is
solution of the linear ordinary differential equation\ $\dot{\boldsymbol{x}}_{_{\boldsymbol{u}}}^{c}(\tau)=-\mathbf{A}_{_{K}}^{^{\mathrm{T}}}\boldsymbol{x}_{_{\boldsymbol{u}}}^{c}(\tau), \ \tau\geq 0\ \
 \text{with initial condition}\ \ \boldsymbol{x}_{_{\boldsymbol{u}}}^{c}(0)=(\rho_{_{\boldsymbol{u}}}\mathbf{I}+\mathbf{G}_{_{K}})^{^{-1}}
 \bar{\boldsymbol{\theta}}_{r}^{^{K}}(\boldsymbol{\lambda}_{_{K}}). $
 \hfill$\nabla$
\end{remark}

It should be noted that $\boldsymbol{u}_{c}^{^{K}}[\boldsymbol{\lambda}]$ is only guaranteed to be feasible when
$\boldsymbol{\lambda}=\boldsymbol{\lambda}^{\mathrm{o}}=(\boldsymbol{\lambda}_{a}^{\mathrm{o}}, \boldsymbol{\lambda}_{b}^{\mathrm{o}})$, \emph{i.e.},
when $\boldsymbol{\lambda}$ is the optimal solution of the dual problem ($Prob.\ D^{^{K}}$). However, a feasible 
$\boldsymbol{u}_{_{R}}^{^{K}}[\hat{\boldsymbol{\lambda}}]$ can be obtained in a natural, heuristic way from an approximation $\hat{\boldsymbol{\lambda}}$ of $\boldsymbol{\lambda}^{\mathrm{o}}$ along the following lines. If $(\boldsymbol{\lambda}_{a}^{\mathrm{o}}, \boldsymbol{\lambda}_{b}^{\mathrm{o}})$ is optimal 
$\boldsymbol{u}_{c}^{^{K}} \in S_{_{\boldsymbol{u}F}}$. Moreover, $\boldsymbol{\lambda}_{a}^{\mathrm{o}}(\tau)=0$ and
$\boldsymbol{\lambda}_{b}^{\mathrm{o}}(\tau)=0$ (hence, $\boldsymbol{\lambda}_{ab}^{\mathrm{o}}(\tau)=0$) whenever 
$\boldsymbol{u}_{c}^{^{K}}[\boldsymbol{\lambda}^{\mathrm{o}}](\tau)\in (\boldsymbol{u}_{a}, \boldsymbol{u}_{b})$ so that, in this case, 
$\check{\boldsymbol{u}}_{_{K}}^{c}[\boldsymbol{\lambda}^{\mathrm{o}}](\tau)$ also belongs to 
$(\boldsymbol{u}_{a}, \boldsymbol{u}_{b})$. When $\boldsymbol{\lambda}_{a}^{\mathrm{o}}(\tau)\neq 0$ (respectively 
$\boldsymbol{\lambda}_{b}^{\mathrm{o}}(\tau)\neq 0$) $\boldsymbol{u}_{c}^{^{K}}[\boldsymbol{\lambda}^{\mathrm{o}}](\tau)=\boldsymbol{u}_{a}$ and 
$\check{\boldsymbol{u}}_{_{K}}^{c}[\boldsymbol{\lambda}^{\mathrm{o}}](\tau)<\boldsymbol{u}_{a}$ (respectively,
$\boldsymbol{u}_{c}^{^{K}}[\boldsymbol{\lambda}^{\mathrm{o}}](\tau)=\boldsymbol{u}_{b}$ and 
$\check{\boldsymbol{u}}_{_{K}}^{c}[\boldsymbol{\lambda}^{\mathrm{o}}](\tau)>\boldsymbol{u}_{a}$). This suggests a heuristic way of obtaining a feasible $\boldsymbol{u}_{_{R}}^{^{K}}[\boldsymbol{\lambda}]$, namely, $\boldsymbol{u}_{_{R}}^{^{K}}[\boldsymbol{\lambda}](\tau)=\check{\boldsymbol{u}}_{_{K}}^{c}[\boldsymbol{\lambda}](\tau)$\  if \ $\check{\boldsymbol{u}}_{_{K}}^{c}[\boldsymbol{\lambda}]\in (\boldsymbol{u}_{a},\boldsymbol{u}_{b})$,\ $\boldsymbol{u}_{_{R}}^{^{K}}[\boldsymbol{\lambda}](\tau)=\boldsymbol{u}_{a}$ if \
$\check{\boldsymbol{u}}_{_{K}}^{c}[\boldsymbol{\lambda}](\tau)\leq \boldsymbol{u}_{a}$ and 
$\boldsymbol{u}_{_{R}}^{^{K}}[\boldsymbol{\lambda}](\tau)=\boldsymbol{u}_{b}$\  if 
$\check{\boldsymbol{u}}_{_{K}}^{c}[\boldsymbol{\lambda}](\tau)\geq \boldsymbol{u}_{b}$. As $\boldsymbol{\lambda}\rightarrow\boldsymbol{\lambda}^{\mathrm{o}}$, $\boldsymbol{u}_{_{R}}^{^{K}}[\boldsymbol{\lambda}]
\rightarrow\boldsymbol{u}_{c}^{^{K}}[\boldsymbol{\lambda}^{\mathrm{o}}]$.\bigskip

Summing up, to compute an approximate solution to $Prob.\ II_{K}$ amounts to:\\
\noindent
\textbf{(1)}  Computing an approximate solution to the dual problem $Prob.\ D^{^{K}}$, say 
$\hat{\boldsymbol{\lambda}}^{\mathrm{o}}=(\hat{\boldsymbol{\lambda}}_{a},\hat{\boldsymbol{\lambda}}_{b})$.\medskip

\noindent
\textbf{(2)} Computing $\boldsymbol{u}_{c}^{^{K}}(\hat{\boldsymbol{\lambda}}^{\mathrm{o}})$ as given by Proposition \ref{prop:5.3}\textbf{(a)} and $\boldsymbol{u}_{_{R}}^{^{K}}(\hat{\boldsymbol{\lambda}}^{\mathrm{o}})$ (to ensure feasibility).

The most demanding step in the duality approach described above is the computation of an approximation $\hat{\boldsymbol{\lambda}}_{_{K}}=(\hat{\boldsymbol{\lambda}}_{aK}, \hat{\boldsymbol{\lambda}}_{bK})$ for the optimal
Lagrange multipliers. This can be accomplished by noting that the dual functional can be explicitly written as a function of the Lagrange multipliers (as in Proposition \ref{prop:5.3}\textbf{(\emph{b})}) and by relying on piecewise-linear continuous classes of Lagrange multipliers which are {linearly-}parametrized by their values on a grid
$\{t_{_{k}}\}\subset[0,t_{_{F}}]$ so that the non-negativeness of $\boldsymbol{\lambda}_{a}(\cdot)$ and
$\boldsymbol{\lambda}_{b}(\cdot)$ on $t\in [0, t_{_{F}}]$ is ensured by the constraints $\forall k$, 
$\boldsymbol{\lambda}_{a}(t_{_{k}})\geq 0$ and $\boldsymbol{\lambda}_{b}(t_{_{k}})\geq 0$. The appropriate duality problem can then be cast as a finite-dimensional, quadratic maximization problem with (coordinate-wise) non-negativeness constraints on the decision variables. This is described in detail in the Appendix.

\section{Examples and numerical results for the HEq}\label{sec:06}

In this section, two simple numerical examples are presented to illustrate the way the results above can be used to 
characterize control signals which aim at steering a solution of a heat equation (HEq, for short) over a given interval $[0, t_{_{F}}]$ towards a prescribed
final state. The PDE considered here is given by
\begin{eqnarray}
\frac{\partial \theta(\boldsymbol{x},t)}{\partial t}&=&
 \alpha\sum_{i=1}^{m_{_{x}}} \frac{\partial^{2}\theta(\boldsymbol{x},t)}{\partial x_{_{i}}^{2}}+f(\boldsymbol{x},t) \ \ \ \ \ \ \ \ \ \ \
 \ \ \ \ \ \ \ \ \ \ \ \ \  \forall \ \boldsymbol{x}\in \Omega,  \ \forall \ t\in (0,t_{F} )\ \label{ch6:eq-01}\\
 \theta (\boldsymbol{x},t)&=&0 \ \ \ \ \ \ \ \  \  \ \ \ \  \ \   \ \ \ \  \ \ (\text{Boundary Conditions}) \ \ \forall  \ t\in [0,t_{F} ],
 \ \forall \boldsymbol{x}\in \partial \Omega \label{ch6:eq-02}\\
\theta (\boldsymbol{x},0)&=&g(\boldsymbol{x}) \ \ \ \ \ \ \ \ \ \ \ \ \ \  \ \ \ \ \ \  \ (\text{Initial Condition})   \ \ \  \ \ \ \ \  \ \ \ \
\ \ \  \ \ \ \ \ \ \ \ \forall \boldsymbol{x}\in \Omega \label{ch6:eq-03}
\end{eqnarray}
with $\Omega=(0, L_{x_{_{1}}})\times\ \cdots\ \times (0, L_{x_{m_{_{x}}}})$. In this case, the eigenfunctions
$\phi_{_{\underline{\boldsymbol{k}}}}$, $\underline{\boldsymbol{k}}=(k_{_{1}},\mathellipsis,k_{_{m_{_{x}}}})$ of the operator $A$ are given by
\begin{equation}\label{eq:2.11New}
\phi_{_{\underline{\boldsymbol{k}}}}(\boldsymbol{x})=\prod_{i=1}^{m_{_{x}}}\sqrt{\frac{2}{L_{x_{_{i}}}}}\sin\left[\frac{k_{_{i}}\pi x_{_{i}}}{L_{x_{_{i}}}}\right]
\end{equation}
and the corresponding eigenvalues are $\lambda_{\underline{\boldsymbol{k}}}=-\sum_{i=1}^{m}\left[\frac{k_{i}\pi}{L_{x_{i}}}\right]^{2}$. Letting $\mathcal{S}_{I}^{K}=\{\underline{\boldsymbol{k}}=(k_{1},\mathellipsis,k_{m}):k_{i}\leq K\}$ the semigroup $S_{A}(\cdot)$ is given by
$$S_{A}(t)[\phi]=\sum_{k=1}^{\infty}\sum_{\underline{\boldsymbol{k}}\in \mathcal{S}_{I}^{K}}e^{\lambda_{\underline{\boldsymbol{k}}} t}\langle \phi, \phi_{\underline{\boldsymbol{k}}}\rangle\phi_{\underline{\boldsymbol{k}}}.$$

In this case, $X_{K}$ is the span of $\{\phi_{\underline{\boldsymbol{k}}}(x_{i}), i=1, \mathellipsis,m:\underline{\boldsymbol{k}}\in \mathcal{S}_{I}^{K}\}$ and
$$S_{K}(t)[\phi]=\sum_{\underline{\boldsymbol{k}}\in \mathcal{S}_{I}^{K}}e^{\lambda_{\underline{\boldsymbol{k}}}t}\langle\phi, \phi_{\underline{\boldsymbol{k}}}\rangle \phi_{\underline{\boldsymbol{k}}}.$$

Note that these eigenfunctions constitute a $L_{2}-$orthonormal set and hence,
$\{\phi_{\underline{\boldsymbol{k}}}:\underline{\boldsymbol{k}} \in \mathcal{S}_{I}^{K}\}$
is a $L_{2}-$orthonormal bases for $X_{K}$.

It is of particular interest here to illustrate the role of the coefficient $\rho_{_{\boldsymbol{u}}}$ or  $\rho_{_{F}}$ in improving final-state approximation, the
effect of imposing a peak-value constraint on the control signals (vis-\'{a}-vis the unconstrained ones) and the way piecewise-linear multipliers
yield approximation to the optimal control signals under peak-value constraints.

In Subsection \ref{subsec:061}, the one-dimensional HEq is considered under the action of a single scalar control signal (\emph{i.e.}, $m=1$). To
facilitate reading (and for concreteness) some of the relevant symbol definitions 
$(\mathbf{A}_{_{K}}, \bar{\boldsymbol{\beta}}_{_{\boldsymbol{S}K}}^{^{\mathrm{T}}}, \underline{\boldsymbol{\theta}}_{r\mathrm{o}}^{^{K}})$ are re-stated now for the 
basis functions $\left\{\sqrt{\frac{2}{L_{_{x}}}}\sin\left[\frac{k \pi x}{L_{_{x}}}\right]\right\}$, $k=1, \mathellipsis, K$. Exploiting the simple case at hand, an explicit upper bound is presented on the $L_{2}-$norm of the approximation error to the final state of the HEq as a function of the corresponding approximation error in the truncated (ODE in $\mathbb{R}^{^{K}}$) problem. In Subsection \ref{subsec:062}, numerical results are presented for the one-dimensional example of Subsection \ref{subsec:061} with a given temperature distribution taken as desired final state and one actuator located at the mid-point of the interval $(0, L_{_{x}})$.
In Subsection \ref{subsec:063}, numerical results are presented for the two-dimensional  HEq with one scalar control signal; for one desired final state, numerical experiments were carried with two different values of $\rho_{_{F}}$.

\subsection{A One-Dimensional Example}\label{subsec:061}

Let $\Omega=(0, L_{_{x}})$, consider the \emph{one-dimensional} heat equation with homogeneous Dirichlet boundary  conditions and
single-point control $\boldsymbol{u}: [0, t_{_{F}}] \rightarrow \mathbb{R}$, \emph{i.e.},
\begin{eqnarray*}
 \ \frac{\partial \theta}{\partial t}(x,t)&=& k_{_{\alpha}}\frac{\partial^2\theta}{\partial x^{2}}(x, t) + 
 \boldsymbol{\beta}_{_{\boldsymbol{S}}}(x)\boldsymbol{u}(t)\ \ \ \  \ \ \ \ \ \  \ \ \ \ \ \ \ \  \ \ \ \ \  \forall \ t\in (0,t_{F} ), \forall \ x\in \Omega, \\
 \theta (x,0)&=&0 \ \ \ \   \ \ \ \ \ \ \ \ \ \ \ \ \ \ \  \ \ \ \ \ \ \ \ \mbox{(zero initial condition)}\ \  \ \ \ \ \ \ \ \ \
 \ \ \ \ \forall x\in \Omega, \\
\theta (0,t)&=&\theta (L_{_{x}},t)=0  \ \ \ \ \ \ \ \ \ \ \ \ \ \  \mbox{(boundary conditions)}\  \ \ \ \ \ \ \ \ \forall  \ t\in [0,t_{F}]
\end{eqnarray*}
and the corresponding Galerkin approximations given by
\begin{eqnarray*}
\forall\ k =1, 2, \mathellipsis, K, \ \ \  \left\langle \frac{\partial \theta}{ \partial t} (\cdot, t), \phi_{_{k}}\right\rangle&=&
-k_{_{\alpha}}\left\langle \frac{\partial \theta}{\partial x}(\cdot, t), \frac{\partial \phi_{_{k}}}{\partial x}\right\rangle
+\left\langle \boldsymbol{\beta}_{_{\boldsymbol{S}}}, \phi_{_{k}}\right\rangle\boldsymbol{u}(t)\\ \\
\left\langle\theta (\cdot,0), \phi_{_{k}}\right\rangle&=&0, 
\end{eqnarray*}
where $\phi_{_{k}}:[0, L_{_{x}}]\rightarrow \mathbb{R}$ is given by
$\phi_{_{k}}(x)=\displaystyle\sqrt{\frac{2}{L_{_{x}}}}\sin\left[\frac{k \pi x}{L_{_{x}}}\right]$  and $X_{_{K}}=\operatorname{span}\{\phi_{1}, \mathellipsis,\phi_{K}\}$.
Approximate solutions $\boldsymbol{u}_{_{K}}$ and $\boldsymbol{u}_{c}^{^{K}}$ are sought to the problems
$$\text{\underline{\emph{Prob.} $I:$}} \ \displaystyle\min_{\boldsymbol{u} \in L_{2}(0, t_{_{F}})} 
\check{\mathcal{J}}(\boldsymbol{u}; \rho_{_{F}}) \ \ \ \ \text{or}\ \ \ \ \text{\underline{\emph{Prob.} $I_{c}:$}} \ 
\displaystyle\min_{\boldsymbol{u}\in S_{_{\boldsymbol{u}F}}} \check{\mathcal{J}}(\boldsymbol{u}; \rho_{_{F}}),$$
where $\check{\mathcal{J}}(\boldsymbol{u}; \rho_{_{F}})=\|\boldsymbol{u} \|_{_{L_{2}(0, t_{_{F}})}}^{^{2}}
+ \rho_{_{F}}\|\mathcal{T}_{_{\theta}}[\boldsymbol{u}]-\theta_{r\mathrm{o}}\|_{_{2}}^{^{2}}$, $\theta_{r\mathrm{o}}$ is the final state to be approximately reached, 
$\rho_{_{F}}=\rho_{_{\boldsymbol{u}}}^{^{-1}}$ and
$$S_{_{\boldsymbol{u}F}}=\left\{\boldsymbol{u}\in L_{\infty}(0, t_{_{F}}): \|\boldsymbol{u} \|_{_{L_{\infty}(0, t_{_{F}})}} \leq \mu_{_{\boldsymbol{u}}}\right\}.$$

In this case, 
$\{\mathbf{A}_{_{K}}\}_{_{k\ell}}=-k_{_{\alpha}}\left\langle \sqrt{\frac{2}{L_{_{x}}}}\left[-\frac{k \pi}{L_{_{x}}}\right]\cos\left[\frac{k \pi\ (\cdot)}{L_{_{x}}}\right],
[\sqrt{\frac{2}{L_{_{x}}}}\left[-\frac{\ell \pi}{L_{_{x}}}\right]\cos\left[\frac{\ell \pi\ (\cdot)}{L_{_{x}}}\right]\right\rangle$, \emph{i.e.},
$$\mathbf{A}_{_{K}}=diag\left\{-k_{_{\alpha}}\left[\displaystyle\frac{k \pi}{L_{_{x}}}\right]^{^{2}}\right\}\ \ \ \text{and}\ \ \
\bar{\boldsymbol{\beta}}_{_{\boldsymbol{S}K}}^{^{\mathrm{T}}}=
\left[\left\langle \boldsymbol{\beta}_{_{\boldsymbol{S}}}, \sqrt{\frac{2}{L_{_{x}}}}\sin\left[\frac{1\pi\ (\cdot)}{L_{_{x}}}\right]\right\rangle
\cdots \left\langle \boldsymbol{\beta}_{_{\boldsymbol{S}}}, \sqrt{\frac{2}{L_{_{x}}}}\sin\left[\frac{K\pi\ (\cdot)}{L_{_{x}}}\right]\right\rangle \right].$$

The optimal solution of \emph{Prob. I} is given by, $\forall \tau \in [0, t_{_{F}}]$
$$\boldsymbol{u}(\tau)=
\bar{\boldsymbol{\beta}}_{_{\boldsymbol{S}K}}^{^{\mathrm{T}}}\exp\{\mathbf{A}_{_{K}}^{^{\mathrm{T}}}(t_{_{F}}-\tau)\}\bar{\boldsymbol{\alpha}}_{_{K}},$$
where \ \ 
$
\bar{\boldsymbol{\alpha}}_{_{K}}=(\mathbf{I}+\rho_{_{F}}\mathbf{G}_{_{K}})^{^{-1}}\rho_{_{F}}\bar{\boldsymbol{\theta}}_{r\mathrm{o}}^{^{K}}$\ \ and\\
$$(\bar{\boldsymbol{\theta}}_{r\mathrm{o}}^{^{K}})^{^{\mathrm{T}}}=\left[\left\langle\theta_{r\mathrm{o}},
\sqrt{\frac{2}{L_{_{x}}}} \sin\left[\frac{1\pi\ (\cdot)}{L_{_{x}}}\right] \right\rangle\ \mathellipsis\ 
\left\langle \theta_{r\mathrm{o}},
\sqrt{\frac{2}{L_{_{x}}}} \sin\left[\frac{K\pi\ (\cdot)}{L_{_{x}}}\right] \right\rangle\right].$$

Moreover, in this case $\mathbf{A}_{_{K}}\in \mathbb{R}^{n_{_{K}}}$ is diagonal which allows for the Lyapunov equation to be solved term by term,
\begin{eqnarray*}
&&\{\mathbf{A}_{_{K}}\mathbf{G}_{_{K}}\}_{_{k\ell}}+\{\mathbf{G}_{_{K}}\mathbf{A}_{_{K}}^{^{\mathrm{T}}}\}_{_{k\ell}}=\{\check{\mathbf{M}}_{_{K}}\}_{_{k\ell}} \ \ \ \ \ \ \ \ \ \ \ \Leftrightarrow\\
&&\{\mathbf{A}_{_{K}}\}_{_{kk}}\{\mathbf{G}_{_{K}}\}_{_{k\ell}}+\{\mathbf{G}_{_{K}}\}_{_{k\ell}}\{\mathbf{A}_{_{K}}\}_{_{\ell \ell}}
=\{\check{\mathbf{M}}_{_{K}}\}_{_{k\ell}} \ \Leftrightarrow\\
&&\{\mathbf{G}_{_{K}}\}_{_{k\ell}}=\{\check{\mathbf{M}}_{_{K}}\}_{_{k\ell}}/\left(\{\mathbf{A}_{_{K}}\}_{_{kk}}+\{\mathbf{A}_{_{K}}\}_{_{\ell \ell}}\right)\ \ \ \ \ \  \Leftrightarrow\\
&&\{\mathbf{G}_{_{K}}\}_{_{k \ell}}=\frac{1}{\{\mathbf{A}_{_{K}}\}_{_{kk}}+\{\mathbf{A}_{_{K}}\}_{_{_{\ell \ell}}}}
\left[\exp\left\{\left(\{\mathbf{A}_{_{K}}\}_{_{kk}}+
\{\mathbf{A}_{_{K}}\}_{_{\ell\ell}}\right)t_{_{F}}\right\}^{-1}\right]\left\{\mathbf{M}_{_{\boldsymbol{\beta}}}^{^{K}}(\mathbf{M})_{_{\boldsymbol{\beta}}}^{^{K}})^{^{\mathrm{T}}}\right\}_{_{k\ell}}.
\end{eqnarray*}
Note also that 
$\left\|\mathcal{T}_{_{\theta}}[\boldsymbol{u}]-\mathcal{T}_{_{\theta}}^{^{K}}[\boldsymbol{u}]\right\|_{_{2}}^{^{2}}
=\left\|\displaystyle\sum_{k=K+1}^{\infty}\boldsymbol{c}_{_{k}}(t_{_{F}};\boldsymbol{u})\phi_{_{k}}\right\|_{_{2}}^{^{2}}
=\displaystyle\sum_{k=K+1}^{\infty}\boldsymbol{c}_{_{k}}(t_{_{F}}; \boldsymbol{u})^{^{2}}$, and\\
$\boldsymbol{c}_{_{k}}(t_{_{F}}, \boldsymbol{u})=
\displaystyle\int_{0}^{t_{_{F}}}\exp\left[-k_{_{\alpha}}\left[\frac{k \pi}{L_{_{x}}}\right]^{^{2}}(t_{_{F}}-\tau)\right]\boldsymbol{\beta}_{_{\boldsymbol{S}k}}\boldsymbol{u}(\tau)d\tau$,
where $\check{\boldsymbol{\beta}}_{_{\boldsymbol{S}k}} \triangleq \langle\boldsymbol{\beta}_{_{\boldsymbol{S}}}, \phi_{_{K}}\rangle$, so that
(in the light of Cauchy-Schwarz inequality)
\begin{eqnarray*}\Rightarrow \ \ \boldsymbol{c}_{_{k}}(t_{_{F}}; \boldsymbol{u})^{^{2}}
&\leq& \left|\check{\boldsymbol{\beta}}_{_{\boldsymbol{S}k}}\right|^{^{2}}
\left\|\exp\left[-k_{_{\alpha}}\left[\frac{k\pi}{L_{_{x}}}\right]^{^{2}}(t_{_{F}}-\cdot)\right]\right\|_{_{L_{2}(0, t_{_{F}})}}^{^{2}}
\|\boldsymbol{u}\|_{_{L_{2}(0, t_{_{F}})}}^{^{2}}\\ \\
\Rightarrow \ \ \boldsymbol{c}_{_{k}}(t_{_{F}}; \boldsymbol{u})^{^{2}}&\leq& \left|\check{\boldsymbol{\beta}}_{_{\boldsymbol{S}k}}\right|^{^{2}}
\frac{1}{k_{_{\alpha}} \left[\frac{k\pi}{L_{_{x}}}\right]^{^{2}}}\left\{1-\exp\left[-k_{_{\alpha}}\left[\frac{k\pi}{L_{_{x}} }\right]^{^{2}}t_{_{F}}\right]\right\}
\|\boldsymbol{u}\|_{_{L_{2}(0, t_{_{F}})}}^{^{2}}\\ \\
&\leq&\left|\check{\boldsymbol{\beta}}_{_{\boldsymbol{S}k}}\right|^{^{2}}\frac{1}{k_{_{\alpha}} \left[\frac{k\pi}{L_{_{x}}}\right]^{^{2}}}
\|\boldsymbol{u}\|_{_{L_{2}(0, t_{_{F}})}}^{^{2}}.
\end{eqnarray*}
It then follows that 
\begin{equation}\label{eq:36}
\|\mathcal{T}_{_{\theta}}[\boldsymbol{u}]-\mathcal{T}_{_{\theta}}^{^{K}}[\boldsymbol{u}]\|_{_{2}}^{^{2}}\leq
\|\boldsymbol{\beta}_{_{\boldsymbol{S}}}-\hat{\boldsymbol{\beta}}_{_{\boldsymbol{S}K}}\|_{_{2}}^{^{2}}
\frac{1}{k_{_{\alpha}} \{(K+1)\frac{\pi}{L_{_{x}}}\}^{^{2}}}\|\boldsymbol{u}\|_{_{L_{2}(0, t_{_{F}})}}^{^{2}},
\end{equation}
where $\hat{\boldsymbol{\beta}}_{_{\boldsymbol{S}K}}\triangleq \sum_{k=1}^{K}\check{\boldsymbol{\beta}}_{_{\boldsymbol{S}k}}\phi_{_{k}}$.\\


Thus, combining (\ref{eq:33}), (\ref{eq:35}) and (\ref{eq:36}) gives an upper bound on 
$\|\mathcal{T}_{_{\theta}}[\boldsymbol{u}]-\theta_{r\mathrm{o}}\|_{_{2}}^{^{2}}$ namely, 
$$\|\mathcal{T}_{_{\theta}}[\boldsymbol{u}]-\theta_{r\mathrm{o}}\|_{_{2}}^{^{2}}
\leq\|\boldsymbol{e}_{_{K}}[\boldsymbol{u}]\|_{_{E}}^{^{2}}
+\left\{\frac{\|\boldsymbol{\beta}_{_{\boldsymbol{S}}}-\hat{\boldsymbol{\beta}}_{_{\boldsymbol{S}K}}\|_{_{2}}}{\sqrt{k_{_{\alpha}}}(K+1)\frac{\pi}{L_{_{x}}}}
\|\boldsymbol{u}\|_{_{L_{2}(0, t_{_{F}})}}+\|\theta_{r\mathrm{o}}-\theta_{r\mathrm{o}}^{^{K}}\|_{_{2}}\right\}^{^{2}},$$
where $\boldsymbol{e}_{_{K}}[\boldsymbol{u}]\triangleq\mathcal{T}_{_{\theta}}^{^{K}}[\boldsymbol{u}]-\theta_{r\mathrm{o}}^{^{K}}$.

Thus, as the approximating property of $span\left\{\sqrt{\frac{2}{L_{_{x}}}}\sin\left[\frac{k\pi x}{L_{_{x}}}\right]:k=1, \mathellipsis, K\right\}$,
$K\geq1$ ensures that\break $\|\boldsymbol{\beta}_{_{\boldsymbol{S}}}-\hat{\boldsymbol{\beta}}_{_{\boldsymbol{S}K}}\|_{_{2}}\rightarrow0$ and
$\|\theta_{r\mathrm{o}}-\theta_{r\mathrm{o}}^{^{K}}\|_{_{2}}\rightarrow0$ as $K\rightarrow \infty$, the $L_{2}-$norm of the approximation error for the final state for 
the HEq (\emph{i.e.}, $\|\mathcal{T}_{_{\theta}}[\boldsymbol{u}]-\theta_{r\mathrm{o}}\|_{_{2}}$) approaches the corresponding error for the $K-$dimensional ODE,
\emph{i.e.}, $\|\boldsymbol{e}_{_{K}}[\boldsymbol{u}]\|_{_{E}}$). In other words, the latter provides progressively more accurate a posteriori estimates for the former, as $K\to\infty$.


\subsection{Numerical Results for the One-dimensional Example}\label{subsec:062}

\begin{figure}[!ht]
\begin{center}
\includegraphics[width=10cm]{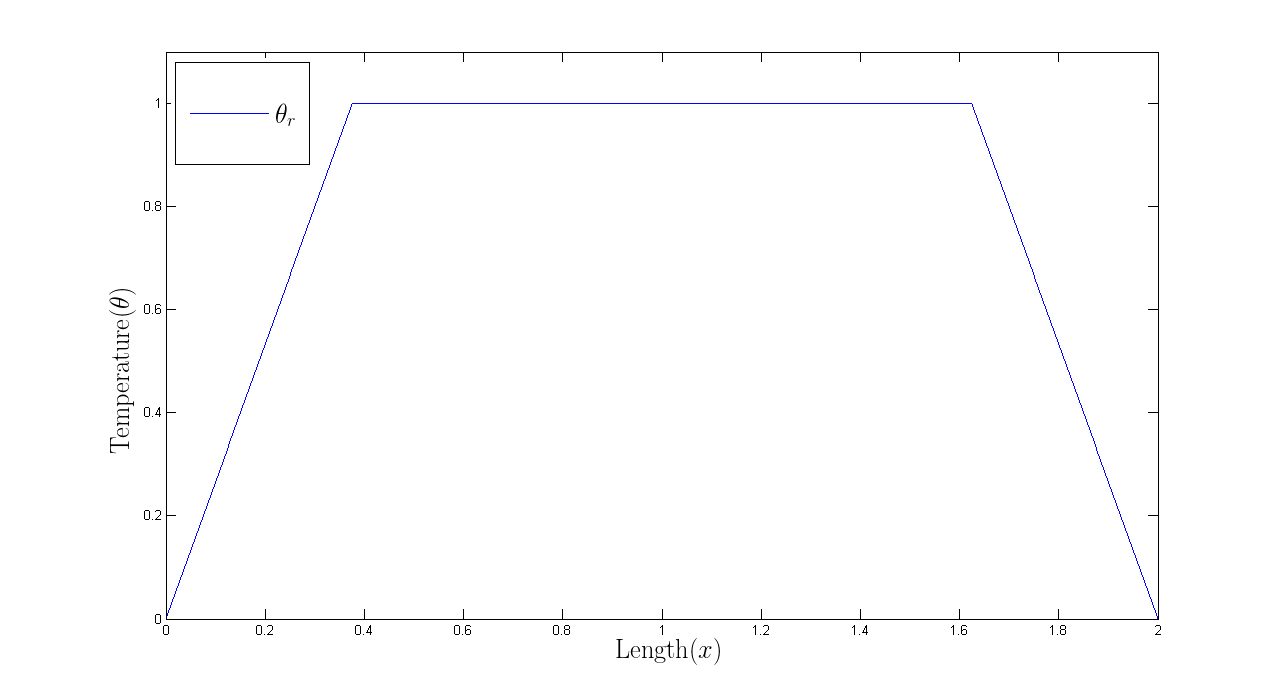}
\end{center}
\caption{$\theta_{r}$: target final state.}
\label{ch6:fig07}
\end{figure}

\begin{figure}[!ht]
\begin{center}
\includegraphics[width=10cm]{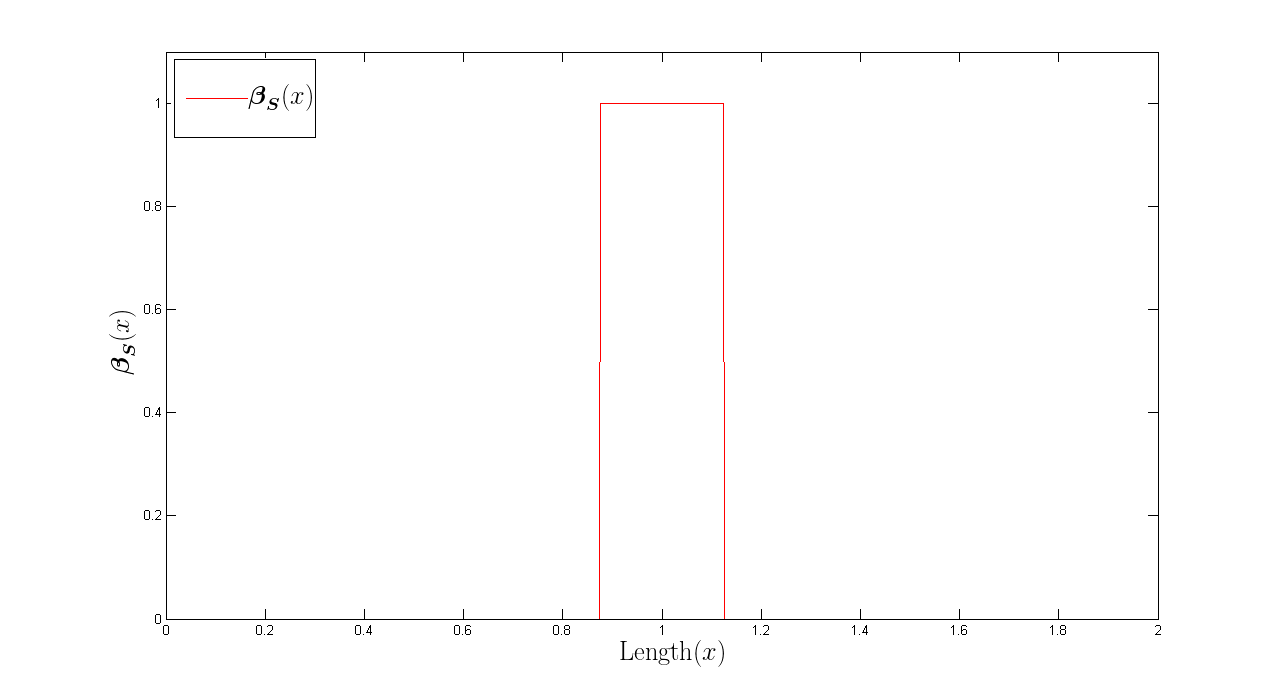}
\end{center}
\caption{$\boldsymbol{\beta}_{_{\boldsymbol{S}}}$: control-to-state actuator.}
\label{ch6:fig08}
\end{figure}

$Prob.\ I_{_{K}}$ and $Prob.\ D_{\boldsymbol{\gamma}}^{^{K}}$ were numerically solved for the pair
$(\theta_{r}, \boldsymbol{\beta}_{_{\boldsymbol{S}}})$ displayed in Figures \ref{ch6:fig07}, \ref{ch6:fig08}, with\
$\rho_{_{F}}=2000$, $K=5$, $L_{_{x}}= 2$, and $N_{_{\boldsymbol{\lambda}}}=30$. 

First, an approximate solution $\boldsymbol{u}_{_{K}}$ was obtained for $Prob.\ I_{_{K}}$ -- see Table \ref{ch6:tab05}  for the values of its 
$L_{2}(0, t_{_{F}})$ and $L_{\infty}(0, t_{_{F}})$ norms and the corresponding values of the cost-functional and the $L_{2}(0, 1)$
norm of the final-state error (projected on $span\{\phi_{_{1}}, \mathellipsis, \phi_{_{K}}\}$).
\begin{table}[!ht]
\begin{center}
\begin{tabular}{|c|c|c|c|}
\hline	$\check{\mathcal{J}}_{_{K}}(\boldsymbol{u}_{_{K}}; \rho_{_{F}})$ & $\|\boldsymbol{u}_{_{K}}\|_{_{2}}$ & $\|\boldsymbol{u}_{_{K}}\|_{\infty}$ & 
$\|\mathcal{T}_{_{\theta}}^{^{K}}[\boldsymbol{u}_{_{K}}]-\boldsymbol{\theta}_{r\mathrm{o}}^{^{K}}\|_{_{2}}$\\
\hline \hline
        283.5120& 13.5254 & 23.5491 & 0.2242 \\
\hline
\end{tabular}
\caption{Unconstrained problem for the pair $(\theta_{r}, \boldsymbol{\beta}_{_{\boldsymbol{S}}})$, $\rho_{_{F}}=2000$.}
\label{ch6:tab05}
\end{center}
\end{table}

A numerical solution $\check{\boldsymbol{u}}_{_{K}}^{^{R}}$ was then obtained for $Prob.\ I_{_{cK}}$  on the basis of piecewise-linear Lagrange multipliers (see Appendix) parametrized by their values on a grid in $[0, t_{F}]$ with $N_{\lambda}$ points, with the prescribed upper limit
$\mu_{_{\boldsymbol{u}}}$ on the $L_{\infty}(0, t_{_{F}})$--norm of $\boldsymbol{u}$ being set at $\mu_{_{\boldsymbol{u}}}=18$. Table \ref{ch6:tab06} exhibits the corresponding assessment data for $\check{\boldsymbol{u}}_{_{K}}^{^{R}}$.
\begin{table}[!ht]
\begin{center}
\begin{tabular}{|c|c|c|c|c|}
\hline	$\check{\mathcal{J}}_{_{K}}(\check{\boldsymbol{u}}_{_{K}}^{^{R}}; \rho_{_{F}})$ & $\varphi_{_{D}}^{^{K}}(\boldsymbol{\lambda}^{^{K}})$ & $\|\check{\boldsymbol{u}}_{_{K}}^{^{R}}\|_{_{2}}$ &
$\|\check{\boldsymbol{u}}_{_{K}}^{^{R}}\|_{\infty}$ & $\|\mathcal{T}_{_{\theta}}^{^{K}}[\check{\boldsymbol{u}}_{_{K}}^{^{R}}]-\boldsymbol{\theta}_{_{r\mathrm{o}}}^{^{R}}\|_{_{2}}$\\
\hline \hline
       300.2274 &  286.3859 &  12.6191 & 18.0000 & 0.2655 \\
\hline
\end{tabular}
\caption{Constrained problem for the pair $(\theta_{r}, \boldsymbol{\beta}_{_{\boldsymbol{S}}})$, $\rho_F=2000$.}
\label{ch6:tab06}
\end{center}
\end{table}

\begin{figure}[!ht]
\begin{center}
\includegraphics[width=10cm]{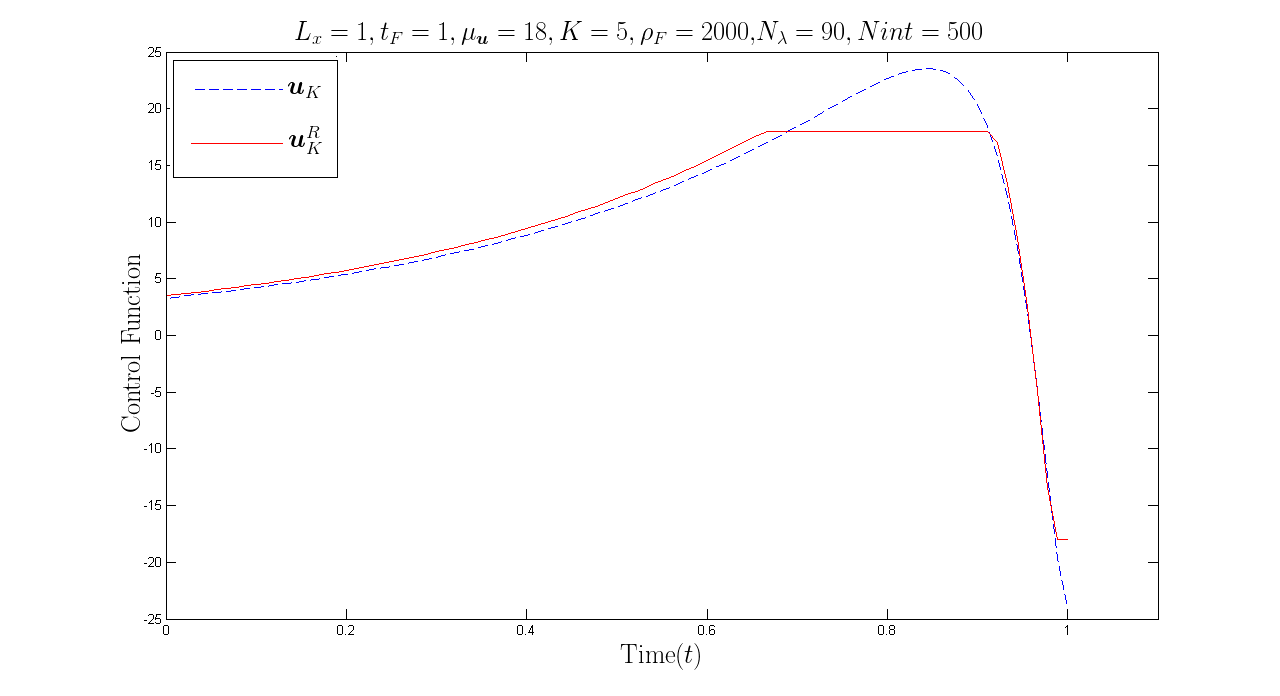}
\end{center}
\caption{Control signals $\boldsymbol{u}_{_{K}}$ (blue dashed), $\boldsymbol{u}_{_{R}}^{^{K}}$ (red solid) for $\rho_F=2000$.}
\label{ch6:fig09}
\end{figure}

\begin{figure}[!ht]
\begin{center}
\includegraphics[width=9.5cm]{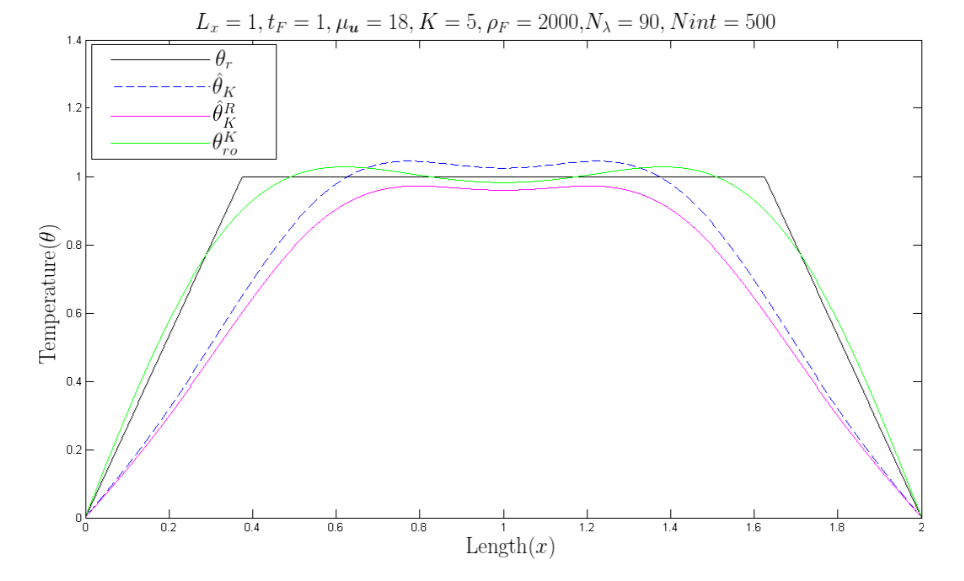}
\end{center}
\caption{Approximations to target final state for $\rho_{_{F}}=2000$.}
\label{ch6:fig10}
\end{figure}

Note that $\check{\mathcal{J}}_{_{K}}(\check{\boldsymbol{u}}_{_{K}}^{^{R}}; \rho_{_{F}})$ may only exceed the optimal value 
$\mathcal{J}_{_{cK}}^{\mathrm{o}}$ of $Prob.\ I_{_{cK}}$ by less than $5\%$ (of $\mathcal{J}_{_{cK}}^{\mathrm{o}}$). Figures \ref{ch6:fig09} and \ref{ch6:fig10} 
respectively display the plots of $\boldsymbol{u}_{_{K}}$ (dashed blue) and $\check{\boldsymbol{u}}_{_{K}}^{^{R}}$ and those of $\theta_{r\mathrm{o}}^{^{K}}$ 
(the projection of $\theta_{r\mathrm{o}}$ on $span\{\phi_{_{1}}, \mathellipsis, \phi_{_{K}}\}$), 
$\check{\theta}_{_{K}}\triangleq \mathcal{T}_{_{\theta}}^{^{K}}[\boldsymbol{u}_{_{K}}]$ (dashed blue) and 
$\check{\theta}_{_{K}}^{R}\triangleq \mathcal{T}_{_{\theta}}^{^{K}}[\boldsymbol{u}_{_{R}}^{^{K}}]$.

\break

\begin{figure}[!ht]
\begin{center}
\includegraphics[width=10cm]{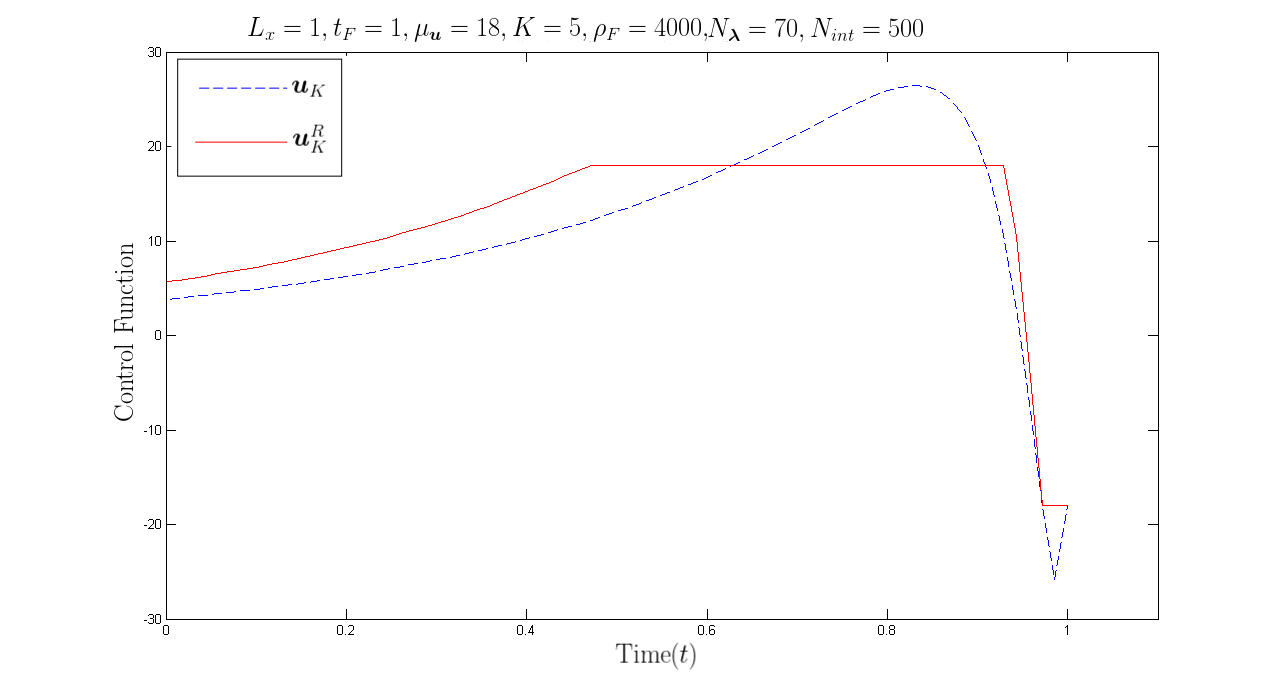}
\end{center}
\caption{Control signals $\boldsymbol{u}_{_{K}}$ (blue dashed), $\boldsymbol{u}_{_{R}}^{^{K}}$ (red solid) for $\rho_{_{F}}=4000$.}
\label{ch6:fig11}
\end{figure}

\begin{figure}[!ht]
\begin{center}
\includegraphics[width=10cm]{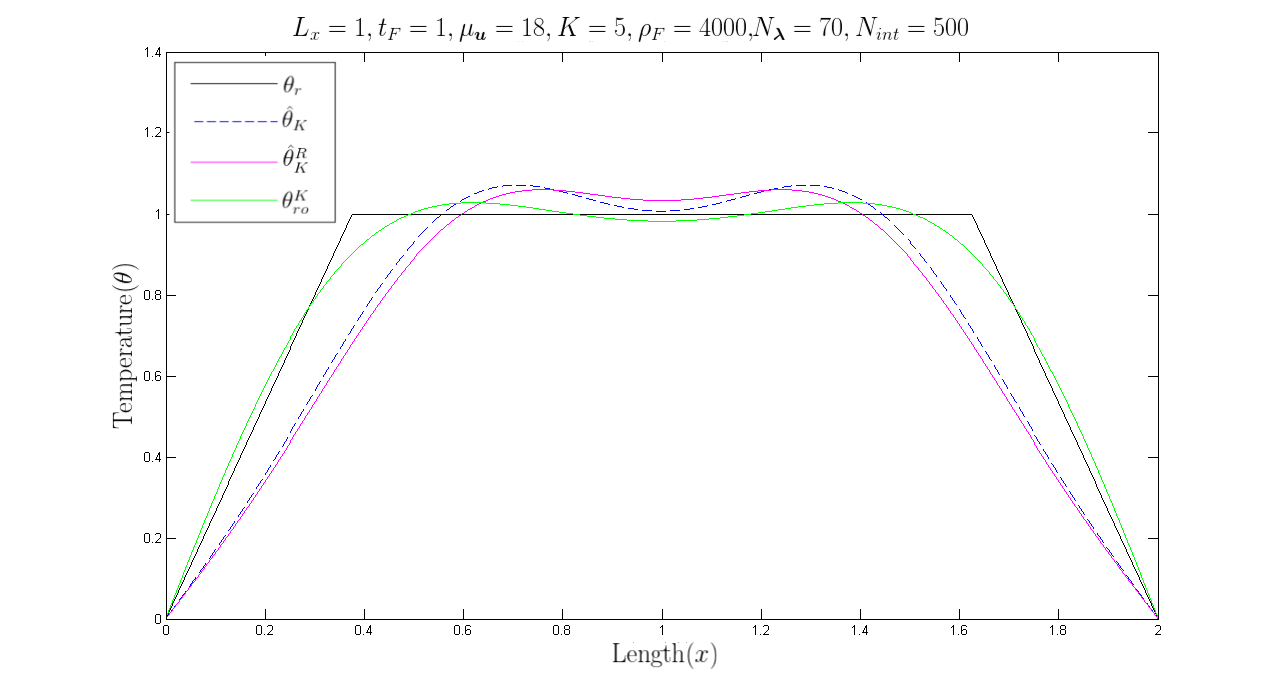}
\end{center}
\caption{Approximations to target final state for $\rho_{_{F}}=4000$.}
\label{ch6:fig12}
\end{figure}

Results were also obtained for the pair $(\theta_{r}, \boldsymbol{\beta}_{_{\boldsymbol{S}}})$ with $\rho_{_{F}}=4000$, as
presented in Tables \ref{ch6:tab07} and \ref{ch6:tab08} and Figures \ref{ch6:fig11} and \ref{ch6:fig12}
\begin{table}[!ht]
\begin{center}
\begin{tabular}{|c|c|c|c|}
\hline	$\check{\mathcal{J}}_{_{K}}(\boldsymbol{u}_{_{K}}; \rho_{_{F}})$ & $\|\boldsymbol{u}_{_{K}}\|_{_{2}}$ & $\|\boldsymbol{u}_{_{K}}\|_{\infty}$ & 
$\|\mathcal{T}_{_{\theta}}^{^{K}}[\boldsymbol{u}_{_{K}}]-\boldsymbol{\theta}_{r\mathrm{o}}^{^{K}}\|_{_{2}}$\\
\hline \hline
        362.0183& 15.3659 & 26.4600 & 0.1774 \\
\hline
\end{tabular}
\caption{Unconstrained problem for the pair $(\theta_{r}, \boldsymbol{\beta}_{_{\boldsymbol{S}}})$, $\rho_{_{F}}=4000$.}
\label{ch6:tab07}
\end{center}
\end{table}
\begin{table}[!ht]
\begin{center}
\begin{tabular}{|c|c|c|c|c|}
\hline	$\check{\mathcal{J}}_{_{K}}(\check{\boldsymbol{u}}_{_{K}}^{^{R}}; \rho_{_{F}})$ & $\varphi_{_{D}}^{^{K}}(\boldsymbol{\lambda}^{^{K}})$ & $\|\check{\boldsymbol{u}}_{_{K}}^{^{R}}\|_{_{2}}$ & $\|\check{\boldsymbol{u}}_{_{K}}^{^{R}}\|_{\infty}$ & $\|\mathcal{T}_{_{\theta}}^{^{K}}[\check{\boldsymbol{u}}_{_{K}}^{^{R}}]-\check{\boldsymbol{\theta}}_{r\mathrm{o}}^{R}\|_{_{2}}$\\
\hline \hline
       387.3645 & 387.2568 &  14.7342 & 18 & 0.2063 \\
\hline
\end{tabular}
\caption{Constrained problem for the pair $(\theta_{r}, \boldsymbol{\beta}_{_{\boldsymbol{S}}})$, $\rho_F=4000$.}
\label{ch6:tab08}
\end{center}
\end{table}

Again, it can be noted that increasing $\rho_{_{F}}$ brings about a better approximation to the desired final state. Note also that
$|\varphi_{_{D}}^{^{K}}(\boldsymbol{\lambda}^{^{K}})-\check{ \mathcal{J}}_{_{K}}(\boldsymbol{u}_{_{R}}^{^{K}}; 4000)|
/\varphi_{_{D}}^{^{K}}(\boldsymbol{\lambda}^{^{K}})\approx 0.11/387.2568\leq 0.03\times 10^{-2}$ and hence $\check{\boldsymbol{u}}_{_{K}}^{^{R}}$ can
be regarded as ``approximately optimal'' for the constrained problem.


\newpage

\newpage

\begin{figure}[!ht]
\begin{center}
\includegraphics[width=10cm]{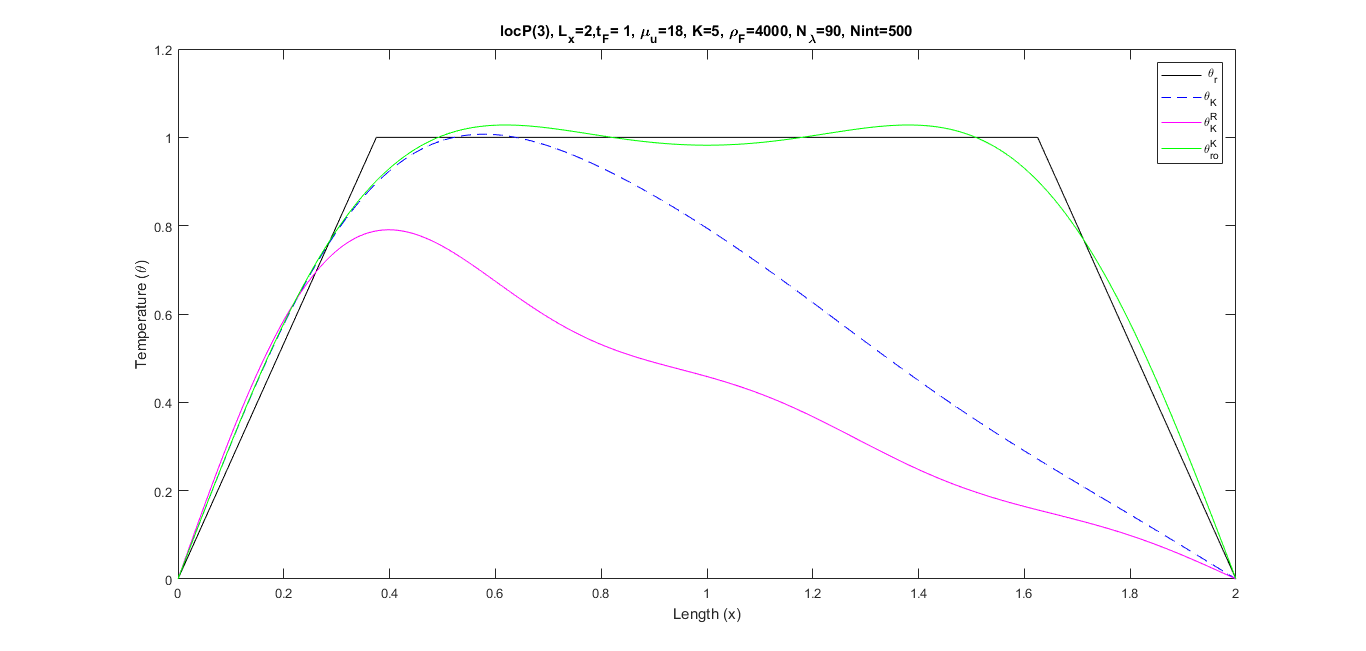}
\end{center}
\caption{Approximations to target final state for $\rho_{_{F}}=4000$, $\ell_{_{x}}=3/10$.}
\label{ch6:fig13}
\end{figure}

\begin{figure}[!ht]
\begin{center}
\includegraphics[width=10cm]{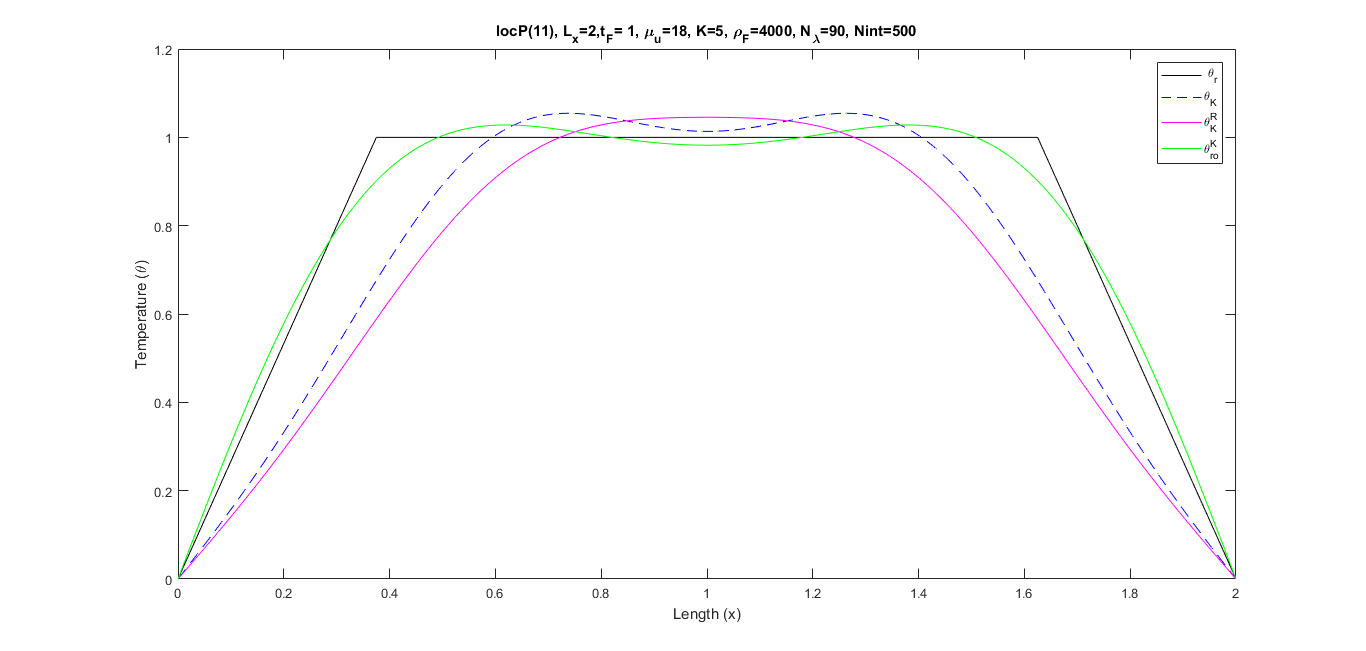}
\end{center}
\caption{Approximations to target final state for $\rho_{_{F}}=4000$,  $\ell_{_{x}}=1$.}
\label{ch6:fig14}
\end{figure}

\begin{figure}[!ht]
\begin{center}
\includegraphics[width=10cm]{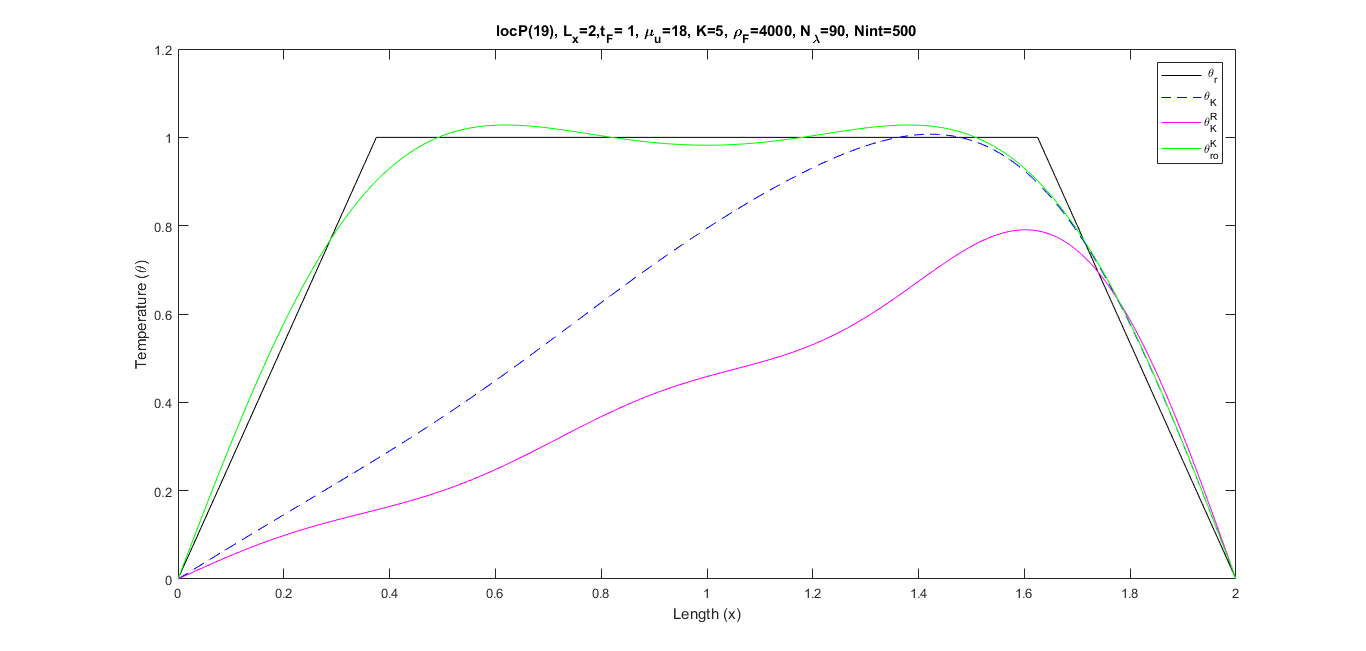}
\end{center}
\caption{Approximations to target final state for $\rho_{_{F}}=4000$,  $\ell_{_{x}}=2-3/10$.}
\label{ch6:fig15}
\end{figure}

Finally, the effect of the location of the ``actuator'' $\boldsymbol{\beta}_{_{\boldsymbol{S}}}$ on the final-state error 
$\mathcal{T}_{_{\theta}}^{^{K}}[\boldsymbol{u}_{c}^{^{K}}]-\theta_{r\mathrm{o}}$ is illustrated by taking $\boldsymbol{\beta}_{_{\boldsymbol{S}}}$ to be centered on $\ell_{_{x}} \in (0,2)$, 
\emph{i.e.}, by
letting $\boldsymbol{\beta}_{_{\boldsymbol{S}}}$ to be given by $\boldsymbol{\beta}_{_{\boldsymbol{S}}}(x)=1$, \ \
$\forall x \in (\ell_{_{x}}-\delta_{_{\boldsymbol{\beta}}}, \ell_{_{x}}+\delta_{_{\boldsymbol{\beta}}})$, \ \ $\boldsymbol{\beta}_{_{\boldsymbol{S}}}(x)=0$ otherwise, and computing the 
resulting $\mathcal{T}_{_{\theta}}^{^{K}}[\boldsymbol{u}_{c}^{^{K}}]$ for several values of $\ell_{_{x}}$ (with $\delta_{_{\boldsymbol{\beta}}}=0.1$), which are displayed in 
Figures \ref{ch6:fig13} -- \ref{ch6:fig15},
respectively for $\ell_{_{x}}=3/10$, $\ell_{_{x}}=1$ and $\ell_{_{x}}=2-3/10$.

\newpage
\subsection{A Two-Dimensional Example}\label{subsec:063}

An example is now presented of an initial/boundary-value problem defined by the heat equation on a rectangle in $\mathbb{R}^{2}$. 
More specifically, let 
$\Omega=(0, L_{_{x}})\times(0, L_{_{y}})$, where $L_{_{x}}, L_{_{y}} \in \mathbb{R}_{+}$ and consider the following equation:
$$ \forall (x, y) \in \Omega, \ \ \frac{\partial \theta}{\partial t}(x,y,t)=
k_{_{\alpha}}\left\{\frac{\partial^{2}\theta}{\partial x^{2}}+\frac{\partial^{2}\theta}{\partial y^{2}}\right\}(x,y,t)+ 
\boldsymbol{\beta}_{_{\boldsymbol{S}}}(x,y)\boldsymbol{u}(t)  
$$
with zero initial conditions, \emph{i.e.}, $\forall(x,y)\in \Omega$,\ \ $\theta(x,y,0)=0$ and homogeneous Dirichlet boundary conditions, 
\emph{i.e.}, 
$$\forall t \in [0, t_{_{F}}], \ \ \forall(x,y)\in \partial \Omega, \ \ \theta(x,y,t)=0,$$
where $\boldsymbol{u}: [0, t_{_{F}}]\rightarrow \mathbb{R}$ and $\boldsymbol{\beta}_{_{\boldsymbol{S}}}:\Omega \rightarrow \mathbb{R}$.

The corresponding weak, ``$K-$th order'', Galerkin version is given by $\forall k=1,\mathellipsis, K$,
$$
\left\langle\frac{\partial \theta}{\partial t}(\cdot, \cdot, t),\phi_{_{k}}\right\rangle=
-k_{_{\alpha}}\left\{\left\langle \frac{\partial \theta}{\partial x}(\cdot, \cdot, t), \frac{\partial \phi_{_{k}}}{\partial x}\right\rangle
+\left\langle \frac{\partial \theta}{\partial y}(\cdot, \cdot, t), \frac{\partial \phi_{_{k}}}{\partial y}\right\rangle\right\} + \boldsymbol{\beta}_{_{\boldsymbol {S}k}}\boldsymbol{u}(t),
$$
where $i=1, \mathellipsis, K_{_{a}}$,\ \ $j=1, \mathellipsis, K_{_{a}}$, $k(i,j)=(i-1)K_{_{a}}+j$,\ \ $K=K_{_{a}}^{^{2}}$,\ \ $\phi_{_{k(i,j)}}(x,y)=\phi_{_{i}}^{^{x}}(x)\phi_{_{j}}^{^{y}}(y)$,\ \
$\phi_{_{i}}^{^{x}}(x)=\sqrt{ \frac{2}{L_{_{x}}}}\sin\left[\frac{i\pi x}{L_{_{x}}}\right]$,\
\ $\phi_{_{j}}^{^{y}}(y)=\sqrt{ \frac{2}{L_{_{y}}}}\sin\left[\frac{j\pi y}{L_{_{y}}}\right]$, $X_{_{K}}=span\{\phi_{i}^{x}\phi_{j}^{y}: i, j=1, \mathellipsis, K_{_{a}}\}$.

As in the previous example, control signals $\boldsymbol{u}_{_{K}}$ and $\boldsymbol{u}_{c}^{^{K}}$ are sought by means of the problems 
$$
\underline{Prob.\ I_{_{K}}:}\ \min_{\boldsymbol{u} \in L_{2}(0,t_{_{F}})} \check{\mathcal{J}}_{_{K}}(\boldsymbol{u};\rho_{_{F}})\ \ \ \ \text{and}\ \ \ \
\underline{Prob.\ I_{_{cK}}:}\ \min_{\boldsymbol{u} \in S_{_{\boldsymbol{u}F}}}\check{\mathcal{J}}_{_{K}}(\boldsymbol{u};\rho_{_{F}}),
$$
where $\check{\mathcal{J}}_{_{K}}(\boldsymbol{u};\rho_F)=\|\boldsymbol{u}\|_{_{L_{2}(0, t_{_{F}})}}^{^{2}}+\rho_F\|\mathcal{T}_{_{\theta}}^{^{K}}[\boldsymbol{u}]-\theta_{r\mathrm{o}}\|_{_{2}}^{^{2}}$,\ \ 
$\mathcal{T}_{_{\theta}}^{^{K}}[\boldsymbol{u}]=\sum_{k=1}^{K}c_{_{k}}(t_{_{F}};\boldsymbol{u})\phi_{_{k}}$,\ \ $\theta_{r}$ is the final state to be \break
``approximately reached'' and, as 
before, $\bar{\boldsymbol{c}}_{_{K}}(t;\boldsymbol{u})=[c_{_{1}}(t;\boldsymbol{u})\ \cdots \ c_{_{K}}(t; \boldsymbol{u})]^{^{\mathrm{T}}}$\ \ is given by\break 
$\bar{\boldsymbol{c}}_{_{K}}(t; \boldsymbol{u})=\displaystyle\int_{0}^{t}\mathbf{F}_{_{K}}(\tau)^{^{\mathrm{T}}}\boldsymbol{u}(\tau)d\tau$ 
with $\mathbf{F}_{_{K}}$ as in (\ref{4eq:19}). In 
this case,
$$
\mathbf{A}_{_{K}}=\operatorname{diag}\{a_{_{k}}: k=k(1,1), \mathellipsis, k(1, K_{a}), k(2,1), \mathellipsis, k(2, K_{a}), \mathellipsis, k(K_{a},1),\mathellipsis,k(K_{a},K_{a})\},
$$
where $a_{_{k(i,j)}}=-k_{_{\alpha}}\left\{\left[\frac{i\pi}{L_{_{x}}}\right]^{^{2}}+\left[\frac{j\pi}{L_{_{y}}}\right]^{^{2}}\right\}$, \ 
$\mathbf{M}_{_{\boldsymbol{\beta}}}^{^{K}}=[\langle\boldsymbol{\beta}_{_{\boldsymbol{S}}},
\phi_{_{1}}\rangle\ \cdots\ \langle\boldsymbol{\beta}_{_{\boldsymbol{S}}}, \phi_{_{k}}\rangle]^{^{\mathrm{T}}}$,\ \ and\ \
$S_{_{\boldsymbol{u}F}}=\{\boldsymbol{u}\in L_{2}(0, t_{_{F}}):\ \text{a.e.},\ |\boldsymbol{u}(t)|\leq \mu_{_{\boldsymbol{u}}} \}$.

Note that
$\check{\mathcal{J}}_{_{K}}(\boldsymbol{u};\rho_{_{F}})=\|\boldsymbol{u}\|_{_{L_{2}(0,t_F)}}^{^{2}}+\rho_{_{F}}\|\mathcal{T}_{_{\theta}}^{^{K}}[\boldsymbol{u}]-\theta_{r\mathrm{o}}^{^{K}}\|_{_{L_{2}(\Omega)}}^{^{2}}
+\|\theta_{r\mathrm{o}}-\theta_{r\mathrm{o}}^{^{K}}\|_{_{L_{2}(\Omega)}}^{^{2}}$, where $\theta_{r\mathrm{o}}^{^{K}}$ is the orthogonal projection of $\theta_{r\mathrm{o}}$ on the span of $\{\phi_{_{1}}, \mathellipsis, \phi_{_{K}}\}$. The constrained problem is tackled by Lagrange duality, as illustrated in the previous section, with the same class of piecewise-linear multipliers.

The numerical results shown in Tables \ref{ch6:tab09} -- \ref{ch6:tab12} were obtained with the following problem data:\ $k_{_{\alpha}}=1$,\ $L_{_{x}}=L_{_{y}}=1$,\ \ $t_{_{F}}=1$,\ \ $\rho_{_{F}}=8000$ and $20000$,\ \
$\mu_{_{\boldsymbol{u}}}=100$,\ \ $K_{a}=5$,\ \ $\theta_{r}(x,y)=0$\ $\forall (x,y)\in \partial \Omega$,\ \ $\theta_{r}(x,y)=2$
$\forall(x,y)\in[L_{_{x}}/10, 9L_{_{x}}/10]\times[L_{_{y}}/10, 9L_{_{y}}/10]$, \emph{i.e.}, the graph of $\theta_{r}$ is the frustum of a rectangular pyramid with 
$[0, L_{_{x}}]\times[0, L_{_{y}}]$ as its basis,
$\|\theta_{r\mathrm{o}}^{^{K}}\|_{_{2}}=1.7289$, $N_{_{\boldsymbol{\lambda}}}=30$, and $\boldsymbol{\beta}_{_{\boldsymbol{S}}}$ is given by\\
$
\left\{
\begin{array}{rl}
\boldsymbol{\beta}_{_{\boldsymbol{S}}}=1 & \text{for } (x,y)\in [L_{_{x}}/4, 3L_{_{x}}/4]\times [L_{_{y}}/4, 3L_{_{y}}/4]\\
\boldsymbol{\beta}_{_{\boldsymbol{S}}}=0 & \text{otherwise}
\end{array} \right.
$.


\begin{table}[!ht]
\begin{center}
\begin{tabular}{|c|c|c|c|c|}
\hline	$\check{\mathcal{J}}_{_{K}}(\boldsymbol{u}_{_{K}}; \rho_{_{F}})$ & $\|\boldsymbol{u}_{_{K}}\|_{_{2}}$ & $\|\boldsymbol{u}_{_{K}}\|_{\infty}$ & 
$\|\mathcal{T}_{_{\theta}}^{^{K}}[\boldsymbol{u}_{_{K}}]-\boldsymbol{\theta}_{r\mathrm{o}}^{^{K}}\|_{_{2}}^{^{2}}$\\
\hline \hline
       4978.00 & 45.6636 &  192.5735 & 0.6037 \\
\hline
\end{tabular}
\caption{Unconstrained problem with $\rho_F=8000$.}
\label{ch6:tab09}
\end{center}
\end{table}

\begin{table}[!ht]
\begin{center}
\begin{tabular}{|c|c|c|c|c|}
\hline	$\check{\mathcal{J}}_{_{K}}(\boldsymbol{u}_{c}^{^{K}}; \rho_{_{F}})$ & $\varphi_{_{D}}^{^{K}}(\boldsymbol{\lambda}^{^{K}})$ & $\|\boldsymbol{u}_{c}^{^{K}}\|_{_{2}}$ &
$\|\boldsymbol{u}_{c}^{^{K}}\|_{\infty}$ & $\|\mathcal{T}_{_{\theta}}^{^{K}}[\boldsymbol{u}_{c}^{^{K}}]-\boldsymbol{\theta}_{r\mathrm{o}}^{^{K}}\|_{_{2}}^{^{2}}$\\
\hline \hline
       5668.10 & 5485.00 & 33.0038 & 100 & 0.7565 \\
\hline
\end{tabular}
\caption{Constrained problem with $\rho_F=8000$.}
\label{ch6:tab10}
\end{center}
\end{table}


\begin{table}[!ht]
\begin{center}
\begin{tabular}{|c|c|c|c|c|}
\hline	$\check{\mathcal{J}}_{_{K}}(\boldsymbol{u}_{_{K}}; \rho_{_{F}})$ & $\|\boldsymbol{u}_{_{K}}\|_{_{2}}$ & $\|\boldsymbol{u}_{c}^{^{K}}\|_{\infty}$ & 
$\|\mathcal{T}_{_{\theta}}^{^{K}}[\boldsymbol{u}_{_{K}}]-\boldsymbol{\theta}_{r\mathrm{o}}^{^{K}}\|_{_{2}}^{^{2}}$\\
\hline \hline
       8127.40 & 64.4017 &  265.37 & 0.4485 \\
\hline
\end{tabular}
\caption{Unconstrained problem with $\rho_F=20000$.}
\label{ch6:tab11}
\end{center}
\end{table}

\begin{table}[!ht]
\begin{center}
\begin{tabular}{|c|c|c|c|c|}
\hline	$\check{\mathcal{J}}_{_{K}}(\boldsymbol{u}_{c}^{^{K}}; \rho_{_{F}})$ & $\varphi_{_{D}}^{^{K}}(\boldsymbol{\lambda}^{^{K}})$ & $\|\boldsymbol{u}_{c}^{^{K}}\|_{_{2}}$ & 
$\|\boldsymbol{u}_{c}^{^{K}}\|_{\infty}$ & $\|\mathcal{T}_{_{\theta}}^{^{K}}[\boldsymbol{u}_{c}^{^{K}}]-\boldsymbol{\theta}_{r\mathrm{o}}^{^{K}}\|_{_{2}}^{^{2}}$\\
\hline \hline
       12281.00 & 11195.00 & 37.8125 & 100 & 0.7366 \\
\hline
\end{tabular}
\caption{Constrained problem with $\rho_F=20000$.}
\label{ch6:tab12}
\end{center}
\end{table}


Similarly to the results in the case of a one-dimensional spatial domain, Tables \ref{ch6:tab09} -- \ref{ch6:tab11} illustrate the
effect of increasing $\rho_{_{F}}$ on the decrease of the approximation 
errors\ $\|\mathcal{T}_{_{\theta}}^{^{K}}[\boldsymbol{u}_{_{K}}]-\theta_{r\mathrm{o}}^{^{K}}\|_{_{2}}$ (from 0.6037 in Table \ref{ch6:tab09} to 0.4484 in 
Table \ref{ch6:tab11}) and 
$\|\mathcal{T}_{_{\theta}}^{^{K}}[\boldsymbol{u}_{c}^{^{K}}]-\theta_{r\mathrm{o}}^{^{K}}\|_{_{2}}$ (from 0.7565 in Table \ref{ch6:tab10} to 0.7366 in 
Table \ref{ch6:tab12}). Note that in the latter case, increasing $\rho_{_{F}}$ from 8000 to 20000 had a small effect on the approximation 
error - this is due to the fact that the
maximum magnitude of $\boldsymbol{u}$ was kept at the same value ($\mu_{_{\boldsymbol{u}}}=100$).

Again, as observed in the 1D-case, the ``relatively small'' difference between $\varphi_{_{D}}^{^{K}}(\boldsymbol{\lambda}^{^{K}})$ and 
$\check{\mathcal{J}}_{_{K}}(\boldsymbol{u}_{c}^{^{K}}; \rho_{_{F}})$ (3.2\% for $\rho_{_{F}}=8000$ and 8.8\% for $\rho_{_{F}}=20000$) indicates that $\boldsymbol{u}_{c}^{^{K}}$ is ``nearly 
optimal'' for the constrained problem - recall that $\varphi_{_{D}}^{^{K}}(\boldsymbol{\lambda}^{^{K}})$ is a lower bound on $\check{\boldsymbol{u}}_{_{K}}(\boldsymbol{u};\rho_{_{F}})$ for
any $\boldsymbol{u}\in S_{_{\boldsymbol{u}F}}$.

\break

\begin{figure}[!ht]
\begin{center}
\includegraphics[width=10cm]{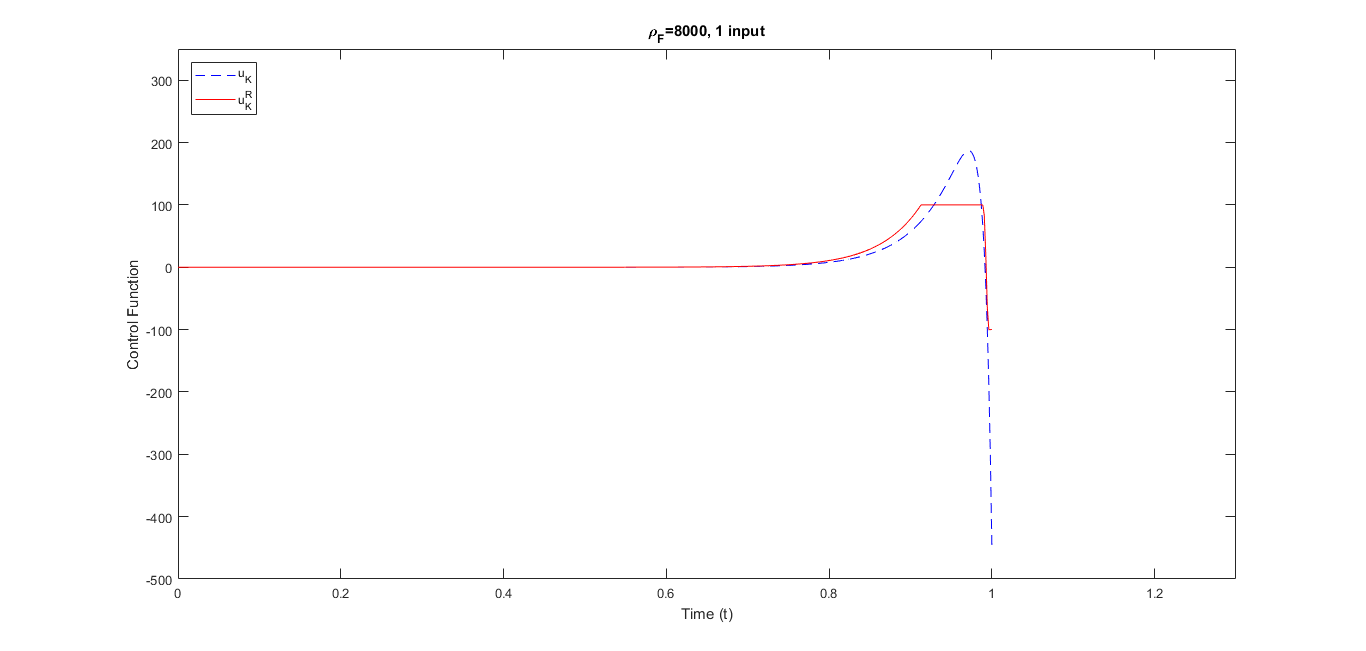}
\end{center}
\caption{Graphs of $\boldsymbol{u}_{_{K}}$ and $\boldsymbol{u}_{c}^{^{K}}$ for $\rho_{_{F}}=8000$.}
\label{ch6:fig19}
\end{figure}

\begin{figure}[!ht]
\begin{center}
\includegraphics[width=10cm]{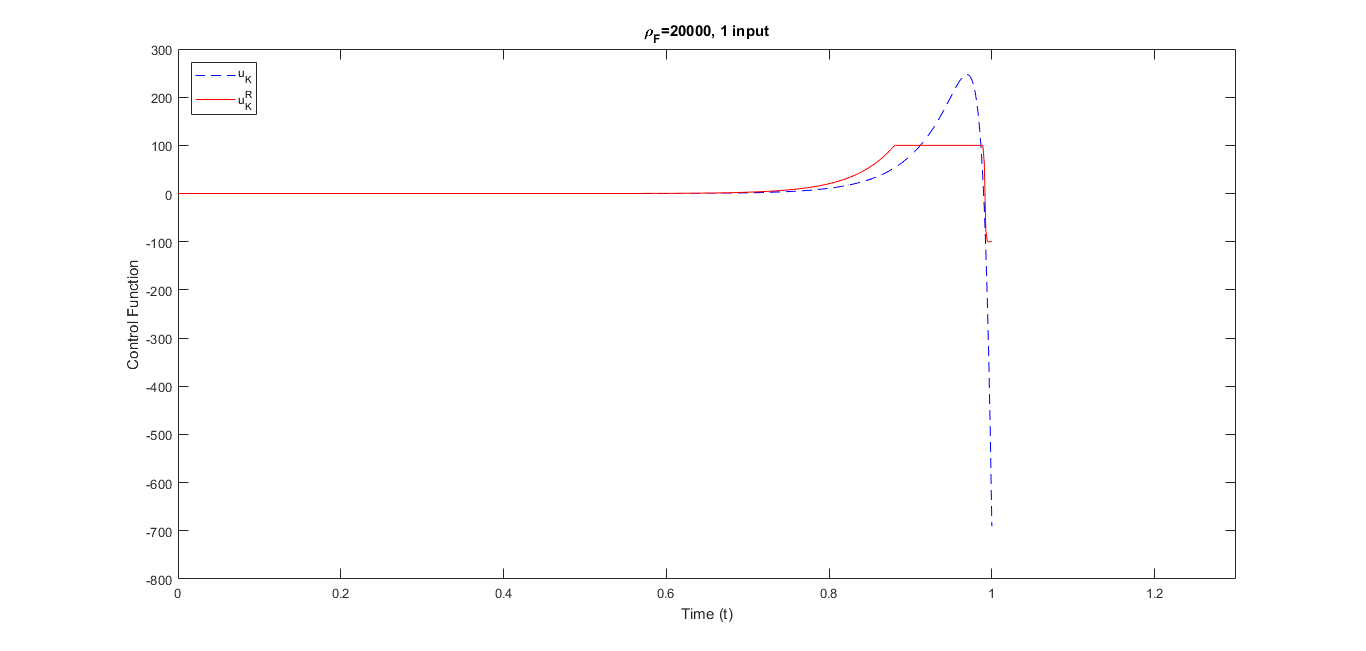}
\end{center}
\caption{Graphs of $\boldsymbol{u}_{_{K}}$ and $\boldsymbol{u}_{c}^{^{K}}$ for $\rho_{_{F}}=20000$.}
\label{ch6:fig20}
\end{figure}

Figures \ref{ch6:fig19} -- \ref{ch6:fig20} 
display $\boldsymbol{u}_{_{K}}$ and $\boldsymbol{u}_{c}^{^{K}}$ for $\rho_{_{F}}=8000$ and 20000, respectively.

\section{Concluding remarks}

In this work, two types of open-loop control problems were addressed in connection
with parabolic PDEs with homogeneous Dirichlet
boundary conditions in which the control function (depending only on time)
constitutes a source term. In both cases, the main objective is to impose a 
prescribed state at the final instant of a given time-
interval. Control signals are to be selected on the basis of two optimization 
problems, one unconstrained and the other one involving constraints on the maximum 
magnitudes of the values taken by the control signals on the time-interval in 
question. Both problems have the same quadratic cost-functional.

Approximations for the optimal control signals can be obtained on the basis of 
finite-dimensional, semi-discrete spectral Galerkin approximations for the PDEs envolved. As a consequence, the resulting optimal
control signals can be effectively computed. Indeed, in the unconstrained case, they 
are given as the output of an autonomous, finite-dimensional linear system with 
initial state given by the data of the original problem. Whereas, in the constrained 
case, using Lagrangian duality, the resulting control signals are obtained from the
output of a linear finite-dimensional system (as in the unconstrained case but with
a modified initial state which depends on the ``approximately-optimal'' Lagrange 
multipliers) and an additive correction term based on these multipliers. Numerical results for the 1D and 2D linear heat equations were presented to illustrate the results mentioned above.


\section*{References}
 
\noindent
[1]\ Curtain, R.F.; Zwart, H., \emph{An Introduction to Infinite-dimensional Linear Systems Theory}, 1st. ed., 
Springer-Verlag, New York, Inc., 1995.\medskip

\noindent
[2]\ Ekeland, I.; T\'{e}mam, R., \emph{Convex Analysis and Variational Problems}, 1st. ed., Society for Industrial and Applied Mathematics, 1976.\medskip
 
\noindent
[3]\ Evans, L.C., \emph{Partial Differential Equations}, 2nd. ed., American Mathematical Society, 2010.\medskip

\noindent
[4]\ Kogut, P.I.; Leugering, G.R., \emph{Optimal Control Problems for Partial Differential Equations on Reticulated 
Domains: Approximation and Asymptotic Analysis}, Birkh\"{a}user, 2011.\medskip

\noindent
[5]\ Laub, A.J., Matrix Analysis for Scientists \& Engineers, 1st. ed., Society for Industrial and Applied Mathematics, 2005.\medskip

\noindent
[6]\ Lions, J.L., \emph{Exact Controllability, Stabilization and Perturbations for Distributed Parameter Systems}, SIAM Review,vol. \textbf{30}, 1988.\medskip

\noindent
[7]\ Lions, J.L., \emph{Contr\^{o}labilit\'{e} Exacte, Perturbations et Stabilisation de Syst\`{e}mes distribu\'{e}s}, Tome 1,2, RMA, vol. \textbf{8,9}, 1988.\medskip

\noindent
[8]\ L\'{o}pez--Flores, M.M., \emph{Final-State Approximate Control for the Heat Equation}. Ph.D. THESIS, Faculdade de Engenharia-UERJ, Rio de Janeiro, RJ, Brasil, 2018. [Online.] Available:\ \url{https://www.bdtd.uerj.br:8443/handle/1/16985}.\medskip

\noindent
[9]\ Luenberger, G., \emph{Optimization by Vector Space Methods}, 1st. ed.,  John Wiley $\&$ Sons, 1969.\medskip

\noindent
[10]\ Morris, K.A., \emph{Design of Finite-dimensional Controller for Infinite-Dimensional System by Approximation}, Journal of Math. System, Estimation and control, vol. \textbf{4}, no.2, pp. 1--30, 1994.\medskip

\noindent
[11]\ Tr\"{o}ltzsch, F., \emph{Optimal Control of Partial Differential Equations: Theory,  Methods and Applications}, American Mathematical Society, 2010.\medskip

\noindent
[12]\ Zuazua, E., \emph{Controllability of Partial Differential Equations and its Semi-discrete Approximations}, Discrete and Continuous Dynamical Systems, vol. \textbf{8}, no. 2, pp. 469 -- 513, 2002.

\break
\section*{Appendix}
\setcounter{equation}{0}
\renewcommand{\theequation}{A.\arabic{equation}}



\noindent
\underline{\textbf{Proof of Proposition \ref{prop:01}:}} Let $\mathcal{T}_{a}\ :\ L_{2}(0,t_{_{F}})^{^{m}}\rightarrow L_{2}(0,t_{_{F}})^{^{m}}\times L_{2}(\Omega)$  be defined by
$\mathcal{T}_{a}[\boldsymbol{u}]\triangleq (\rho^{^{1/2}}_{_{\boldsymbol{u}}}\boldsymbol{u},\mathcal{T}_{_{\theta}}[\boldsymbol{u}])$.
Then $\mathcal{J}(\boldsymbol{u})=\|\mathcal{T}_{a}[\boldsymbol{u}]-(0,\theta_{r\mathrm{o}})\|^{^{2}}_{_{X_{a}}}$\ , where
$X_{a}\triangleq L_{2}(0,t_{_{F}})^{^{m}}\times L_{2}(\Omega)$, and \emph{Prob. I} is seen as the problem of finding the
minimum-distance approximation to $(0,\theta_{r\mathrm{o}})\in X_{a}$ in $\mathcal{T}_{a}[L_{2}(0,t_{_{F}})^{^{m}}]$ - note that $X_{a}$  is a
Hilbert Space with the inner product
$$\langle (v_{_{1}},w_{_{1}}),(v_{_{2}},w_{_{2}})\rangle_{_{X_{a}}}=\langle v_{_{1}},v_{_{2}}\rangle_{_{L_{2}(0,t_{_{F}})^{^{m}}}}+
\langle w_{_{1}},w_{_{2}}\rangle_{_{L_{2}(\Omega)}}.$$

Moreover, $\mathcal{T}_{a}[L_{2}(0,t_{_{F}})^{^{m}}]$ is closed. Indeed, if $\mathcal{T}_{a}[\boldsymbol{u}_{_{K}}]\rightarrow
\boldsymbol{x}_{_{0}}=(\hat{\boldsymbol{u}}_{_{\mathrm{o}}},\hat{\theta}_{a\mathrm{o}})$ or, equivalently,
$(\rho_{_{\boldsymbol{u}}}^{^{1/2}} \boldsymbol{u}_{_{K}},\mathcal{T}_{_{\theta}}[\boldsymbol{u}_{_{K}}])\rightarrow
(\hat{\boldsymbol{u}}_{_{\mathrm{o}}},\hat{\theta}_{a\mathrm{o}})$ then
$\boldsymbol{u}_{_{K}}\rightarrow \rho_{_{\boldsymbol{u}}}^{^{-1/2}}\hat{\boldsymbol{u}}_{_{\mathrm{o}}}$ and
(since $\mathcal{T}_{_{\theta}}$  is continuous) $\mathcal{T}_{_{\theta}}[\boldsymbol{u}_{_{K}}]\rightarrow
\mathcal{T}_{_{\theta}}[\rho_{_{\boldsymbol{u}}}^{^{-1/2}}\hat{\boldsymbol{u}}_{_{\mathrm{o}}}]=\hat{\theta}_{a\mathrm{o}}$.
Thus,  $\mathcal{T}_{a}(\rho_{_{\boldsymbol{u}}}^{^{-1/2}}\hat{\boldsymbol{u}}_{_{\mathrm{o}}})=(\hat{\boldsymbol{u}}_{_{\mathrm{o}}},
\mathcal{T}_{_{\theta}}[\rho_{_{\boldsymbol{u}}}^{^{-1/2}}\hat{\boldsymbol{u}}_{_{\mathrm{o}}}])=(\hat{\boldsymbol{u}}_{_{\mathrm{o}}},\hat{\theta}_{a\mathrm{o}})=
\boldsymbol{x}_{_{0}} \Rightarrow  \boldsymbol{x}_{_{0}} \in \mathcal{T}_{a}[L_{2}(0,t_{_{F}})^{^{m}}]$.

As $\mathcal{T}_{a}[L_{2}(0,t_{_{F}})^{^{m}}]$ is also convex, it follows from  ([9], Theorem 3.12.1, p. 69) that \emph{Prob. I} has a
unique solution $\boldsymbol{u}_{_{\mathrm{o}}}$ (say).

Note now that $\boldsymbol{u}_{_{\mathrm{o}}}$ is a solution to \emph{Prob. I}
$\Leftrightarrow \forall\delta
\boldsymbol{u}\in L_{2}(0,t_{_{F}})^{^{m}}$,$$\mathcal{J}(\boldsymbol{u}_{_{\mathrm{o}}})\leq \mathcal{J}(\boldsymbol{u}_{_{\mathrm{o}}} + \delta \boldsymbol{u})\
\Leftrightarrow \  \forall \delta \boldsymbol{u}\in L_{2}(0,t_{_{F}})^{^{m}}\ ,
$$
\[
\ 2\rho_{_{\boldsymbol{u}}}\langle \boldsymbol{u}_{_{\mathrm{o}}},\delta \boldsymbol{u}\rangle_{L_{2}(0, t_{_{F}})^{^{m}}} +
\rho_{_{\boldsymbol{u}}}\|\delta \boldsymbol{u} \|_{_{L_{2}(0, t_{_{F}})^{^{m}}}}^{^{2}}+2\langle \mathcal{T}_{_{\theta}}[\boldsymbol{u}_{_{\mathrm{o}}}]-
\theta_{r\mathrm{o}}, \mathcal{T}_{_{\theta}}[\delta \boldsymbol{u}]\rangle +\| \mathcal{T}_{_{\theta}}[\delta \boldsymbol{u}]\|_{_{L_{2}(\Omega)}}^{^{2}}\geq0
\]
\[
\Leftrightarrow \ \ \ \ \forall \delta \boldsymbol{u}\in L_{2}(0,t_{_{F}})^{^{m}}\quad ,\quad \langle \rho_{_{\boldsymbol{u}}}
\boldsymbol{u}_{_{\mathrm{o}}}+\mathcal{T}^{*}_{_{\theta}}\cdot\mathcal{T}_{_{\theta}}[\boldsymbol{u}_{_{\mathrm{o}}}]-\mathcal{T}^{*}_{_{\theta}}[\theta_{r\mathrm{o}}]\ ,
\ \delta \boldsymbol{u}\rangle_{L_{2}(0, t_{_{F}})^{^{m}}} \geq 0
\]
\[
\ \ \ \ \ \ \ \ \ \  \  \ \ \ \ \ \ \Leftrightarrow \rho_{_{\boldsymbol{u}}} \boldsymbol{u}_{_{\mathrm{o}}}+ \mathcal{T}^{*}_{_{\theta}}\cdot
\mathcal{T}_{_{\theta}}[\boldsymbol{u}_{_{\mathrm{o}}}]-\mathcal{T}^{*}_{_{\theta}}[\theta_{r\mathrm{o}}]=0.
\]
(if $v_{_{\mathrm{o}}}\triangleq\rho_{_{\boldsymbol{u}}}\boldsymbol{u}_{_{\mathrm{o}}} +\mathcal{T}_{_{\theta}}^{*}\circ\mathcal{T}_{_{\theta}}[\boldsymbol{u}_{_{\mathrm{o}}}]-
\mathcal{T}[\theta_{r\mathrm{o}}]$ is such that $v_{_{\mathrm{o}}}\neq0$, then it can be seen that $\delta_{_{\boldsymbol{u}}}=-v_{_{\mathrm{o}}}$ violates the optimality condition). Thus, $\boldsymbol{u}_{_{\mathrm{o}}}$ is the unique solution of the linear equation (\ref{4eq:14}). \hfill$\blacksquare$

\vspace*{3mm}
\noindent
\underline{\textbf{Proof of Proposition \ref{prop:4.1}:}}
Note first that
\begin{eqnarray}
 \mathcal{T}_{_{\theta}}^{^{K}}[\boldsymbol{u}]
 &=&\int_{0}^{t_{_{F}}}S_{_{K}}(t_{_{F}}-\tau)
 \left[P_{_{K}}\left[\sum_{i=1}^{m}\boldsymbol{\beta}_{_{\boldsymbol{S}i}}u_{_{i}}(\tau)\right]\right]d\tau \ \ \ \Leftrightarrow \nonumber\\  \nonumber\\
 \mathcal{T}_{_{\theta}}^{^{K}}[\boldsymbol{u}]&=&\int_{0}^{t_{_{F}}}S_{_{K}}(t_{_{F}}-\tau)
 \left[\sum_{i=1}^{m} P_{_{K}}[\boldsymbol{\beta}_{_{\boldsymbol{S}i}}]u_{_{i}}(\tau)\right]d\tau \ \ \ \ \ \Leftrightarrow \nonumber\\ \nonumber\\
 \mathcal{T}_{_{\theta}}^{^{K}}[\boldsymbol{u}]&=&\int_{0}^{t_{_{F}}}\sum_{i=1}^{m} \left\{ S_{_{K}}(t_{_{F}}-\tau)\left[P_{_{K}}[\boldsymbol{\beta}_{_{\boldsymbol{S}i}}]\right]\right\}u_{_{i}}(\tau)d\tau.\label{4eq:16a}
 \end{eqnarray}
Moreover, \ $P_{_{K}}[\boldsymbol{\beta}_{_{\boldsymbol{S}i}}]=\sum_{k=1}^{K}\langle\boldsymbol{\beta}_{_{\boldsymbol{S}i}}, \phi_{_{k}}\rangle\phi_{_{k}}$.

Now, for $\phi=\sum_{k=1}^{K}\gamma_{_{k}}\phi_{_{k}}$, 
\begin{eqnarray*}
S_{_{K}}(t)[\phi]&=&\sum_{\ell=0}^{\infty}\frac{t^{^{\ell}}}{\ell!}\mathcal{A}_{_{K}}^{^{\ell}}[\phi]
=\sum_{\ell=0}^{\infty}\frac{t^{^{\ell}}}{\ell !}\sum_{q=1}^{K}\bar{\gamma}_{_{K}}^{^{\ell}}\phi_{q}
=\sum_{\ell=0}^{\infty}\sum_{q=1}^{K}\frac{t^{^{\ell}}}{\ell !}
\{\boldsymbol{e}_{_{q}}^{^{\mathrm{T}}}\mathbf{A}_{_{K}}^{^{\ell}}\bar{\boldsymbol{\gamma}}\}\phi_{q}\\
&=&\sum_{q=1}^{K}c_{_{q}}^{^{S}}[\phi](t)\phi_{q},
\end{eqnarray*}
where
$$c_{_{q}}^{^{S}}[\phi](t)=\sum_{\ell=0}^{\infty}\frac{t^{^{\ell}}}{\ell!}\boldsymbol{e}_{_{q}}^{^{\mathrm{T}}}\mathbf{A}_{_{K}}^{^{\ell}}\bar{\boldsymbol{\gamma}}
=\boldsymbol{e}_{_{q}}^{^{\mathrm{T}}}\left[\sum_{\ell=0}^{\infty}\frac{t^{^{\ell}}}{\ell!}\mathbf{A}_{_{K}}^{^{\ell}}\right]\bar{\boldsymbol{\gamma}}$$
so that the vector of coefficients \ $\underline{\boldsymbol{c}}_{_{S}}=[c_{_{1}}^{^{S}}[\phi](t)\ \cdots \ c_{_{K}}^{^{S}}[\phi](t)]^{^{\mathrm{T}}}$ is given by
$\underline{\boldsymbol{c}}_{_{S}}[\phi](t)=\exp\left[\mathbf{A}_{_{K}}t\right]\bar{\boldsymbol{\gamma}}$.

It then follows that 
$$S_{_{K}}(t)\left[P_{_{K}}[\boldsymbol{\beta}_{_{\boldsymbol{S}i}}]\right]
=\sum_{q=1}^{K}c_{_{q}}^{^{S}}\left[P_{_{K}}[\boldsymbol{\beta}_{_{\boldsymbol{S}i}}]\right](t)\phi_{q},$$
where
\begin{equation}
\underline{\boldsymbol{c}}_{_{S}}\left[P_{_{K}}[\boldsymbol{\beta}_{_{\boldsymbol{S}i}}]\right](t)=
\exp\left[\mathbf{A}_{_{K}}t\right]\begin{bmatrix}
                                \langle\boldsymbol{\beta}_{_{\boldsymbol{S}i}}, \phi_{_{1}}\rangle\\
                                \vdots\\
                                \langle\boldsymbol{\beta}_{_{\boldsymbol{S}i}}, \phi_{_{K}}\rangle
                                \end{bmatrix}.\label{4eq:16b}
\end{equation}
Thus, taking (\ref{4eq:16b}) into (\ref{4eq:16a})  leads to                  
\begin{eqnarray*}
\mathcal{T}_{_{\theta}}^{^{n_{_{K}}}}[\boldsymbol{u}]&=&\int_{0}^{t_{_{F}}}\sum_{i=1}^{m}\sum_{q=1}^{K}\boldsymbol{e}_{_{q}}^{^{\mathrm{T}}}
\left\{\exp[\mathbf{A}_{_{K}}(t-\tau)]\begin{bmatrix}
                                \langle\boldsymbol{\beta}_{_{\boldsymbol{S}i}}, \phi_{_{1}}\rangle\\
                                \vdots\\
                                \langle\boldsymbol{\beta}_{_{\boldsymbol{S}i}}, \phi_{n_{_{K}}}\rangle
                                \end{bmatrix}u_{_{i}}(\tau)\right\}\phi_{q}d\tau \\\\
&=&\sum_{q=1}^{K}\boldsymbol{e}_{_{q}}^{^{\mathrm{T}}}\left\{\int_{0}^{t_{_{F}}}\exp\left[\mathbf{A}_{_{K}}(t-\tau)\right]
\mathbf{M}_{_{\boldsymbol{\beta}}}^{^{K}}\boldsymbol{u}(\tau) d \tau\right\}\phi_{_{q}}                       
\end{eqnarray*}                                                          
so that $\mathcal{T}_{_{\theta}}^{^{n_{_{K}}}}[\boldsymbol{u}]=\displaystyle\sum_{q=1}^{K}c_{q}(t_{_{F}}; \boldsymbol{u})\phi_{q}$, where 
$\underline{\boldsymbol{c}}_{_{K}}(t; \boldsymbol{u})=[c_{_{1}}(t; \boldsymbol{u}),\mathellipsis, c_{n_{_{K}}}(t; \boldsymbol{u})]^{^{\mathrm{T}}}$ is given by\\
$\underline{\boldsymbol{c}}_{_{K}}(t; \boldsymbol{u})=
\displaystyle\int_{0}^{t}\exp[\mathbf{A}_{_{K}}(t-\tau)]\mathbf{M}_{_{\boldsymbol{\beta}}}^{^{K}}\boldsymbol{u}(\tau)d\tau$,\ \ \ 
$\boldsymbol{\beta}_{_{\boldsymbol{S}}}^{^{\mathrm{T}}}=[\boldsymbol{\beta}_{_{\boldsymbol{S}1}}\ \cdots\ \boldsymbol{\beta}_{_{\boldsymbol{S}m}}]$\ \ \ and 
\[
\mathbf{M}_{_{\boldsymbol{\beta}}}^{^{K}}\triangleq \begin{bmatrix}
                                               \langle\boldsymbol{\beta}_{_{\boldsymbol{S}1}}, \phi_{_{1}}\rangle& \cdots & \langle\boldsymbol{\beta}_{_{\boldsymbol{S}m}}, \phi_{_{1}}\rangle\\
                                               \vdots &  & \vdots\\
                                                \langle\boldsymbol{\beta}_{_{\boldsymbol{S}1}}, \phi_{n_{_{K}}}\rangle& \cdots &  \langle\boldsymbol{\beta}_{_{\boldsymbol{S}m}}, \phi_{n_{_{K}}}\rangle\\
                                              \end{bmatrix}.
\]
\hfill$\blacksquare$

\vspace*{3mm}
\noindent
\underline{\textbf{Proof of Proposition \ref{prop:4.2new}}} To obtain  $\boldsymbol{u}_{_{K}}$ note that it follows from (\ref{4eq:18}) that $\boldsymbol{u}_{_{K}}$ belongs to the image of $(\mathcal{T}_{_{\theta}}^{^{K}})^{*}$,
\emph{i.e.}, there exists 
\[
\phi \in L_{2}(\Omega)\ \mbox{such that}\ \boldsymbol{u}_{_{K}}=(\mathcal{T}_{_{\theta}}^{^{K}})^{*}[\phi]=\mathbf{F}_{_{K}}\bar{\boldsymbol{\phi}}_{_{K}},
\]
\emph{i.e.}, there exists $\bar{\boldsymbol{\alpha}}_{_{K}}\in \mathbb{R}^{^{n}}$ such that
\begin{equation}\label{4eq:20}
\boldsymbol{u}_{_{K}}=\mathbf{F}_{_{K}}\bar{\boldsymbol{\alpha}}_{_{K}}.
\end{equation}
It then follows from (\ref{4eq:18}) that 
\begin{equation}\label{4eq:22}
\rho_{_{\boldsymbol{u}}}\mathbf{F}_{_{K}}\bar{\boldsymbol{\alpha}}_{_{K}}+ \mathbf{F}_{_{K}}\underline{\boldsymbol{c}}_{_{K}}(t_{_{F}};\mathbf{F}_{_{K}}\bar{\boldsymbol{\alpha}}_{_{K}})
-\mathbf{F}_{_{K}}\bar{ \boldsymbol{\theta}}_{r\mathrm{o}}^{^{K}}=0
\end{equation}
a sufficient condition for which being
\begin{equation}\label{4eq:23}
\rho_F\bar{\boldsymbol{\alpha}}_{_{K}}+\underline{\boldsymbol{c}}_{_{K}}(t_{_{F}};\mathbf{F}_{_{K}}\bar{\boldsymbol{\alpha}}_{_{K}})-
\bar{\boldsymbol{\theta}}^{^{K}}_{r\mathrm{o}}=\mathbf{0},
\end{equation}
where
$\bar{ \boldsymbol{\theta}}_{r\mathrm{o}}^{^{K}}\triangleq \left[\langle \phi_{_{1}},\theta_{r\mathrm{o}}\rangle \cdots \langle \phi_{_{n_{_{K}}}},
\theta_{r\mathrm{o}}\rangle \right]^{^{\mathrm{T}}}$.

Thus, as $\underline{\boldsymbol{c}}_{_{K}}(t_{_{F}}; \mathbf{F}_{_{K}}\bar{\boldsymbol{\alpha}}_{_{K}})=\mathbf{G}_{_{K}}\bar{\boldsymbol{\alpha}}_{_{K}}$,  where
$\mathbf{G}_{_{K}}\triangleq \displaystyle\int_{0}^{t_{_{F}}}\mathbf{F}_{_{K}}(\tau )^{^{\mathrm{T}}}\mathbf{F}_{_{K}}(\tau )d\tau$, (\ref{4eq:22}) can be 
rewritten as\\
$\rho_{_{\boldsymbol{u}}}\bar{\boldsymbol{\alpha}}_{_{K}}+\mathbf{G}_{_{K}}\bar{\boldsymbol{\alpha}}_{_{K}}=\bar{\boldsymbol{\theta}}_{r\mathrm{o}}^{^{K}}$ from which it follows
that $\bar{\boldsymbol{\alpha}}_{_{K}}=(\rho_{_{\boldsymbol{u}}}\mathbf{I}+\mathbf{G}_{_{K}})^{^{-1}}\bar{ \boldsymbol{\theta}}_{r\mathrm{o}}^{^{K}}$  and, 
hence,
\begin{equation}
\boldsymbol{u}_{_{K}}(\tau)=\mathbf{F}_{_{K}}(\tau )(\rho_{_{\boldsymbol{u}}}\mathbf{I}+\mathbf{G}_{_{K}})^{^{-1}}\bar{ \boldsymbol{\theta}}_{r\mathrm{o}}^{^{K}}, \ \ \tau \in [0,t_{_{F}}].
\end{equation}
\hfill$\blacksquare$

\vspace*{3mm}
\noindent
\underline{\textbf{Proof of Proposition \ref{prop:02}:}} Let $E_{_{\mathcal{T}}}^{^{K}}\triangleq\mathcal{T}_{_{\theta}}[\boldsymbol{u}]-\mathcal{T}_{_{\theta}}[\boldsymbol{u}]^{^{K}}$\ \ and\\ $E_{_{\mathcal{S}}}^{^{K}}(\tau)\triangleq\mathcal{S}_{_{A}}(t_{F}-\tau)-\mathcal{S}_{_{K}}(t_{F}-\tau)P_{_{K}}$ and note that
\begin{eqnarray*}
E_{_{\mathcal{T}}}^{^{K}}&=&\int_{0}^{t_{F}}E_{_{\mathcal{S}}}^{^{K}}(\tau)[\boldsymbol{\beta}_{_{\boldsymbol{S}}}^{^{\mathrm{T}}}\boldsymbol{u}(\tau)]\, d\tau=\int_{0}^{t_{F}}\sum_{i=1}^{m}E_{_{\mathcal{S}}}^{^{K}}(\tau)[\boldsymbol{\beta}_{_{\boldsymbol{S}i}}\boldsymbol{u}_{_{i}}(\tau)]\, d\tau\\
\Rightarrow\ \ E_{_{\mathcal{T}}}^{^{K}}&=&\sum_{i=1}^{m}\int_{0}^{t_{F}}E_{_{\mathcal{S}}}^{^{K}}(\tau)[\boldsymbol{\beta}_{_{\boldsymbol{S}i}}]\boldsymbol{u}_{_{i}}\, d\tau\ \ \ \text{so that}
\end{eqnarray*}
\begin{eqnarray*}
 \|E_{_{\mathcal{T}}}^{^{K}}\|_{_{L_{2}}(\Omega)}&\leq& \sum_{i=1}^{m}\int_{0}^{t_{F}}\left\|E_{_{\mathcal{S}}}^{^{K}}(\tau)[\boldsymbol{\beta}_{_{\boldsymbol{S}i}}]\boldsymbol{u}_{i}(\tau)\right\|_{_{L_{2}}(\Omega)}\, d\tau\ \ \ \Rightarrow\\
\|E_{_{\mathcal{T}}}^{^{K}}\|_{_{L_{_{2}}(\Omega)}}&\leq& \sum_{i=1}^{m}\int_{0}^{t_{F}} \boldsymbol{h}_{_{i}}^{^{K}}(\tau)|\boldsymbol{u}_{_{i}}(\tau)|\, d\tau,
\end{eqnarray*}
where\ \ $\boldsymbol{h}_{_{i}}^{^{K}}(\tau)\triangleq\left\|E_{_{\mathcal{T}}}^{^{K}}(\tau)[\boldsymbol{\beta}_{_{\boldsymbol{S}i}}]\right\|_{_{L_{2}}(\Omega)}$. It then follows (Cauchy-Schwarz inequality on $L_{_{2}}(0, t_{F})$) that
\begin{eqnarray*}
\|E_{_{\mathcal{T}}}^{^{K}}\|_{_{L_{2}(\Omega)}}&\leq& \sum_{i=1}^{m}\|\boldsymbol{h}_{_{i}}^{^{K}}\|_{_{L_{2}(0,t_{F})}}\|\boldsymbol{u}_{_{i}}\|_{_{L_{2}(0,t_{F})}}\ \ \ \Rightarrow\ \ (\text{idem on}\ \mathbb{R}^{m})\\
\|E_{_{\mathcal{T}}}^{^{K}}\|_{_{L_{2}(\Omega)}}&\leq&\left\{\sum_{i=1}^{m}\|\boldsymbol{h}_{_{i}}^{^{K}}\|_{_{L_{2}(0, t_{F})}}^{^{2}}\right\}^{^{1/2}}\left\{\sum_{i=1}^{m}\|\boldsymbol{u}_{_{i}}\|_{_{L_{2}(0,t_{F})}}^{^{2}}\right\}^{^{1/2}}\ \ \Rightarrow\\
\|E_{_{\mathcal{T}}}^{^{K}}\|_{_{L_{2}(\Omega)}}&\leq&\eta_{_{\mathcal{T}}}^{^{K}}\|\boldsymbol{u}\|_{_{L_{2}(0,t_{F})^{m}}},
\end{eqnarray*}
where\ $\eta_{_{\mathcal{T}}}^{^{K}}\triangleq\left\{\sum_{i=1}^{m}\|\boldsymbol{h}_{_{i}}^{^{K}}\|_{_{L_{2}(0, t_{F})}}^{^{2}}\right\}^{^{1/2}}$. This proves Proposition \ref{prop:02}(\textbf{a}).

Moreover, in the light of [10], Theorem 5.2, $\forall i=1, \mathellipsis,m$, $\boldsymbol{h}_{_{i}}^{^{K}}\to0$ as $K\to\infty$. Thus, $\eta_{_{\mathcal{T}}}^{^{K}}\to 0$ as $K\to \infty$.\hfill$\blacksquare$

\break
\noindent
\underline{\textbf{Proof of Corollary 2.1:}} Note that
$\mathcal{J}_{_{K}}( \boldsymbol{u})=\rho_{_{\boldsymbol{u}}}\|\boldsymbol{u}\|_{_{L_{2}(0, t_{_{F}})}}^{^{2}}+
\|\mathcal{T}_{_{\theta}}[\boldsymbol{u}]-\theta_{r\mathrm{o}}-
(\mathcal{T}_{_{\theta}}[\boldsymbol{u}]-\mathcal{T}_{_{\theta}}^K[\boldsymbol{u}])\|_{_{2}}^{^{2}}\ \ \Longleftrightarrow $\\
$\mathcal{J}_{_{K}}(\boldsymbol{u})=\mathcal{J}(\boldsymbol{u})+\|\mathcal{T}_{_{\theta}}[\boldsymbol{u}]-
\mathcal{T}_{_{\theta}}^K[\boldsymbol{u}]\|_{_{2}}^{^{2}}-
2\langle \mathcal{T}_{_{\theta}}[\boldsymbol{u}]-\theta_{r\mathrm{o}}, \mathcal{T}_{_{\theta}}[\boldsymbol{u}]-
\mathcal{T}_{_{\theta}}^{^{K}}[\boldsymbol{u}]\rangle$.

As a result, with
$E^{^{K}}_{_{\mathcal{J}}}(\boldsymbol{u})\triangleq \mathcal{J}( \boldsymbol{u} )- \mathcal{J}_{_{K}}(\boldsymbol{u})$,  it follows
from Proposition \ref{prop:02} that
\begin{equation}\label{4eq:25}
|E^{^{K}}_{_{\mathcal{J}}}(\boldsymbol{u})|\leq (\eta_{_{\mathcal{T}}}^{^{K}})^{^{2}}\|\boldsymbol{u}\|^{^{2}}_{_{L_{2}(0,t_{_{F}})^{^{m}}}}+
2\|\mathcal{T}_{_{\theta}}[\boldsymbol{u}]-\theta_{r\mathrm{o}}\|_{_{2}}(\eta_{_{\mathcal{T}}}^{^{K}})\|\boldsymbol{u}\|_{_{L_{2}(0, t_{_{F}})^{^{m}}}}.
\end{equation}

\noindent
On the other hand,\\
$\mathcal{J}_{_{K}}(\boldsymbol{u}_{_{K}})\leq \mathcal{J}_{_{K}}(\boldsymbol{u}_{_{\mathrm{o}}})=\mathcal{J}(\boldsymbol{u}_{_{\mathrm{o}}})-
E_\mathcal{J}^{^{K}}(\boldsymbol{u}_{_{\mathrm{o}}}) \Longleftrightarrow $
$\mathcal{J}(\boldsymbol{u}_{_{K}})-E_\mathcal{J}^{^{K}}(\boldsymbol{u}_{_{K}})\leq \mathcal{J}(\boldsymbol{u}_{_{\mathrm{o}}})
-E_\mathcal{J}^{^{K}}(\boldsymbol{u}_{_{\mathrm{o}}})$\\
$\Longrightarrow \mathcal{J}(\boldsymbol{u}_{_{K}})
\leq \mathcal{J}(\boldsymbol{u}_{_{\mathrm{o}}})-E_\mathcal{J}^{^{K}}(\boldsymbol{u}_{_{\mathrm{o}}})+E_\mathcal{J}^{^{K}}(\boldsymbol{u}_{_{K}})$ $\Longrightarrow $
$$
\mathcal{J}(\boldsymbol{u}_{_{K}})\leq \mathcal{J}(\boldsymbol{u}_{_{\mathrm{o}}})+|E_\mathcal{J}^{^{K}}(\boldsymbol{u}_{_{\mathrm{o}}})|
+|E_\mathcal{J}^{^{K}}(\boldsymbol{u}_{_{K}})|
$$
$\Longrightarrow$ (since $\mathcal{J}(\boldsymbol{u}_{_{K}})\geq \mathcal{J}(\boldsymbol{u}_{_{\mathrm{o}}})$)
\begin{equation}\label{4eq:26}
 0\leq \mathcal{J}(\boldsymbol{u}_{_{K}})-\mathcal{J}(\boldsymbol{u}_{_{\mathrm{o}}})\leq |E_{\mathcal{J}}^{^{K}}(\boldsymbol{u}_{_{\mathrm{o}}})|+|E_{\mathcal{J}}^{^{K}}(\boldsymbol{u}_{_{K}})|.
\end{equation}

Note also that, as $\eta_{_{\mathcal{T}}}^{^{K}}\rightarrow 0$ (Proposition \ref{prop:02}(b)), it follows from (\ref{4eq:25}) that $|E_{\mathcal{J}}^{^{K}}(\boldsymbol{u}_{_{\mathrm{o}}})|\rightarrow0$.
Moreover, $\{\boldsymbol{u}_{_{K}}\}$ is a bounded sequence -- indeed,
$\|\boldsymbol{u}_{_{K}}\|_{_{L_{2}(0, t_{_{F}})^{^{m}}}}^{^{2}} \leq
\|\boldsymbol{\theta}_{r\mathrm{o}}\|_{_{L_{2}(\Omega)}}^{^{2}}\rho_{_{\boldsymbol{u}}}^{^{-1}}$ for,
if $\|\boldsymbol{u}_{_{K}}\|^{^{2}}>\rho_{_{\boldsymbol{u}}}^{^{-1}}\|{\boldsymbol\theta}_{r\mathrm{o}}\|_{_{L_{2}(\Omega)}}^{^{2}}$ then
$\mathcal{J}_{_{K}}(\boldsymbol{u}_{_{K}})>\|\boldsymbol{\theta}_{r\mathrm{o}}\|_{_{L_{2}(\Omega)}}^{^{2}}=\mathcal{J}_{_{K}}(0)$
in which case $\boldsymbol{u}_{_{K}}$ would not be optimal for \emph{Prob. $I_{_{K}}$}. Thus, as\break
$\mathcal{T}_{_{\theta}}[\boldsymbol{u}]=\displaystyle\int_{0}^{t_{_{F}}}S_{_{A}}(t_{_{F}}-\tau)\left\{\sum_{i=1}^{m}\boldsymbol{\beta}_{_{\boldsymbol{S}i}}
\boldsymbol{u}(\tau)\right\}\, d\tau$, \ \
$\left\{\mathcal{T}_{_{\theta}}[\boldsymbol{u}_{_{K}}]\right\}$ is also bounded and, hence, it follows from (\ref{4eq:25}) that
(as $\eta_{_{\mathcal{T}}}^{^{K}}\rightarrow 0$) $E_{\mathcal{J}}^{^{K}}(\boldsymbol{u}_{_{K}})\rightarrow 0$. Thus,
\begin{equation}\label{4eq:26a}
 \left\{|E_{\mathcal{J}}^{^{K}}(\boldsymbol{u}_{_{\mathrm{o}}})| + |E_{\mathcal{J}}^{^{K}}(\boldsymbol{u}_{_{K}})|\right\}\rightarrow 0
\end{equation}
which together with (\ref{4eq:26}) implies that
$\mathcal{J}(\boldsymbol{u}_{_{K}})\rightarrow \mathcal{J}(\boldsymbol{u}_{_{\mathrm{o}}})$. \hfill$\blacksquare$

\vspace*{3mm}
\noindent
\underline{\textbf{Proof of Proposition \ref{prop:03}:}} Note first that (since $\boldsymbol{u}_{_{\mathrm{o}}}$ is an optimal solution of \emph{Prob. I})
$$
\mathcal{J}(\boldsymbol{u}_{_{K}})=\mathcal{J}(\boldsymbol{u}_{_{\mathrm{o}}}+(\boldsymbol{u}_{_{K}}-\boldsymbol{u}_{_{\mathrm{o}}}))
=\mathcal{J}(\boldsymbol{u}_{_{\mathrm{o}}})+\rho_{_{\boldsymbol{u}}}\|\boldsymbol{u}_{_{K}}-\boldsymbol{u}_{_{\mathrm{o}}}\|_{L_{2}(0, t_{_{F}})^{^{m}} }^2
+\|\mathcal{T}_{_{\theta}}[(\boldsymbol{u}_{_{K}}-\boldsymbol{u}_{_{\mathrm{o}}})]\|_{L_{2}(\Omega) }^{2}.
$$
It then follows from (\ref{4eq:26}) that
\[
\rho_{_{\boldsymbol{u}}}\|\boldsymbol{u}_{_{K}}-\boldsymbol{u}_{_{\mathrm{o}}}\|_{L_{2}(0, t_{_{F}})^{^{m}} }^2
+\|\mathcal{T}_{_{\theta}}[(\boldsymbol{u}_{_{K}}-\boldsymbol{u}_{_{\mathrm{o}}})]\\|_{_{L_{2}(\Omega)}}^{^{2}}\leq |E_{\mathcal{J}}^{^{K}}(\boldsymbol{u}_{_{\mathrm{o}}})|
+|E_{\mathcal{J}}^{^{K}}(\boldsymbol{u}_{_{K}})|\Rightarrow
\]
\[
\rho_{_{\boldsymbol{u}}}\|\boldsymbol{u}_{_{K}}-\boldsymbol{u}_{_{\mathrm{o}}}\|_{_{L_{2}(0, t_{_{F}})^{^{m}}}}^{^{2}}
\leq |E_{\mathcal{J}}^{^{K}}(\boldsymbol{u}_{_{\mathrm{o}}})|+|E_{\mathcal{J}}^{^{K}}(\boldsymbol{u}_K)|.
\]
Thus, in the light of (\ref{4eq:26a}), $\boldsymbol{u}_{_{K}}\rightarrow\boldsymbol{u}_{_{\mathrm{o}}}$ in $L_{2}(0, t_{_{F}})^{^{m}}$.\hfill$\blacksquare$



\vspace*{3mm}
\noindent
\textbf{\underline{Proof of Proposition  \ref{prop:07}:}} \textbf{(\emph{a})} It was established in the proof of Proposition 3.1 that 
$S_{_{\boldsymbol{u}F}}$ is convex and closed. Then, as done in the proof of Proposition 2.1, \emph{Prob. $II_{_{K}}$} is cast as a minimum distance
problem to a convex and closed set so that the existence of $\boldsymbol{u}_{c}^{^{K}}$ follows from ([9], Theorem 3.12.1, p. 69).

\noindent
\textbf{(\emph{b})} Proceeding as in the proof of Proposition \ref{prop:03}, write
\begin{align}\label{5eq:A.5}
\mathcal{J}(\boldsymbol{u}_{c}^{^{K}})=\mathcal{J}(\boldsymbol{u}_{c}+(\boldsymbol{u}_{c}^{^{K}}-\boldsymbol{u}_{c}))
=\mathcal{J}(\boldsymbol{u}_{c})+2\langle \rho_{_{\boldsymbol{u}}}\boldsymbol{u}_{c}
+Z_{_{a}}[\boldsymbol{u}_{c}],(\boldsymbol{u}_{c}^{^{K}}-\boldsymbol{u}_{c})\rangle +\|\rho_{_{\boldsymbol{u}}}(\boldsymbol{u}_{c}^{^{K}}-\boldsymbol{u}_{c})\|_{_{2}}^{^{2}}
+\|\mathcal{T}_{_{\theta}}[\boldsymbol{u}_{c}^{^{K}}-\boldsymbol{u}_{c}]\|_{_{2}}^{^{2}},
\end{align}
where\ $Z_{_{a}}[\boldsymbol{u}]=\mathcal{T}_{_{\theta}}^{*}[\mathcal{T}_{_{\theta}}[\boldsymbol{u}]-
\boldsymbol{\theta}_{r\mathrm{o}}]$\ and note that (as in the derivation of (\ref{4eq:25}))
\begin{align}
\mathcal{J}_{_{K}}(\boldsymbol{u}_{c}^{^{K}})&\leq \mathcal{J}_{_{K}}(\boldsymbol{u}_{c})=\mathcal{J}(\boldsymbol{u}_{c})-E_{_{\mathcal{J}}}^{^{K}}(\boldsymbol{u}_{c}) \Leftrightarrow \nonumber \\
\mathcal{J}(\boldsymbol{u}_{c}^{^{K}})-E_{_{\mathcal{J}}}^{^{K}}(\boldsymbol{u}_{c}^{^{K}})&\leq \mathcal{J}(\boldsymbol{u}_{c})-E_{_{\mathcal{J}}}^{^{K}}(\boldsymbol{u}_{c})\Rightarrow\label{5eq:A.6}\\
\mathcal{J}(\boldsymbol{u}_{c}^{^{K}})&\leq \mathcal{J}(\boldsymbol{u}_{c})-E_{_{\mathcal{J}}}^{^{K}}(\boldsymbol{u}_{c})+E_{_{\mathcal{J}}}^{^{K}}(\boldsymbol{u}_{c}^{^{K}})\Rightarrow \label{5eq:A.7}\\
\mathcal{J}(\boldsymbol{u}_{c}^{^{K}})&\leq \mathcal{J}(\boldsymbol{u}_{c})+|E_{_{\mathcal{J}}}^{^{K}}(\boldsymbol{u}_{c})|+|E_{_{\mathcal{J}}}^{^{K}}(\boldsymbol{u}_{c}^{^{K}})|.\label{5eq:A.8}
\end{align}
Combining (\ref{5eq:A.5}) and (\ref{5eq:A.8}) leads to
\begin{eqnarray*}
\|\rho_{_{\boldsymbol{u}}}(\boldsymbol{u}_{c}^{^{K}}-\boldsymbol{u}_{c})\|_{_{2}}^{^{2}}+\|\mathcal{T}_{_{\theta}}[\boldsymbol{u}_{c}^{^{K}}-\boldsymbol{u}_{c}]\|_{_{2}}^{^{2}}+2\langle \rho_{_{\boldsymbol{u}}}\boldsymbol{u}_{c}+Z_{_{a}}[\boldsymbol{u}_{c}],(\boldsymbol{u}_{c}^{^{K}}-\boldsymbol{u}_{c})\rangle
\leq |E_{_{\mathcal{J}}}^{^{K}}(\boldsymbol{u}_{c})|+|E_{_{\mathcal{J}}}^{^{K}}(\boldsymbol{u}_{c}^{^{K}})|
\end{eqnarray*}
$\Rightarrow$ (in the light of the optimality condition of Proposition  \ref{prop:05})
\[
\rho_{_{\boldsymbol{u}}}\|\boldsymbol{u}_{c}^{^{K}}-\boldsymbol{u}_{c}\|_{_{2}}^{^{2}}\leq |E_{_{\mathcal{J}}}^{^{K}}(\boldsymbol{u}_{c})|+|E_{_{\mathcal{J}}}^{^{K}}(\boldsymbol{u}_{c}^{^{K}})|.
\]
Now it follows from (\ref{4eq:25}) and the fact that $\eta_{_{\mathcal{T}}}^{^{K}}\rightarrow0$ as $K\rightarrow\infty$
(Proposition \ref{prop:02}(b)) that 
$|E_{_{\mathcal{J}}}^{^{K}}(\boldsymbol{u}_{c})|\rightarrow0$ as $K\rightarrow\infty$. Moreover, as $\boldsymbol{u}_{c}^{^{K}}$ is bounded (since 
$\boldsymbol{u}_{c}^{^{K}}\in S_{_{\boldsymbol{u}F}}$ and hence $\|\boldsymbol{u}_{c}^{^{K}}\|_{_{L_{2}(0, t_{_{F}})^{^{m}}}}\leq \left(\sum_{i=1}^{m}\mu_{_{i}}^{^{2}}\right)^{^{1/2}}t_{_{F}})$, (\ref{4eq:25}) and 
``$\eta_{_{\mathcal{T}}}^{^{K}}\rightarrow0$'' also imply that $|E_{_{\mathcal{J}}}^{^{K}}(\boldsymbol{u}_{c}^{^{K}})|\rightarrow0$ as $K\rightarrow\infty$.
 Hence, $\|\boldsymbol{u}_{c}^{^{K}}-\boldsymbol{u}_{c}\|_2\rightarrow0$ as $K\rightarrow \infty$. \hfill $\blacksquare$
 
\vspace*{3mm}
\noindent
\textbf{\underline{Proof of Proposition \ref{prop:5.3}:}}  The optimality condition satisfied by 
$\boldsymbol{u}_{c}^{^{K}}(\boldsymbol{\lambda} )$  is given by 
\begin{align}
&\forall \delta_{_{\boldsymbol{u}}}\in L_{2}(0, t_{_{F}})^{^{m}}, \ Lag_{_{K}}(\boldsymbol{u}, \boldsymbol{\lambda})
\leq Lag_{_{K}}(\boldsymbol{u}_{c}^{^{K}}+\delta_{_{\boldsymbol{u}}}, \boldsymbol{\lambda})\ \ \Leftrightarrow\nonumber\\
&\forall \delta_{_{\boldsymbol{u}}}\in L_{2}(0, t_{_{F}})^{^{m}},\ \langle\rho_{_{\boldsymbol{u}}}\boldsymbol{u}_{c}^{^{K}}, \delta_{_{\boldsymbol{u}}}\rangle
 +\langle\mathcal{T}_{_{\theta}}^{^{K}}[\boldsymbol{u}_{c}^{^{K}}]-\boldsymbol{\theta}_{r\mathrm{o}}, \mathcal{T}_{_{\theta}}^{^{K}}[\delta_{_{\boldsymbol{u}}}]\rangle+
 \langle\boldsymbol{\lambda}_{a}, -\delta_{_{\boldsymbol{u}}}\rangle+\langle\boldsymbol{\lambda}_{b}, \delta_{_{\boldsymbol{u}}}\rangle=0\ \ 
 \Leftrightarrow\nonumber\\ 
&\ \ \ \ \ \ \ \ \ \ \ \ \ \ \ \ \ \ \ \ \ \ \ \ \ \rho_{_{\boldsymbol{u}}}\boldsymbol{u}+(\mathcal{T}_{_{\theta}}^{^{K}})^{*}[\mathcal{T}_{_{\theta}}^{^{K}}[\boldsymbol{u}]
-\boldsymbol{\theta}_{r\mathrm{o}}]+(\boldsymbol{\lambda}_{b}-\boldsymbol{\lambda}_{a})=0\label{5eq:A.9}
\end{align}
or, equivalently, taking orthogonal projections  $\boldsymbol{u}^{^{1}}$ and $\boldsymbol{u}^{^{2}}$  of $\boldsymbol{u}$ on 
$(\mathcal{T}_{_{\theta}}^{^{K}})^{*}[L_2(\Omega)]$ and on its orthogonal complement,
\[
\rho_{_{\boldsymbol{u}}}\boldsymbol{u}^{^{1}}+(\mathcal{T}_{_{\theta}}^{^{K}})^{\ast}[\mathcal{T}_{_{\theta}}^{^{K}}[\boldsymbol{u}^{^{1}}+\boldsymbol{u}^{^{2}}]]-(\mathcal{T}_{_{\theta}}^{^{K}})^{*}[\boldsymbol{\theta}_{r\mathrm{o}}]-\boldsymbol{\lambda}_{ab}^{^{1}}=0
\]
and $\rho_{_{\boldsymbol{u}}}\boldsymbol{u}^{^{2}}-\boldsymbol{\lambda}_{ab}^{^{2}}=0$ where  $\boldsymbol{\lambda}_{ab}\triangleq  \boldsymbol{\lambda}_{a}-\boldsymbol{\lambda}_{b}$, $\boldsymbol{\lambda}_{ab}^{^{1}}$ and 
$\boldsymbol{\lambda}_{ab}^{^{2}}$ are the corresponding projections of $\boldsymbol{\lambda}_{ab}$.

Noting further that $\mathcal{T}_{_{\theta}}^{^{K}}[\boldsymbol{u}^{^{2}}]=0$  ($\boldsymbol{u}^{^{2}}$ is orthogonal to the range space of 
$(\mathcal{T}_{_{\theta}}^{^{K}})^{\ast}$ and hence is in the null space of  $\mathcal{T}_{_{\theta}}^{^{K}}$)  the equations above can be 
rewritten as
\[
\rho_{_{\boldsymbol{u}}}\boldsymbol{u}^{^{1}}+(\mathcal{T}_{_{\theta}}^{^{K}})^{\ast}[\mathcal{T}_{_{\theta}}^{^{K}}[\boldsymbol{u}^{^{1}}]]
-(\mathcal{T}_{_{\theta}}^{^{K}})^{\ast}[\boldsymbol{\theta}_{r\mathrm{o}}]-\boldsymbol{\lambda}_{ab}^{^{1}}=0
\]
and\ \ \  $\rho_{_{\boldsymbol{u}}}\boldsymbol{u}^{^{2}}=\boldsymbol{\lambda}_{ab}-\boldsymbol{\lambda}_{ab}^{^{1}}$.

Now, $\mathcal{T}_{_{\theta}}^{^{K}}[\boldsymbol{u}]=\displaystyle{\sum^K_{k=1}}c_{_{k}}(t_{_{F}};\boldsymbol{u}) \phi_{_{k}}$ 
and $(\mathcal{T}_{_{\theta}}^{^{K}})^{\ast}[\boldsymbol{w}](\tau )=\mathbf{F}_{_{K}}(\tau)\bar{\boldsymbol{w}}^{^{K}}$,\
where  $\{\phi_{_{k}} ; k=1, \mathellipsis, n_{_{K}}\}$  is an orthogonal basis for $X_{_{K}}$,\
$c_{_{k}}(t_{_{F}};\boldsymbol{u})\triangleq \boldsymbol{e}_k(n_{_{K}})^{^{\mathrm{T}}}\displaystyle\int_{0}^{t_{_{F}}}\mathbf{F}_{_{K}}(\tau)^{^{\mathrm{T}}}\boldsymbol{u}(\tau)d\tau$,
where $\mathbf{F}_{_{K}}(\tau)\triangleq(\mathbf{M}_{_{\boldsymbol{\beta}}}^{^{K}})^{^{\mathrm{T}}}\exp[\mathbf{A}_{_{K}}^{^{\mathrm{T}}}(t_{_{F}}-\tau)]$,\break
$\bar{\boldsymbol{w}}^{^{K}}\triangleq[\langle \boldsymbol{w},\phi_{_{1}} ,\rangle \cdots \langle \boldsymbol{w},\phi_{_{n_{_{K}}}}\rangle]$
and
\[
\mathbf{M}_{_{\boldsymbol{\beta}}}^{^{K}}\triangleq \begin{bmatrix}
                                               \langle\boldsymbol{\beta}_{_{\boldsymbol{S}1}}, \phi_{_{1}}\rangle& \cdots & \langle\boldsymbol{\beta}_{_{\boldsymbol{S}m}}, \phi_{_{1}}\rangle\\
                                               \vdots &  & \vdots\\
                                                \langle\boldsymbol{\beta}_{_{\boldsymbol{S}1}}, \phi_{_{n_{_{K}}}}\rangle& \cdots &  \langle\boldsymbol{\beta}_{_{\boldsymbol{S}m}}, \phi_{_{n_{_{K}}}}\rangle\\
                                              \end{bmatrix}.
\]

It follows that $\boldsymbol{u}^{^{1}}=\mathbf{F}_{_{K}}\bar{\boldsymbol{\alpha}}_{c}^{^{K}}$  and 
$\boldsymbol{\lambda}_{ab}^{^{1}}=\mathbf{F}_{_{K}}\bar{\boldsymbol{\alpha}}_{_{\boldsymbol{\lambda}}}^{^{K}}$ and, hence, the equation involving 
$\boldsymbol{u}^{^{1}}$  above can be written as
\begin{equation}\label{5eq:A.10}
\mathbf{F}_{_{K}}\left\{\rho_{_{\boldsymbol{u}}}\bar{\boldsymbol{\alpha}}_{c}^{^{K}}
+\bar{\boldsymbol{w}}^{^{K}}_{a}[\bar{\boldsymbol{\alpha}}_{c}^{^{K}}]-\bar{\boldsymbol{\theta}}_{r\mathrm{o}}^{^{K}}-\bar{\boldsymbol{\alpha}}_{_{\boldsymbol{\lambda}}}^{^{K}}\right\}=0,
\end{equation}
where $\bar{\boldsymbol{\theta}}_{r\mathrm{o}}^{^{K}}\triangleq
[\langle \boldsymbol{\theta}_{r\mathrm{o}},\theta_{_{1}}\rangle \cdots \langle \boldsymbol{\theta}_{r\mathrm{o}},\theta_{n_{_{K}}}\rangle ]^{^{\mathrm{T}}}$  and
\[
\bar{\boldsymbol{w}}^{^{K}}_{a}[\bar{\boldsymbol{\alpha}}_{c}^{^{K}}]\triangleq \left[\langle \mathcal{T}_{_{\theta}}^{^{K}}[\boldsymbol{u}^{^{1}}],\phi_{_{1}}\rangle \cdots \langle \mathcal{T}_{_{\theta}}^{^{K}}[\boldsymbol{u}^{^{1}}],\phi_{_{n_{_{K}}}}\rangle \right]^{^{\mathrm{T}}}
\]
\emph{i.e.},\\
$\bar{\boldsymbol{w}}[\bar{\boldsymbol{\alpha}}_{c}^{^{K}}]
=[c_{_{1}}(t_{_{F}};\boldsymbol{u}^{^{1}})\cdots c_{_{n_{_{K}}}}(t_{_{F}};\boldsymbol{u}^{^{1}})]^{^{\mathrm{T}}}
=\displaystyle\int_{0}^{t_{_{F}}}\mathbf{F}_{_{K}}(\tau)^{^{\mathrm{T}}}\boldsymbol{u}^{^{1}}(\tau )d\tau=\left\{\displaystyle\int_{0}^{t_{_{F}}}\mathbf{F}_{_{K}}(\tau)^{^{\mathrm{T}}}\mathbf{F}_{_{K}}(\tau)d\tau\right\}\bar{\boldsymbol{\alpha}}_{c}^{K}$\\
$\Leftrightarrow \bar{\boldsymbol{w}}^{^{K}}_{a}[\bar{\boldsymbol{\alpha}}_{c}^{^{K}}]=\mathbf{G}_{_{K}}\bar{\boldsymbol{\alpha}}_{c}^{^{K}}$ and
$\mathbf{G}_{_{K}}\triangleq \displaystyle\int_{0}^{t_{_{F}}}\mathbf{F}_{_{K}}(\tau)^{^{\mathrm{T}}}\mathbf{F}_{_{K}}(\tau)d\tau$.

A sufficient condition for (\ref{5eq:A.10}) to be satisfied is then given by
\[
\rho_{_{\boldsymbol{u}}}\bar{\boldsymbol{\alpha}}_{c}^{^{K}}+\mathbf{G}_{_{K}}\bar{\boldsymbol{\alpha}}_{c}^{^{K}}
=\bar{\boldsymbol{\theta}}_{r\mathrm{o}}^{^{K}}+\bar{\boldsymbol{\alpha}}_{_{\boldsymbol{\lambda}}}^{^{K}}\ \Leftrightarrow \ 
\bar{\boldsymbol{\alpha}}_{c}^{^{K}}=(\rho_{_{\boldsymbol{u}}}\mathbf{I}+\mathbf{G}_{_{K}})^{^{-1}}(\bar{\boldsymbol{\theta}}_{r\mathrm{o}}^{^{K}}
+\bar{\boldsymbol{\alpha}}_{_{\boldsymbol{\lambda}}}^{^{K}}).
\]
It then follows that $\boldsymbol{u}_{c}^{^{K}}[\boldsymbol{\lambda} ]$ is given by (since $\rho_{_{\boldsymbol{u}}}\boldsymbol{u}^{^{2}}=\boldsymbol{\lambda}_{ab}^{^{2}}$ )
\begin{eqnarray*}
\boldsymbol{u}_{c}^{^{K}}[\boldsymbol{\lambda} ]&=&\mathbf{F}_{_{K}}\bar{\boldsymbol{\alpha}}_{c}^{^{K}}
+\rho_{_{\boldsymbol{u}}}^{^{-1}}(\boldsymbol{\lambda}_{ab}-\mathbf{F}_{_{K}}\bar{\boldsymbol{\alpha}}_{_{\boldsymbol{\lambda}}}^{^{K}}) \Leftrightarrow \nonumber \\
\boldsymbol{u}_{c}^{^{K}}[\boldsymbol{\lambda} ](\tau )&=&\mathbf{F}_{_{K}}(\tau)(\bar{\boldsymbol{\alpha}}_{c}^{^{K}}-\rho_{_{\boldsymbol{u}}}^{^{-1}}\bar{\boldsymbol{\alpha}}_{_{\boldsymbol{\lambda}}}^{^{K}})+\rho_{_{\boldsymbol{u}}}^{^{-1}}\boldsymbol{\lambda}_{ab}(\tau ) \Leftrightarrow \nonumber \\
\boldsymbol{u}_{c}^{^{K}}[\boldsymbol{\lambda} ](\tau )&=&\boldsymbol{u}_{K}(\tau )+
\mathbf{F}_{_{K}}(\tau)\left\{(\rho_{_{\boldsymbol{u}}}\mathbf{I}+\mathbf{G}_{_{K}})^{^{-1}}-\rho_{_{\boldsymbol{u}}}^{^{-1}}\mathbf{I}\right\}\bar{\boldsymbol{\alpha}}_{_{\boldsymbol{\lambda}}}^{^{K}}+\rho_{_{\boldsymbol{u}}}^{^{-1}}\boldsymbol{\lambda}_{ab}(\tau ).\nonumber 
\end{eqnarray*}
\hfill $\blacksquare$

\bigskip

\noindent
\underline{\textbf{Piecewise-linear Lagrange multipliers}}

To compute approximate solutions to $Prob.\ I_{c}$, consider the truncated problem\\
$\text{\underline{\emph{Prob.}}\ $I_{_{cK}}$}:\displaystyle\min_{\boldsymbol{u}\in S_{_{\boldsymbol{u}F}}}
\check{\mathcal{J}}_{_{K}}(\boldsymbol{u}; \rho_{_{F}})$ and the corresponding dual problem,\\ 
$\text{\underline{\emph{Prob.}}\ $D_{_{K}}$}:\displaystyle\max_{\boldsymbol{\lambda}_{a}, \boldsymbol{\lambda}_{b}} 
\varphi_{_{D}}^{^{K}}(\boldsymbol{\lambda}_{a}, \boldsymbol{\lambda}_{b}; \rho_{_{F}})$ subject to $\forall t$ a.e. in $(0, t_{_{F}})$, 
$\boldsymbol{\lambda}_{a}\geq 0$, $\boldsymbol{\lambda}_{b}\geq 0$,\\
where $\varphi_{_{D}}^{^{K}}(\boldsymbol{\lambda}_{a}, \boldsymbol{\lambda}_{b})
=\inf\{Lag_{_{K}}(\boldsymbol{u}; \boldsymbol{\lambda}_{a}, \boldsymbol{\lambda}_{b}): \boldsymbol{u}\in L_{2}(0, t_{_{F}})\}$,\\
$Lag_{_{K}}(\boldsymbol{u}; \boldsymbol{\lambda}_{a}, \boldsymbol{\lambda}_{b})= 
\check{\mathcal{J}}_{_{K}}(\boldsymbol{u}; \rho_{_{F}})+2\langle\boldsymbol{\lambda}_{a}, \boldsymbol{u}_{a}-\boldsymbol{u}\rangle
+2\langle\boldsymbol{\lambda}_{b}, \boldsymbol{u}-\boldsymbol{u}_{b}\rangle$ and $\boldsymbol{u}_{b}=\mu_{_{\boldsymbol{u}}}$ and
$\boldsymbol{u}_{a}=-\mu_{_{\boldsymbol{u}}}$, and\break
$S_{_{\boldsymbol{u}F}}=\{\boldsymbol{u}\in L_{2}(0, t_{_{F}}): \forall t \ \text{a.e. in} 
(0, t_{_{F}}), -\mu_{_{\boldsymbol{u}}}\leq \boldsymbol{u}(t)\leq \mu_{_{\boldsymbol{u}}} \}$.

The unique solution to the problem $\displaystyle\min_{\boldsymbol{u}\in L_{2}(0, t_{_{F}})} 
Lag_{_{K}}(\boldsymbol{u}; \boldsymbol{\lambda}_{a}, \boldsymbol{\lambda}_{b})$ is given by\
$\boldsymbol{u}_{c}^{^{K}}[\boldsymbol{\lambda}]=\hat{\boldsymbol{u}}_{_{K}}^{c}+ \boldsymbol{\lambda}_{ab}$,\ where\ 
$\check{\boldsymbol{u}}_{_{K}}^{c}[\boldsymbol{\lambda}](\tau)=
\mathbf{H}_{_{K}}^{^{\mathrm{T}}}(t_{_{F}} - \tau)\left\{\bar{\boldsymbol{\alpha}}_{_{K}}
-(\mathbf{I}+\rho_{_{F}}\mathbf{G}_{_{K}})^{^{-1}}\rho_{_{F}}\boldsymbol{\xi}_{_{\boldsymbol{\lambda}}}^{^{K}}\right\}$,\ 
$\boldsymbol{\lambda}=(\boldsymbol{\lambda}_{a}, \boldsymbol{\lambda}_{b})$, 
$\boldsymbol{\lambda}_{ab}=\boldsymbol{\lambda}_{a}-\boldsymbol{\lambda}_{b}$ and\break
$\boldsymbol{\xi}_{_{\boldsymbol{\lambda}}}^{^{K}}
=\displaystyle\int_{0}^{t_{_{F}}}\mathbf{H}_{_{K}}(t_{_{F}}-\tau)\boldsymbol{\lambda}_{ab}(\tau)d\tau$.

The corresponding value for the dual functional is given by 
$$\varphi_{_{D}}^{^{K}}(\boldsymbol{\lambda}_{a}, \boldsymbol{\lambda}_{b})
= Lag_{_{K}}(\boldsymbol{u}_{c}^{^{K}}[\boldsymbol{\lambda}]; \boldsymbol{\lambda}_{a}, \boldsymbol{\lambda}_{b})
=\rho_{_{F}}\|\theta_{r\mathrm{o}}\|_{_{2}}^{^{2}}+\rho_{_{F}}\langle\mathcal{T}_{_{\theta}}^{^{K}}[\boldsymbol{u}_{c}^{^{K}}], -\theta_{r\mathrm{o}} \rangle
+\hat{\varphi}_{_{D}}^{^{K}}(\boldsymbol{\lambda}_{a}, \boldsymbol{\lambda}_{b}).$$
Note that for any non-negative $\boldsymbol{\lambda}_{a}$ and $\boldsymbol{\lambda}_{b}$,
$\varphi_{_{D}}^{^{K}}(\boldsymbol{\lambda}_{a}, \boldsymbol{\lambda}_{b})$ is a lower bound for the optimal value of 
$Prob. \ I_{_{cK}}$.

To obtain approximate solutions to $Prob.\ D_{_{K}}$, piecewise linear classes of multipliers are considered, \emph{i.e.}, let 
$N_{_{\boldsymbol{\lambda}}}\in \mathbb{Z}_{+}$, $\delta_{_{t}}=t_{_{F}}/N_{_{\boldsymbol{\lambda}}}$, 
$\mathcal{I}_{_{k}}=[(k-1)\delta_{_{t}}, k\delta_{_{t}}]$, $\boldsymbol{\gamma}=[\gamma_{_{1}} \cdots \gamma_{_{N_{_{\boldsymbol{\lambda}}}+1}}]$ 
and define $\forall k=1, \mathellipsis, N_{_{\boldsymbol{\lambda}}}$, \ \ $\forall t \in \mathcal{I}_{_{k}}$,\ \ 
$\boldsymbol{\lambda}(t; \boldsymbol{\gamma})=\gamma_{_{k}}+(1/\delta_{_{t}})(\gamma_{_{k+1}}-\gamma_{_{k}})\Delta t_{_{k}}$, \
where $\Delta t_{_{k}}= t-(k-1)\delta_{_{t}}$ (note that $\gamma_{_{k}}$ and $\gamma_{_{k+1}}$ are respectively the values of 
$\boldsymbol{\lambda}(t, \boldsymbol{\gamma})$ at the lower and upper extreme points of the interval $\mathcal{I}_{_{k}}$). Such 
multipliers can then be written as a function of $\boldsymbol{\gamma}$ as follows:
$$ \forall t\in \mathcal{I}_{_{k}}, \ \ \boldsymbol{\lambda}(t; \boldsymbol{\gamma})
=\boldsymbol{h}_{_{kab}}^{^{\mathrm{T}}}(t) \boldsymbol{E}_{_{k}}\boldsymbol{\gamma},$$
where $\boldsymbol{h}_{_{kab}}^{^{\mathrm{T}}}(t)=[h_{_{ka}}(t)\ \  \vdots\ \  h_{_{kb}}(t)]$, \  
$\boldsymbol{E}_{_{k}}^{^{\mathrm{T}}}=[e_{k}(m_{_{\gamma}})\ \  \vdots\ \  e_{k+1}(m_{_{\gamma}})], \ \ m_{_{\gamma}}=N_{_{\boldsymbol{\lambda}}}+1$, 
$h_{_{ka}}: \mathcal{I}_{_{k}}\rightarrow \mathbb{R}$,\break $h_{_{ka}}(t)=1-h_{_{kb}}(t)$, \
$h_{_{kb}}: \mathcal{I}_{_{k}}\rightarrow \mathbb{R}$, \  $h_{_{kb}}(t)=(1/\delta_{_{t}})(t-a_{_{k}})$, where $a_{_{k}}=(k-1)\delta_{_{t}}$.

As a result, $\boldsymbol{\xi}_{_{\boldsymbol{\lambda}}}^{^{K}}=
\boldsymbol{T}_{\boldsymbol{\xi}\boldsymbol{\gamma}}(\boldsymbol{\gamma}_{a}-\boldsymbol{\gamma}_{b})$, where 
$\boldsymbol{T}_{\boldsymbol{\xi}\boldsymbol{\gamma}}=
\left\{\displaystyle\sum_{k=1}^{N_{_{\boldsymbol{\lambda}}}}
\int_{\mathcal{I}_{_{k}}}\mathbf{H}_{_{K}}(t_{f}-\tau) \boldsymbol{h}_{_{kab}}^{^{\mathrm{T}}}(\tau)d\tau \right\}\boldsymbol{E}_{_{k}}$ 
and
$$-\hat{\varphi}_{_{D}}^{^{K}}(\boldsymbol{\lambda}_{a}, \boldsymbol{\lambda}_{b})
=\boldsymbol{\gamma}_{ab}^{^{\mathrm{T}}}\left(\boldsymbol{P}_{_{\boldsymbol{\gamma}}}-\boldsymbol{T}_{_{\boldsymbol{\xi \gamma}}}^{^{\mathrm{T}}}
\rho_{_{F}}(\mathbf{I}+\rho_{_{F}}\mathbf{G}_{_{K}})^{^{-1}}\boldsymbol{T}_{\boldsymbol{\xi\gamma}}\right)\boldsymbol{\gamma}_{ab}
+2\bar{\boldsymbol{\alpha}}_{_{K}}^{^{\mathrm{T}}}\boldsymbol{T}_{_{\boldsymbol{\xi \gamma}}} \boldsymbol{\gamma}_{ab}
-2\boldsymbol{r}_{_{\boldsymbol{\gamma}a}}^{^{\mathrm{T}}}\boldsymbol{\gamma}_{a}
+2\boldsymbol{r}_{_{\boldsymbol{\gamma}b}}^{^{\mathrm{T}}}\boldsymbol{\gamma}_{b},$$
where\ 
$\boldsymbol{\gamma}_{ab}\triangleq\boldsymbol{\gamma}_{a}-\boldsymbol{\gamma}_{b}$, \ \ 
$\boldsymbol{P}_{\boldsymbol{\gamma} }\triangleq\displaystyle\sum_{k=1}^{N_{_{\boldsymbol{\lambda}}}}\boldsymbol{E}_{_{k}}^{^{\mathrm{T}}}
\int_{\mathcal{I}_{_{k}}}\boldsymbol{h}_{_{kab}}(t) \boldsymbol{h}_{_{kab}}^{^{\mathrm{T}}}(t)dt\boldsymbol{E}_{_{k}}$,\ 
$\boldsymbol{r}_{_{\boldsymbol{\gamma}a}}^{^{\mathrm{T}}}
=\displaystyle\sum_{k=1}^{N_{_{\boldsymbol{\lambda}}}}
\left\{\left[\int_{\mathcal{I}_{_{k}}}\boldsymbol{u}_{a}(t)\boldsymbol{h}_{_{kab}}^{^{\mathrm{T}}}(t)dt \right]\boldsymbol{E}_{_{k}}\right\}$,\ and \
$\boldsymbol{r}_{_{\boldsymbol{\gamma}b}}^{^{\mathrm{T}}}
=\displaystyle\sum_{k=1}^{N_{_{\boldsymbol{\lambda}}}}
\left\{\left[\int_{\mathcal{I}_{_{k}}}\boldsymbol{u}_{b}(t)\boldsymbol{h}_{_{kab}}^{^{\mathrm{T}}}(t)dt \right]\boldsymbol{E}_{_{k}}\right\}$.\\

The problem to be numerically solved is then
\begin{equation}
 \text{\underline{\emph{Prob.} $D_{\boldsymbol{\gamma}}^{^{K}}:$}}\ 
 \displaystyle\max_{\boldsymbol{\gamma}_{a}, \boldsymbol{\gamma}_{b} \in \mathbb{R}^{^{N_{_{\boldsymbol{\lambda}}}+1}}}
 \varphi_{_{D}}^{^{K}}(\boldsymbol{\lambda}_{a}(\boldsymbol{\gamma}_{a}), \boldsymbol{\lambda}_{b}(\boldsymbol{\gamma}_{b}); \rho_{_{F}})\ \text{subject to:}\ \boldsymbol{\gamma}_{a}>0, \boldsymbol{\gamma}_{b}>0.
\end{equation}

\end{document}